\documentclass[showpacs,superscriptaddress,reprint,onecolumn,nofootinbib]{article}
\pdfoutput=1
\usepackage{amsmath}
\usepackage{amsbsy}
\usepackage{amssymb}
\usepackage{graphicx}
\usepackage{chngcntr} 
\counterwithout{figure}{section}

\usepackage{geometry}
\usepackage{rotating}
\usepackage[citestyle=numeric-comp, backend=bibtex,sorting=none]{biblatex}
\bibliography{biblio_draft_v2}

\usepackage[affil-it]{authblk}

\usepackage[english]{babel}
\usepackage[T1]{fontenc}
\usepackage[utf8]{inputenc}
\IfFileExists{lmodern.sty}{\usepackage{lmodern}}{}
\usepackage{xcolor}
\definecolor{darkblue}{rgb}{0,0,0.6}
\definecolor{darkred}{rgb}{0.6,0,0}
\definecolor{darkgreen}{rgb}{0,0.6,0}
\usepackage{setspace}
\usepackage{environ}
\usepackage{wasysym}
\usepackage{times}
\numberwithin{equation}{section}
\usepackage[colorlinks=true,urlcolor=darkblue,citecolor=darkblue,linkcolor=darkblue,hyperindex=true,hyperfootnotes=false]{hyperref}
\usepackage[makeroom]{cancel}
\usepackage{multirow}
\usepackage{enumerate}

\newcommand{\bfLL}{\mathbf{\Lambda}}

\newcommand{\bi}{\hat{\bf{i}}}

\newcommand{\cD}{\mathcal{D}}

\newcommand{\cH}{\mathcal{H}}

\newcommand{\cL}{\mathcal{L}}
\newcommand{\cP}{\mathcal{P}}

\newcommand{\cO}{\mathcal{O}}

\newcommand{\bet}{\boldsymbol{\eta}}

\newcommand{\eps}{\varepsilon}

\newcommand\bfr{{\bf r}}

\newcommand\bfq{{\bf q}}
\newcommand\bfu{{\bf u}}
\newcommand\bfm{{\bf m}}
\newcommand\bfe{{\bf e}}
\newcommand{\bfa}{{\bf{a}}}
\newcommand\bfv{{\bf v}}

\usepackage{tikz}

\newcommand{\mbS}{\mathbb{S}}
\newcommand{\rmd}{\mathrm{d}}
\newcommand{\wh}[1]{\widehat{#1}}
\newcommand{\bI}{{\bf{I}}}

\newcommand\llangle{\left\langle}
\newcommand\rrangle{\right\rangle}

\newcommand{\bPi}{\mathbf{\Pi}}
\newcommand{\br}{\bold{r}}
\newcommand{\ba}{\bold{a}}
\newcommand{\bu}{\bold{u}}

\newcommand{\be}{\bold{e}}
\newcommand{\bm}{\bold{m}}
\newcommand{\bS}{\bold{S}}
\newcommand{\bT}{\bold{T}}
\newcommand{\bb}{\bold{b}}
\newcommand{\bA}{\bold{A}}
\newcommand{\bY}{\bold{Y}}
\newcommand{\bxi}{\boldsymbol{\xi}}

\newcommand{\bLambda}{\boldsymbol{\Lambda}}

\newcommand{\gradr}{\nabla_{\br}}
\newcommand{\grade}{\nabla_{\be}}
\newcommand{\divr}{\nabla_{\br}\cdot}
\newcommand{\dive}{\nabla_{\be}\cdot}
\newcommand{\lape}{\Delta_{\be}}
\newcommand{\lapu}{\Delta_{\bu}}

\newcommand{\mbN}{\mathbb{N}}
\newcommand{\mbZ}{\mathbb{Z}}
\newcommand{\mbR}{\mathbb{R}}
\newcommand{\mbL}{\mathbb{L}}

\newcommand{\cG}{\mathcal{G}}
\newcommand{\mcH}{\mathcal{H}}
\newcommand{\mcO}{\mathcal{O}}
\newcommand{\mcP}{\mathcal{P}}

\newcommand{\ie}{\textit{i.e. }}
\newcommand{\hu}[1]{\wh{\bu^{\otimes #1}}}

\newcommand{\harm}[1]{\widehat{\bfu^{\otimes #1}}}
\newcommand{\BFE}[1]{\bfe^{\otimes #1}}

\usepackage{pdfpages}

\begin{document}


PURPOSE-LED
PUBLISHING™

PAPER

# Fluctuating hydrodynamics of active particles interacting via taxis and quorum sensing: static and dynamics

To cite this article: Alberto Dinelli et al 2024 J. Phys. A: Math. Theor. **57** 395002

View the article online for updates and enhancements.

## You may also like

- New cluster algebras from old: integrability beyond Zamolodchikov periodicity
  Andrew N W Hone, Wookyung Kim and Takafumi Mase

- Entropic uncertainty relations and entanglement detection from quantum designs
  Yundu Zhao, Shan Huang and Shengjun Wu

- Exact approaches on the string worldsheet
  Saskia Demulder, Sibylle Driezen, Bob Knighton et al.

# Fluctuating hydrodynamics of active particles interacting via taxis and quorum sensing: static and dynamics

Alberto Dinelli[1,*]●, Jérémy O'Byrne[1,2]● and Julien Tailleur[1,3]●

[1] Université Paris Cité, Laboratoire Matière et Systèmes Complexes (MSC), UMR 7057 CNRS, F-75205, 75205 Paris, France
[2] Sorbonne Université, Laboratoire Jean Perrin, 4 place Jussieu, 75005 Paris, France
[3] Department of Physics, Massachusetts Institute of Technology, Cambridge, MA 02139, United States of America

E-mail: alberto.dinelli@u-paris.fr

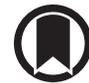

## Abstract

In this article we derive and test the fluctuating hydrodynamic description of active particles interacting via *t*axis and quorum sensing, both for monodisperse systems and for mixtures of co-existing species of active particles. We compute the average steady-state density profile in the presence of spatial motility regulation, as well as the structure factor and intermediate scattering function for interacting systems. By comparing our predictions to microscopic numerical simulations, we show that our fluctuating hydrodynamics correctly predicts the large-scale static and dynamical properties of the system. We also discuss how the theory breaks down when structures emerge at scales smaller or comparable to the persistence length of the particles. When the density field is the unique hydrodynamic mode of the system, we show that active Brownian particles, run-and-tumble particles and active Ornstein–Uhlenbeck particles, interacting via quorum-sensing or chemotactic interactions, display undistinguishable large-scale properties. This form of universality implies an interesting robustness of the predicted physics but also that large-scale observations of patterns are insufficient to assess their microscopic origins. In particular, our results predict that chemotaxis-induced and motility-induced phase separation should share strong qualitative similarities at the macroscopic scale.

Keywords: active matter, fluctuating hydrodynamics, motility regulation, coarse-graining, non-equilibrium statistical mechanics

* Author to whom any correspondence should be addressed.



\newpage

\setcounter{page}{2}
\tableofcontents

\newpage

\section{Introduction}\label{sec:Intro}

Active systems comprise particles able to convert an internal or
ambient source of energy into non-conservative, self-propulsion
forces.  A large variety of self-propulsion mechanisms have been
discovered in nature or engineered in the lab, from self-phoretic
colloids that are powered by chemical or electrostatic energy
sources~\cite{howse2007self,ke2010motion,mano2005bioelectrochemical,ebbens2010pursuit},
to swimming or migrating cells that rely on ATP
consumption~\cite{liu2011sequential,basu2005synthetic,budrene1991complex}. What
makes active matter a unified field is that all these systems
ultimately lead to non-equilibrium persistent random walks that, in
turn, share a common---and rich!---emerging phenomenology. Indeed,
from the emergence of static phase separation in the absence of
attractive
forces~\cite{tailleur2008statistical,liu2011sequential,cates2015motility,bauerle2018self}
to pattern
formation~\cite{cates2010arrested,curatolo2020cooperative,arlt2018painting,frangipane2018dynamic},
laning~\cite{thutupalli2018boundary}, or collective
motion~\cite{vicsek1995novel,bricard2013emergence}, active systems
have access to a much richer set of phases than their passive
counterparts. Understanding---and eventually controlling---the
self-organization of active entities is thus an important open challenge.

From a theoretical point of view, phenomenological approaches provide
generic tools to study the accessible phases of active
systems~\cite{cross_pattern_1993,caussin_emergent_2014,bergmann2018active,saha2020scalar,you2020nonreciprocity,fruchart2021non,frohoff2021suppression},
by postulating symmetry-based field-theoretical descriptions to
describe the system at large scales. Such approaches have been
spectacularly successful in uncovering new phenomena and in assessing
their large scale 
properties~\cite{marchetti_hydrodynamics_2013}. However, purely
macroscopic treatments cannot provide control on actual systems,
lacking a direct relation with the underlying microscopic dynamics. An
important challenge is thus to bridge the gap between microscopic
models and macroscopic behaviors, both to obtain a satisfactory level
of control of engineered systems as well as to assess the scope of
phenomenological field theories. {Establishing such a connection between microscopic and macroscopic scales in active systems is a challenging task, often requiring a different set of tools from one problem to the next. In this work, we build on existing methods in the literature and present a general framework to coarse-grain the dynamics of a large class of `dry scalar active systems'. 

\textit{Dry} systems are those 
where the total conservation of momentum when describing the active particles and their environment does not impact the physics of the active subsystems itself. For instance, dry active systems are naturally relevant to the modelling of shaken grains~\cite{deseigne2010collective} or crawling cells on solid substrates~\cite{sepulveda2013collective}. On the contrary, wet active systems are defined as systems for which the coupling to a momentum-conserving environment has to be explicitly taken into account because the total conservation of momentum plays an important in the observed phenomenology~\cite{marchetti_hydrodynamics_2013}. We note that this terminology can be confusing at times because systems immersed in a viscous solvent need not always be modelled as wet active matter. For instance, dry active matter has been very successful in modelling Quincke rollers~\cite{bricard2013emergence}. The study of wet active systems goes beyond the scope of this work, but we point out that coarse-graining techniques also exist in this case~\cite{saintillan2008instabilities,subramanian2009critical,gao2017analytical,weady2022thermodynamically}. \textit{Scalar} systems are those whose sole large-scale hydrodynamic modes are the conserved density fields. As such, they exclude the ordered phases of systems in which collective motion emerges due to aligning interactions. 

In  this article, we focus on \textit{run-and-tumble particles} (RTPs)~\cite{schnitzer_theory_1993}, \textit{active Brownian particles} (ABPs)~\cite{fily2012athermal} and \textit{active Ornstein-Uhlenbeck particles} (AOUPs)~\cite{sepulveda2013collective,szamel2014self} and consider mediated interactions like \textit{quorum sensing} (QS)~\cite{tailleur2008statistical} and \textit{taxis}~\cite{saha2014clusters}. We restrict our analysis to cases in which these interactions do not lead to long-range order of the particle orientations or velocities. Starting from the microscopic dynamics, we provide a generic framework to describe their large-scale fluctuating hydrodynamics. We note that a lot has been done on the coarse-graining of such systems, especially in low dimensions and for quorum-sensing interactions, and separate accounts can be found in the literature~\cite{tailleur2008statistical,cates_when_2013,solon_active_2015,martin_PRE_2021}.  Here we provide a unified derivation of the fluctuating hydrodynamics of these different models in dimensions $d>1$, for both taxis and quorum-sensing interactions, at the single-species level as well as for mixture. In addition to this unifying perspective, we actually test the predictions of the derived fluctuating hydrodynamics. We do so both at the single-particle steady-state level, but also by computing static and dynamic correlation functions. To the best of our knowledge, this is the first time that the predictive power of such active fluctuating hydrodynamics is demonstrated at the dynamical level. 

The article is organised as follows. In Sec.~\ref{sec:Intro}, we introduce all the models at the single-particle level and then define QS and tactic interactions. In Sec.~\ref{sec:diffusive_limit} we review and generalize the coarse-graining method for non-interacting particles with position-dependent motility parameters. To this purpose, in Sec.~\ref{sec:harmonic-tensors}, we review the basic properties of harmonic tensors, a mathematical tool that we use throughout our coarse-graining. Under the assumption of a large-scale diffusive scaling, we then bridge the gap between the microscopic dynamics and their fluctuating hydrodynamic descriptions. Finally, we test the validity of our approximations against microscopic simulations. In Sec.~\ref{sec:interacting} we restore the interactions between the particles and derive the corresponding fluctuating hydrodynamics. We then test our assumptions and predictions by comparing theoretical expressions for the mean-squared displacement, the static structure factor, and the intermediate scattering function to the results of numerical measurements carried out for active particles interacting via QS. {Finally, in Sec.~\ref{sec:mixtures} we extend our coarse-graining procedure to active mixtures, namely systems composed of a number $S>1$ of interacting active species.}

\subsection{Run-and-tumble particles}
The run-and-tumble dynamics alternates between running phases, during which particles move with a self-propulsion speed $v$ along their orientation $\bfu$, and tumbling phases, during which propulsion stops and the particles randomize their orientations. In the absence of any external perturbation, the self-propulsion speed $v$ is a constant. This simple dynamics is commonly used to model the motion of swimming bacteria such as \textit{E. coli}~\cite{schnitzer_theory_1993,berg2004coli,wilson2011differential,kurzthaler2022characterization}. 

The transition from running to tumbling occurs with a tumbling rate that we denote by $\alpha$. The particle then resumes running with a running rate that we denote by $\beta$, along a new orientation $\bfu'$ that we assume to be sampled uniformly on the unit sphere $\mbS^{d-1}$, where $d$ denote the number of spatial dimensions. For \textit{E. coli}, $\beta \simeq 10 \alpha$~\cite{berg2004coli} and the tumbling phases are typically much shorter than the running ones. In the following we thus consider instantaneous tumbling events, \textit{i.e.} we take the limit $\beta/\alpha \to \infty$. We note that many results derived below can be generalized to finite-duration tumbles by rescaling the propulsion speed as~\cite{tailleur2008statistical,chatterjee2011chemotaxis}:
\begin{equation}
    v \to \frac{v}{\sqrt{1+\alpha/\beta}}\;.
\end{equation}
Accounting also for the possibility of translational noise, the dynamics of a single RTP can thus be modelled as an It\=o-Langevin equation for the position $\bfr$ coupled to a jump process for $\bfu$:
\begin{eqnarray}
    \label{eq:RTPs_dyn}
    \dot{\bfr}(t) &=& v \bfu + \sqrt{2 D_t} \,\boldsymbol{\xi} (t) \\
    \bfu &\overset{\alpha}{\longrightarrow}& \bfu' \in \mbS^{d-1}\;,
    \label{eq:RTPs_u_dyn}
\end{eqnarray}
where $\mbS^{d-1}$ is the unit sphere of $\mbR^d$, $D_t$ is the translational diffusivity, and $\boldsymbol{\xi}(t)$ is a Gaussian white noise satisfying:
\begin{equation}
    \langle \boldsymbol{\xi}(t) \rangle = 0 \quad \text{and} \quad  \langle \xi_i (t) \> \xi_j (t') \rangle = \delta_{ij} \delta(t-t') \ .
\end{equation}
The stochastic dynamics~\eqref{eq:RTPs_dyn} and~\eqref{eq:RTPs_u_dyn} are associated to a master equation for the probability $\cP(\bfr,\bfu,t)$ of finding the particle at position $\bfr$ with an orientation $\bfu$ that reads:
\begin{equation}
\partial_t\cP (\bfr,\bfu,t) = -\nabla_\bfr\cdot \left[ v \bfu \cP(\bfr,\bfu,t) - D_t \nabla_\bfr \cP(\bfr,\bfu,t) \right] - \alpha  \cP(\bfr,\bfu,t) + \frac{1}{\Omega} \int \alpha \cP(\bfr,\bfu,t) \rmd \bfu 
\label{eq:MasterEq_RTP}
\end{equation}
where $\Omega$ is the area of $\mbS^{d-1}$. Finally, the active nature of the dynamics is characterized by the persistence time and length, which are given by
\begin{equation}
    \tau = \frac 1 \alpha \qquad\text{and}\qquad \ell_p= \frac v \alpha\;.
\end{equation}

\subsection{Active Brownian particles}
The sudden reorientations of run-and-tumble particles have often been used to model the dynamics of bacteria and crawling cells. For synthetic active particles~\cite{golestanian2007designing,jiang2010active,deseigne2010collective,theurkauff2012dynamic,palacci2013living,bricard2013emergence}, the evolution of the particle orientation is generally smoother and often modelled using active Brownian particles (ABPs)~\cite{fily2012athermal}, whose orientation undergoes Brownian motion on the sphere $\mbS^{d-1}$. In $2d$, the  dynamics of an ABP reads:
\begin{eqnarray}
    \label{eq:ABPs_2d_dyn}
    \dot{\bfr}(t) &=& v \bfu + \sqrt{2 D_t} \,\boldsymbol{\xi} (t) \\
    \dot{\theta}(t)  &=& \sqrt{2 \Gamma} \eta(t)
    \label{eq:ABPs_u_2d_dyn}
\end{eqnarray}
with $\bfu = (\cos \theta, \sin \theta)$. Here, $\eta, \boldsymbol\xi$ are both Gaussian unitary white noises with zero mean. We note that, for self-diffusiophoretic Janus colloids, the approximation of a constant speed has been shown to be experimentally relevant~\cite{ginot2018sedimentation}.
The corresponding master equation, valid in $d>1$ dimension, reads:
\begin{equation}
\partial_t\cP (\bfr,\bfu,t) = -\nabla_\bfr\cdot \left[ v \bfu \cP(\bfr,\bfu,t) - D_t \nabla_\bfr \cP(\bfr,\bfu,t) \right] + \Delta_\bfu \left[ \Gamma \cP(\bfr,\bfu,t) \right]\;,
\label{eq:MasterEq_ABP}
\end{equation}
where $\bfu\in \mbS^{d-1}$. The persistence time and length of ABPs are given by:
\begin{equation}
    \tau= \frac 1{(d-1)\Gamma}\qquad\text{and}\qquad \ell_p=\frac{v}{\Gamma}\;.
\end{equation}

\subsection{Active Ornstein-Uhlenbeck particles}
\label{sec:AOUP_intro}
In the absence of external perturbations, the self-propulsion speeds of RTPs and ABPs are  constant in time. While this is relevant for many active systems~\cite{berg2004coli,bricard2013emergence,ginot2018sedimentation}, it is sometime important to account for intrinsic fluctuations of the self-propulsion speed, for instance when modelling crawling cells~\cite{sepulveda2013collective}. Active Ornstein-Uhlenbeck particles (AOUPs) have been introduced to this purpose and have since become a workhorse model of active matter~\cite{szamel2014self,szamel2015glassy,marconi2015towards,wittmann2017effective1,wittmann2017effective2,martin_PRE_2021}. The self-propulsion velocity of a single AOUP evolves according to an Ornstein-Uhlenbeck process so that the overall dynamics in $d$ dimensions read:
\begin{eqnarray}
    \label{eq:AOUPs_dyn}
    \dot{\bfr}(t) &=& \bfv(t) + \sqrt{2 D_t} \boldsymbol{\xi} (t) \\
    \tau \dot{\bfv}(t)  &=& - \bfv(t) + \sqrt{2 D_a}\boldsymbol{\eta}(t) \;.
    \label{eq:AOUPs_v_dyn}
\end{eqnarray}
The associated master equation for $\cP(\bfr, \bfv,t)$ is then:
\begin{equation}
\partial_t\cP (\bfr,\bfv,t) = -\nabla_\bfr \cdot \left[ \bfv \cP(\bfr,\bfv,t) - D_t \nabla_\bfr \cP(\bfr,\bfv,t) \right] - \nabla_\bfv \cdot \left\{-\frac{\bfv}{\tau} \cP(\bfr,\bfv,t) - \nabla_\bfv \left[\frac{D_a}{\tau^2}   \cP(\bfr,\bfv,t) \right] \right\} \;.
\label{eq:MasterEq_AOUPs}
\end{equation}
Note that, when $\tau$ and $D_a$ are constants, $\bfv(t)$ is a Gaussian colored noise whose correlations are given by:
\begin{equation}
    \langle v_i(t) v_j(t') \rangle = \delta_{ij} \frac{D_a}{\tau} e^{-\frac{|t-t'|}{\tau}}\;,
\end{equation}
which reduces to a $\delta$-correlated white noise in the limit $\tau \to 0$. In other words, the finite persistence time $\tau$ is responsible for the non-equilibrium nature of AOUP dynamics \cite{martin_PRE_2021}. 

The parameters $\tau$ and $D_a$ that characterize the self-propulsion velocity in Eqs~\eqref{eq:AOUPs_dyn} and~\eqref{eq:AOUPs_v_dyn} have a simple interpretation. $\tau$ is the persistence time of the dynamics and $D_a$ the contribution of the active force to the large-scale translational diffusivity of the particle. In this form, however, the direct comparison to ABPs and RTPs is not immediate. Introducing the typical speed
\begin{equation}
  v=\sqrt{\langle |\bfv|^2\rangle}= \sqrt{\frac{d D_a}\tau}
\end{equation}
and rewriting the velocity as $\bfv = v \be$, one obtains an equivalent formulation of AOUP dynamics:
\begin{eqnarray}\label{eq:dynAOUPsJO}
    \dot{\br} &=& v\be+\sqrt{2D_t}\boldsymbol{\xi} \\
    \dot{\be} &=& - \frac{\be}{\tau} + \sqrt{\frac{2}{d \tau}}\bet\;.
\end{eqnarray}
In this form, the contribution of the active force to the large scale
diffusivity reads $D=v^2\tau/d$, as for ABPs and RTPs. Albeit less
transparent, Eq.~\eqref{eq:dynAOUPsJO} allows us to derive a
universal form below for the large-scale diffusive dynamics of ABPs,
RTPs and AOUPs. Note that, while it is tempting to refer to $\be$ as
an orientation, the magnitude of this dimensionless vector fluctuates {around $\sqrt{\langle |\be|^2\rangle}=1$}.

\subsection{Motility regulation: from directed control to interacting systems}
\label{sec:interactions}

In the simple models introduced above, active particles self-propel
forever with time-translation invariant dynamics. This is of course an
approximate description of any real active system. Indeed, energy
sources may fluctuate in time or be inhomogeneous in space, and the
presence of an active particle typically impacts the propulsion
statistics of its neighbors. We refer to these effects as ``motility
regulation''. In this article, we consider cases in which the dynamics
of active particles are given by the ABP, RTP, and AOUP models but
we allow the microscopic parameters that define these dynamics to vary
in space and time. We refer to these parameters, which include
persistence times or self-propulsion speeds, as ``motility parameters'' and denote them collectively by $\{\gamma\}$. 

In experiments, $\{\gamma\}$ can be controlled externally, for
instance in the case of
synthetic~\cite{buttinoni2013dynamical,palacci2013living} or
biological~\cite{arlt2018painting,frangipane2018dynamic}
light-powered active particles, or as a result of interactions. Here,
we consider cases in which the motility parameters are determined
by some field $c(\bfr)$ that may be imposed externally or produced by
the particles. We denote by `kinesis' the response of the motility
parameters to the value of $c(\bfr)$:
\begin{equation}
  \gamma(\bfr)=\gamma(c(\bfr))\;,
\end{equation}
and by `taxis' their response to $\nabla c(\bfr)$. In
practice, we consider smoothly varying fields and thus restrict taxis to a linear response to $\nabla c(\bfr)$:
\begin{equation}
  \gamma(\bfr,\bfu) \equiv \gamma_0 + \gamma_1 \bfu \cdot \nabla_\bfr c(\bfr) \;.
\end{equation}
Kinesis and taxis have been implemented for light-controlled
self-propelled
colloids~\cite{bauerle2018self,lavergne_group_2019} and
they are frequently met in biological systems, for instance in the
form of quorum sensing~\cite{miller2001QS,hammer2003biofilm,daniels2004swarming},
phototaxis~\cite{polin2009chlamydomonas,arrieta2017phototaxis,drescher2010fidelity}, or
chemotaxis~\cite{berg2004coli,budrene1991complex}. Note that, even in
the biological world, the response of active particles can be designed
by experimentalists, thanks to the progress of synthetic
biology~\cite{liu2011sequential,curatolo2020cooperative}.

To model interacting systems, we consider the case in which $c(\bfr)$ is a chemical field produced by the particles, which can diffuse and degrade in the
environment. The dynamics of $c$ can then be modeled
as~\cite{obyrne_lamellar_2020,obyrne_introduction_2022}:
\begin{equation}
\partial_t c(\bfr, t) = \beta \rho(\bfr,t) + D_c \Delta c(\bfr, t) - \chi c(\bfr, t) \;,
\label{eq:chem_dyn}
\end{equation}
where $\rho$ is the particle density, and $\beta$, $D_c$, $\chi$ are
the production rate, diffusivity and degradation rate of $c$,
respectively. Since $\rho(\bfr, t)$ is a conserved field, its
evolution occurs on a slow, diffusive timescale $T\sim \cO(L^2)$, $L$
being the linear system size. On the contrary, the chemical field $c$
undergoes a fast relaxation with rate $\chi \sim \cO(1)$. Due to this
timescale separation, the field $c(\bfr, t)$ is effectively enslaved
to $\rho(\bfr, t)$, to which it adapts adiabatically. We can thus set
$\partial_t c = 0$ in Eq.~\eqref{eq:chem_dyn} and solve for the
chemical field as:
\begin{equation}\label{eq:Greens}
c(\bfr, t) = \int \rmd^d G(\bfr-\bfr') \rho(\bfr', t) \;,
\end{equation}
where $G$ is the Green's function associated with the linear operator: $\cL \equiv\beta^{-1} \left(\chi - D_c \Delta \right)$. {The equation $\cL G = \delta(\bfr)$ corresponds to a screened Poisson equation, with a solution $G(\bfr)$ that decays to zero with a finite screening length $r_0 \equiv \sqrt{D_c/\chi}$. As a consequence, chemically-mediated interactions are expected to have a finite interaction range when the signalling field evolves according to Eq.~\eqref{eq:chem_dyn}. To provide a biological example, let us consider acyl-homoserine lactones (AHLs), a class of small signalling molecules responsible for quorum-sensing motility regulation in bacteria. Refs.~\cite{pai2009optimal,marenda2016modeling} estimate the decay rate of several AHL molecules to be of the order of $\chi \sim 0.1$---$1$ hr$^{-1}$, while Ref.~\cite{marenda2016modeling} provides a value for the associated diffusivity of the order of $D_c \sim 10^5$---$10^6~\mu$m$^2$/hr. All in all, this gives roughly a screening length $r_0 \equiv \sqrt{D_c/\chi} \sim 0.1$---$1~$mm. Compared to the typical length of an \textit{E. Coli} bacterium $\lessapprox 10~\mu$m~\cite{berg2004coli}, the QS-interaction range $r_0$ can be large compared to the cell size, but small in comparison with the size of a colony or a biofilm~$L \sim 1$---$10$ cm. We note that some signalling fields (e.g. oxygen) will not be degraded~\cite{schnitzer_theory_1993}, hence leading to a power-law decay of the Green's function. The finite detection threshold of the bacteria chemotactic circuits however make these weak power-law tails irrelevant. We thus expect the fast field approximation to apply to a much broader context than the one discussed above.} 

{All in all,} this fast-variable treatment thus allows us to express $c(\bfr)$ as a
functional of the density field: $c(\bfr, [\rho])$. Then, the particle dynamics are biased by the density
field itself, and taxis and kinesis are
respectively modeled as:
\begin{eqnarray}
\gamma(\bfr) &=& \gamma_0 + \gamma_1 \bfu \cdot \nabla_\bfr c(\bfr, [\rho]) \label{eq:CT} \\
\gamma(\bfr) &=& \gamma(\bfr, [\rho])\;, \label{eq:QS} \hspace{3.2cm} 
\end{eqnarray}
Following the biological literature, we will refer to
Eqs~\eqref{eq:CT} and~\eqref{eq:QS} as chemotaxis and quorum sensing,
respectively, even though they describe broader situations. 

\newpage

\if{
micro-organisms interact between themselves and the environment through chemical, electrical, mechanical signals. gradients of chemicals can drive the motion of chemotactic bacteria~\cite{berg2004coli,budrene1991complex}; phototactic \textit{algae} adapt their movements according to light gradients~\cite{witman1993chlamydomonas,drescher2010fidelity}; durotactic and galvanotactic cells are driven by rigidity~\cite{sunyer2020durotaxis} and electric potential gradients~\cite{rapp1988galvanotaxis,allen2013electrophoresis}, respectively. Beyond these biological examples, synthetic systems such as self-propelled colloids can also be engineered in the lab to exhibit tactic behaviours~\cite{You2018intelligent, lozano2016phototaxis}.

The interactions between active particles can lead to the emergence of collective phases without any counterpart at equilibrium. In this review, we study the case of active particles that can regulate each other's motility via \textit{taxis} or \textit{quorum sensing}, two prototypical examples of mediated interactions.

In the biological world, chemically-mediated interactions are the rule: micro-organisms search for nutrients following chemical gradients, or escape regions with high concentrations of toxins. To communicate with each other, bacteria like E. Coli can exchange auto-inducers like acylated homoserine lactones, which in turn regulate the genetic expression in other bacteria.

Tactic particles can adapt their motility parameters $\gamma$ (persistence time, diffusivity, self-propulsion speed...) according to spatial gradients of an external field $\nabla c$. For a particle at position $\bfr$, this means:
\begin{equation}
    \gamma = \gamma(\nabla c(\bfr)) \; .
\end{equation}
In the biological world, taxis can manifest itself in a large variety of forms:
gradients of chemicals - such as food or toxins - can drive the motion of chemotactic bacteria~\cite{berg2004coli,budrene1991complex}; phototactic \textit{algae} adapt their movements according to light gradients~\cite{witman1993chlamydomonas,drescher2010fidelity}; durotactic and galvanotactic cells are driven by rigidity~\cite{sunyer2020durotaxis} and electric potential gradients~\cite{rapp1988galvanotaxis,allen2013electrophoresis}, respectively. Beyond these biological examples, synthetic systems such as self-propelled colloids can also be engineered in the lab to exhibit tactic behaviours~\cite{You2018intelligent, lozano2016phototaxis}. \\
Importantly, there is a separation of timescale between the fast dynamics of the chemical field $c$ dynamics and the slow, conserved dynamics of the density field $\rho$. As a consequence, the chemical field $c$ adiabatically adapts to the density field $\rho$ through $c[\rho]$ \cite{obyrne_lamellar_2020}. Eventually, since the tactic bias is controlled by the density field, we will refer to this mechanism as \textit{taxis}.

quorum sensing (QS), instead, is a mechanism by which active particles regulate their behaviour according the local density of their peers. For one particle at position $\bfr$,
the regulation of motility parameters $\gamma$ via QS takes the form:
\begin{equation}
    \gamma = \gamma(\bfr, [\rho]) \;.
\end{equation}
QS is another example of mediated interaction, where particles communicate through some chemical field. For instance, certain species of bacteria can exchange signalling molecules---so-called autoinducers---to regulate the production of luminescent enzymes \cite{nealson1970luminescence,engebrecht1984lux}. At large bacterial densities, the concentration of autoinducers $c(\bfr)$ is sufficiently large to trigger the transcription of those genes responsible for bio-luminescence. \\
Motility regulation via QS can indeed be observed in bacterial ecosystems, for example in the control of twitching motility~\cite{glessner1999roles} or swarming migration~\cite{daniels2004swarming} in \textit{Pseudomonas aeruginosa}. Furthermore, QS motility control can be engineered in the lab, via orthogonal circuits in clonal bacterial strains \cite{liu2011sequential,curatolo2020cooperative} or via optical feedback systems \cite{bauerle2018self}.}\fi

\if{
Both for taxis and QS, the dynamics of the external field $c$ is typically faster than the particles' one, leading to an adiabatic adaptation of the external field $c$ to the density field $\rho$. For instance, in the case of bacteria interacting with a chemical field $c$, chemicals are produced and consumed by the particles themselves, resulting in an effective coupling between the bacterial and chemical dynamics. A natural model \cite{obyrne_lamellar_2020} for the chemical field dynamics is then to consider diffusion, degradation and production by the bacteria:
\begin{equation}
    \partial_t c(\mathbf{r}) = D_c \Delta c - \lambda c - \beta \sum_i \delta(\mathbf{r}-\mathbf{r_i})
    \label{eq:taxis_chem_field_dyn}
\end{equation}
where $D_c$ is the chemical diffusivity, $\lambda$ is the rate of degradation in the environment, and $\beta$ the production rate by bacteria. Assuming that the chemical dynamics is much faster than the particle's, we can set $\partial_t c = 0$ in (\ref{eq:taxis_chem_field_dyn}) and obtain the concentration field $c(\mathbf{r})$ as a functional of the particle density $\rho(\mathbf{x})$:
\begin{equation}
    c(\mathbf{r}) = \int \> \mathrm{d}^d y \> G(\mathbf{x}-\mathbf{y}) \> \rho(\mathbf{y})
    \label{eq:rho_cr}
\end{equation}
where $G(\mathbf{r})$ is the Green function solution of $(D_c \Delta - \lambda) G(\mathbf{r}) = \beta \delta(\mathbf{r})$. Eventually, at the particle's scale, this means that tactic and QS motility regulations depend on the particle's density field $\rho$ through Eq.~\eqref{eq:rho_cr}: both taxis and quorum sensing can thus be seen as many-body, mediated interactions between the particles. In the following, we will therefore use the term \textit{auto}-taxis (AT) to remark how the tactic bias is controlled by the particles themselves.
}\fi

\section{Non-interacting particles in motility-regulating fields, from micro to macro}
\label{sec:diffusive_limit}

In this section we derive the coarse-grained dynamics of active particles experiencing taxis and kinesis induced by an external field $c(\bfr)$. For such a coarse-graining to make sense, we consider the case where $c(\bfr)$ varies over lengthscales much larger than the particle persistence length $\ell_p$. For clarity, we first derive the diffusive approximation to the large-scale dynamics of ABPs and RTPs in two space dimensions, where the expansion of the angular dependence of the probability density on Fourier modes allows for a simple and transparent treatment. Then, in Section~\ref{sec:ABP-RTP_d}, we consider the general $d$-dimensional case, which relies on using an expansion on spherical harmonic tensors. Section~\ref{sec:AOUP_dDim} discusses the case of AOUPs in $d$ dimensions. At this stage, our diffusive coarse-graining approximates the large-scale dynamics of ABPs, RTPs and AOUPs as effective Langevin equations; in Section~\ref{sec:Dean} we then use It\=o calculus to derive the associated fluctuating hydrodynamics for the density field. Finally, the theoretical predictions of the coarse-grained theory are tested against particle-based simulations in Section~\ref{sec:micro-simul-noninteracting}}.

\subsection{ABPs \& RTPs in two space dimensions}\label{sec:ABP-RTP_2d}
Before tackling the general $d$-dimensional problem, we consider RTPs and ABPs in 2 dimensions. We consider a single active particle endowed with both run-and-tumble~\eqref{eq:RTPs_dyn}-\eqref{eq:RTPs_u_dyn} and active Brownian~\eqref{eq:ABPs_2d_dyn}-\eqref{eq:ABPs_u_2d_dyn} dynamics. In $d=2$ the polarization vector $\bfu$ can be directly expressed as $(\cos\theta, \sin\theta)$, so that the master equation for $\cP(\bfr,\theta)$ reads: 
\begin{equation}
    \partial_t\cP (\bfr,\theta) = -\nabla_\bfr\cdot \left[ v \begin{pmatrix} \cos\theta \\ \sin\theta\end{pmatrix} \cP(\bfr,\theta) - D_t \nabla_\bfr \cP(\bfr,\theta) \right] - \alpha  \cP(\bfr,\theta) + \frac{1}{2\pi} \int_0^{2\pi} \alpha  \cP(\bfr,\theta) \> \rmd \theta + \partial_\theta^2 \left[ \Gamma \cP(\bfr,\theta) \right]\;.
    \label{eq:FP_2d_ABP-RTP}
\end{equation}
Besides, we assume that the parameters $v, \alpha, \Gamma$ appearing in Eq.~\eqref{eq:FP_2d_ABP-RTP} depend both on the position $\bfr$ of the particle and on its orientation $\bfu$ with respect to an external chemical gradient through:
\begin{eqnarray}
  v      &=& v_0(\bfr) - v_1 \> \bfu \cdot \nabla_\bfr c(\bfr) \\
  \alpha &=& \alpha_0(\bfr) + \alpha_1 \>  \bfu \cdot \nabla_\bfr c(\bfr) \\
  \Gamma &=& \Gamma_{0}(\bfr) + \Gamma_{1} \> \bfu \cdot \nabla_\bfr c(\bfr)\;.
\end{eqnarray}
When $v_1, \alpha_1, \Gamma_1$ are positive the field $c(\bfr)$ acts as a chemorepellent, since the particle's persistence length is decreased when moving up the gradients of $c(\bfr)$. On the contrary, negative values of $v_1, \alpha_1, \Gamma_1$ correspond to chemoattraction. 

First, we expand the angular dependence of $\mathcal{P}$ in Fourier series:
\begin{equation}
    \mathcal{P}(\bfr, \theta, t) = \frac{1}{2 \pi} \sum_{n=-\infty}^{+\infty} a_n(\bfr, t) \> e^{i n \theta} \qquad \text{ with } a_n = {a_{-n}}^* \ .
    \label{eq:Fourier_P}
\end{equation}
The zeroth-order harmonics $a_0(\bfr,t)$ corresponds  to the marginalized probability of finding a particle in position $\bfr$ at time $t$, irrespective of its orientation. Multiplying Eq.~\eqref{eq:FP_2d_ABP-RTP} by $e^{-i n\theta}$ and integrating over $\theta$, one can obtain a hierarchy of coupled equations for the Fourier modes. We now introduce some useful notation:
\begin{equation}
    \bi_{\pm} \equiv \begin{pmatrix}
            1 \\
            \pm i
            \end{pmatrix}
            \ , \qquad
            \partial_z^\pm f \equiv \partial_x f \pm i \partial_y f \;,
\end{equation}
and define the scalar product $\langle f | g \rangle = \int d \theta f^*(\theta) g(\theta)$. We will make use of the following results, which can be easily proved by direct calculation:
\begin{eqnarray}
\langle e^{i n \theta} | \cP \rangle &=& a_{n} \\[0.2cm]
\langle e^{i n \theta} | \bfu \cP \rangle &=& \frac{1}{2} (
            \bi_+ \> a_{n+1} + \bi_- \> a_{n-1} )\\[0.2cm]
\langle e^{i n \theta} | \partial_\theta^2 \cP \rangle &=& -n^2 a_{n} \\[0.2cm]
\langle e^{i n \theta} | (\bfu \cdot \nabla_\bfr c) \> \cP \rangle &=& \frac{1}{2} (a_{n+1} \> \partial_z^+ c  +  a_{n-1} \> \partial_z^- c ) \\[0.2cm]
\langle e^{i n \theta} | \bfu (\bfu \cdot \nabla_\bfr c) \> \cP \rangle &=& \frac{1}{2} a_n \nabla c  + \frac{1}{4} \left(a_{n+2} \> \bi_+ \> \partial_z^+ c  + a_{n-2}  \> \bi_- \> \partial_z^- c \> \right) \\[0.2cm]
\langle e^{i n \theta} | \partial_\theta^2 \left[ \left( \bfu \cdot \nabla_\bfr c \right) \cP \right] \rangle &=& -\frac{n^2}{2} (a_{n+1} \> \partial_z^+ c + a_{n-1} \> \partial_z^- c)\;.
\end{eqnarray}
By projecting Eq.~\eqref{eq:FP_2d_ABP-RTP} onto $\langle e^{i n \theta} \vert$ we then write the dynamics of the $n$-th harmonics as:
\begin{equation}
    \begin{split}
        \partial_t a_n = & -\nabla_{\bfr} \cdot
        \left[ \frac{v_0}{2} (
            \bi_+ \> a_{n+1} + \bi_- \> a_{n-1} ) - D_t \nabla_{\bfr} a_n - \frac{v_1}{2} a_n \> \nabla c - \frac{v_1}{4} \left(a_{n+2} \> \bi_+ \> \partial_z^+ c  + a_{n-2}  \> \bi_- \> \partial_z^- c \> \right)
            \right]\\[0.1cm]
            & - \alpha_0 (1 - \delta_{n,0}) \left[ a_n + \frac{\alpha_1}{2 \alpha_0}  (a_{n+1} \> \partial_z^+ c  +  a_{n-1} \> \partial_z^- c ) \right] - n^2 \Gamma_{0} \left[a_n + \frac{\Gamma_{1}}{2 \Gamma_{0}}  (a_{n+1} \> \partial_z^+ c  +  a_{n-1} \> \partial_z^- c ) \right]\;.
    \end{split}
    \label{eq:Cn_dyn}
\end{equation}
Explicitating Eq.~\eqref{eq:Cn_dyn} for the first three harmonics we obtain:
\begin{eqnarray}
    \label{eq:C0dyn}
    &&\partial_t a_0 = - \nabla_{\bfr} \cdot
        \left[ \frac{v_0}{2} (
            \bi_+ \> a_{1} + \bi_- \> a_{-1} ) - D_t \nabla_{\bfr} a_0 - \frac{v_1}{2} a_0 \> \nabla c - \frac{v_1}{4} \left(a_{2} \> \bi_+ \> \partial_z^+ c  + a_{-2}  \> \bi_- \> \partial_z^- c \> \right)
            \right]\\[0.2cm]
    \label{eq:C1dyn}
    &&\begin{split}
        \partial_t a_{\pm1} = &- \nabla_{\bfr} \cdot
        \left[ \frac{v_0}{2} (
            \bi_\pm \> a_{\pm 2} + \bi_\mp \> a_{0} )- D_t \nabla_{\bfr} a_{\pm 1} - \frac{v_1}{2} \> a_{\pm 1} \nabla c \>  - \frac{v_1}{4} (a_{2 \pm 1} \> \bi_+ \> \partial_z^+ c \>  + a_{-2 \pm 1}\> \bi_- \> \partial_z^- c)
            \right]\\[0.1cm]
            & - \left(\alpha_0 + \Gamma_{0}\right) a_{\pm 1} - \frac{1}{2} \left(\alpha_1 + \Gamma_{1}\right) \left[a_0 \>  \partial_z^\mp c  + a_{\pm 2} \>  \partial_z^\pm c \right]
    \end{split}\\[0.2cm]
    \label{eq:C2dyn}
    && \begin{split}
    \partial_t a_{\pm2} = & - \nabla_{\bfr} \cdot
    \left[ \frac{v_0}{2} (
            \bi_\pm \> a_{\pm 3} + \bi_\mp \> a_{\pm 1} ) - D_t \nabla_{\bfr} a_{\pm 2} -  \frac{v_1}{2} \> a_{\pm 2} \nabla c \>  - \frac{v_1}{4} (a_{2 \pm 2} \> \bi_+ \> \partial_z^+ c \>  + a_{-2 \pm 2}\> \bi_- \> \partial_z^- c)
            \right] \\[0.1cm]
        & - \left(\alpha_0  + 4 \Gamma_{0}\right) a_{\pm 2} - \frac{1}{2} \left(\alpha_1 + 4 \Gamma_{1}\right) \left[a_{\pm1} \>  \partial_z^\mp c  + a_{\pm 3} \>  \partial_z^\pm c \right]\;.
    \end{split}
\end{eqnarray}
In order to close the above hierarchy of equations, we first observe from Eq.~\eqref{eq:C0dyn} that $a_0$ is a conserved field, evolving on a slow diffusive scale. On the contrary, the higher-order harmonics undergo both a large-scale diffusive dynamics ($\sim \nabla_{\bfr}$) and a fast exponential relaxation. Subsequently, in the limit of large system size $L \to \infty$ we can safely assume that all modes $\{a_{\pm n}\}_{n>0}$ relax instantaneously to values enslaved to that of $a_0(\bfr,t)$. Given this timescale separation we can take $\partial_t a_{\pm n} = 0$ in Eqs.~\eqref{eq:C1dyn}, \eqref{eq:C2dyn}, thus concluding:
\begin{eqnarray}
    \label{eq:C1_trunc}
    \partial_t a_{\pm 1} = 0 \quad &\Rightarrow& \quad a_{\pm 1} = - \frac{\nabla_{\bfr} \cdot
         (\> v_0 \> \bi_{\mp} \> a_{0} \>)}{2(\alpha_0 + \Gamma_{0})} -  a_0 \frac{\alpha_1 + \Gamma_{1}}{2 (\alpha_0 + \Gamma_{0})} \partial_z^\mp c   + \cO(\nabla_\bfr^2)\\[0.2cm]    \label{eq:Cn_trunc}
    \partial_t a_{\pm n} = 0 \quad &\Rightarrow& \quad a_{\pm n} \sim \cO(\nabla_\bfr^2) \qquad \forall \> n > 1 \;.
\end{eqnarray}
Finally, to impose closure to Eq.~\eqref{eq:C0dyn} we resort to the so-called \textit{diffusion-drift approximation}, i.e. we truncate the expansion up to terms of order $\mathcal{O}(\nabla_\bfr^2)$. This relies on the fact that large-scale hydrodynamic modes are assumed to satisfy $\nabla^k \sim \frac{1}{L^k}$, thus justifying the gradient truncation in the large $L$ limit.

In conclusion, we can inject Eqs.~\eqref{eq:C1_trunc}, \eqref{eq:Cn_trunc} into Eq.~\eqref{eq:C0dyn}, obtaining a Fokker-Planck equation for the marginalized probability $a_0(\bfr, t)$:
\begin{equation}
    \partial_t a_0 = -\nabla_{\bfr} \cdot \left[ \mathbf{V} a_0 - \cD \nabla_{\bfr} a_0 \right]
    \label{eq:FokkerPlanck_meso}
\end{equation}
where we introduced the drift velocity $\mathbf{V}$ and the diffusivity $\cD$:
\begin{equation}
    \mathbf{V} = -\frac{v_0 \nabla v_0}{2 \left(\alpha_0  + \Gamma_{0}\right)} - \frac{1}{2} \left[ v_1 + v_0 \frac{\alpha_1 + \Gamma_{1}}{\alpha_0 + \Gamma_{0}}\right] \nabla_{\bfr} c \;,\qquad\qquad \cD = \frac{v_0^2}{2 \left(\alpha_0  + \Gamma_{0}\right)} + D_t \;.
    \label{eq:drift_diff}
\end{equation}
As a result, at large space- and time-scales, one can associate to Eq.~\eqref{eq:FokkerPlanck_meso} an It\=o-Langevin dynamics for the position $\bfr_i$ of particle $i$:
\begin{equation}
    \dot{\bfr}_i = \mathbf{V}(\bfr_i) + \nabla_{\bfr_i} \cD(\bfr_i) + \sqrt{2  \cD(\bfr_i)} \bxi_i(t) \;,
    \label{eq:meso_Langevin-2d}
\end{equation}
where $\bxi_i(t)$ is a delta-correlated Gaussian white noise with zero mean and unit variance.

\bigskip

\subsection{ABPs \& RTPs in $d$ space dimensions}
\label{sec:ABP-RTP_d}

\subsubsection{Harmonic tensors}
\label{sec:harmonic-tensors}
In higher dimensions, we need an alternative to the Fourier expansion used above. Harmonic tensors prove the relevant tools and we briefly review below their definition and properties. For further mathematical details and derivations of the results presented here, we refer the interested reader
to~\cite{ehrentraut1998symmetric,schouten1989tensor,spencer1970note,zee2016group,sakurai_napolitano_2017}.

Let us start by a definition: a tensor $\bT$ of order $p$ is said to be harmonic whenever it is symmetric and traceless. This reads, in an arbitrary basis:
\begin{eqnarray}
  &i) & \qquad T_{\cdots  i_l \cdots i_m \cdots} = T_{\cdots i_m \cdots i_l \cdots} \;, \quad 1 \leq l, m \leq p \; , \\[0.1cm] 
  &ii)& \qquad \delta_{i_1 i_2} \> T_{\>i_1 i_2 \cdots i_p} = 0 \; ,
\end{eqnarray}
where summation over repeated indices is implied. Consider a vector $\bfu$ on the unit sphere $\mathbb{S}^{d-1}$, from which we build the order-$p$ tensor:
\begin{equation}
  \bfu^{\otimes p} = \underbrace{\bfu \otimes \bfu \otimes \dots \otimes \bfu}_{p \text{ times}}\; , \quad \text{with} \quad |\bfu|^2 = \sum_{i=1}^{d} u_i^2 = 1 \; .
\end{equation}
We denote by $\harm{p}$ the orthogonal projection\footnote{The orthogonality is here understood with respect to the Euclidean inner product.} of $\bu^{\otimes p}$ onto the subspace of harmonic tensors of order $p$.
For example:
\begin{eqnarray}
\label{eq:harmonic0}
  & \harm{0} &\quad=\quad 1 \\
  & \harm{1} &\quad=\quad \bfu \\
  & \harm{2} &\quad=\quad \bfu^{\otimes 2} - \frac{\bI}{d} \label{eq:harmonic2} \\
  & \harm{3} &\quad= \quad\bfu^{\otimes 3} - \frac{3}{d+2} \bI\odot \bu \label{eq:harmonic3} \\
  & \harm{4} &\quad= \quad\bfu^{\otimes 4} - \frac{6}{d+4} \bI\odot \bu^{\otimes 2} + \left[ \frac{6}{d(d+4)}-\frac{3}{d(d+2)}\right]\bI^{\odot 2}
\label{eq:harmonic4}
\end{eqnarray}
where $\bI$ is the identity tensor and $\odot$ denotes the symmetrized
tensor product operation\footnote{More precisely, if $\bS=S^{i_1...i_p}\bb_{i_1}\otimes...\otimes \bb_{i_p}$ and $\bT=T^{i_1...i_q}\bb_{i_1}\otimes...\otimes \bb_{i_q}$ in an arbitrary basis $(\bb_1,...,\bb_d)$ of $\mbR^d$, then
\begin{equation}
    \bS\odot\bT \equiv \frac{1}{(p+q)!}\sum_{\sigma\in\cG_{p+q}} S^{i_{\sigma(1)}...i_{\sigma(p)}}T^{i_{\sigma(p+1)}...i_{\sigma(p+q)}} \bb_{i_1}\otimes ...\otimes\bb_{i_{p+q}} \ ,
\end{equation}
where $\cG_{p+q}$ is the permutation group of $p+q$ elements.}. For instance, $\bI\odot\bu$ can be written
in an orthonormal basis as: $(\bI\odot\bu)_{ijk} =
\frac{1}{3}(\delta_{ij}u_k + \delta_{ki} u_j + \delta_{jk} u_i)$. 

We refer to the family $\{\harm{p}\}_{p\in\mbN}$ as the  \textit{spherical harmonic tensors}. As will be clear in Sec.~\ref{subsec:difflim}, this family is particularly convenient to decompose any function on the unit sphere $\mathbb{S}^{d-1}$, and in particular $\cP(\bfr,\bfu)$. {To begin with, it is useful to introduce the product
$\langle \mathbf{T} | \mathbf{S} \rangle$ between two tensors $\mathbf{T}(\bfu)$ and $\mathbf{S}(\bfu)$ of arbitrary rank, which denotes the integral $\int_{\mbS^{d-1}} \mathbf{T}(\bfu) \otimes \mathbf{S}(\bfu) \rmd \bfu$ on the unit sphere $\mbS^{d-1}$ and extends the product $\langle \cdot | \cdot \rangle$ that we used in the $2d$ case. Then,} we
list the results on the $\harm{p}$ that will
prove most useful in the rest of the paper:
\begin{enumerate}
\item \textit{Orthogonality:}
  \begin{equation}
    {\llangle \harm{p} \middle| \harm{q} \rrangle} \equiv \int_{\mathbb{S}^{d-1}}  \rmd\bfu \> \harm{p} \otimes \harm{q} = \Omega \frac{p!(d-2)!!}{(d-2+2p)!!} \delta_{pq} \mathbf{\Pi}_{H_p}\;
    \label{harmonicOrthog}
  \end{equation}
      where $\Omega$ is the surface of $\mathbb{S}^{d-1}$, and
      $\mathbf{\Pi}_{H_p}$ is a tensor of rank $2p$ whose contraction
      with any tensor of order $p$ gives the orthogonal projection of
      the latter on the subspace of harmonic tensors. 
    \item \textit{Completeness:} \\ Harmonic tensors form a complete
      basis of $\mathbb{L}^2(\mathbb{S}^{d-1}, \mathbb{R})$. Any
      square-integrable function $f(\bfu)$ on the unit sphere can thus be
      expressed as:
      \begin{equation}
        f(\bfu) = \frac{1}{\Omega} \sum_{p=0}^\infty \frac{(d-2+2p)!!}{p!(d-2)!!} \bfa^p \cdot \harm{p} \; ,
        \label{harmonicTensorDecomp}
      \end{equation}
      where  
\begin{equation}
    \bfa^p = \int_{\mbS^{d-1}} \wh{\bfu^{\otimes p}} f(\bfu) 
 \: \rmd \bfu = {\langle  \harm{p}  | f\rangle} \ ,
    \label{ABP-RTP_dDim:harmonicPartDef}
\end{equation}
and `$\cdot$' denotes the (full) tensor contraction\footnote{If $\bS$
and $\bT$ are tensors of order $p$ and $q$, respectively, the
coordinates of their full tensor contraction $\bS\cdot\bT$ in an
orthonormal basis read
\begin{equation*}
[\bS\cdot\bT]^{i_1\dots i_{p-q}}\equiv \sum_{j_1\dots j_q}S^{i_1\dots i_{p-q}j_1\dots j_q}T^{j_1\dots j_q} 
\quad \text{if $p\geq q$, \   and} \quad  
[\bS\cdot\bT]^{i_1\dots i_{q-p}}\equiv \sum_{j_1\dots j_p}S^{j_1\dots j_p} T^{j_1\dots j_pi_1\dots j_{q-p}} \quad \text{otherwise.}
\end{equation*}
}, a convention we keep throughout this article.  Note that~\eqref{ABP-RTP_dDim:harmonicPartDef} can be shown as follows:
        \begin{eqnarray}
{\llangle \wh{\bu^{\otimes p}}  \middle| f \rrangle}  = \sum_{q\in \mbN} \frac{1}{\Omega} \frac{(d-2+2q)!!}{q!(d-2)!!} \bfa^q \cdot {\llangle \wh{\bu^{\otimes p}} \middle|  \wh{\bu^{\otimes q}}  \rrangle} =   \ba_p \cdot \bPi_{H_p} =  \ba_p  \ ,
\label{harmonicDecompProof}
\end{eqnarray}
      where the second equality in~\eqref{harmonicDecompProof} comes from~\eqref{harmonicOrthog}, while the last one is a consequence of $\ba^p$ being harmonic and $\bPi_{H_p}$ being self--adjoint.
    \item \textit{Eigenvectors of the Laplacian on the unit sphere}:
      \begin{equation}
        \Delta_{\bu} \harm{p} = -p(p+d-2) \harm{p}
        \label{eq:eigvec_laplacian}
      \end{equation}
      where $\Delta_{\bu}$ is the Laplace-Beltrami
      operator on the unit sphere $\mathbb{S}^{d-1}$, defined as:
      $\Delta_\bu f(\bfr) \equiv \nabla_\bfr^2 f(\bfr/|\bfr|)
      \>$ with $\bfr \in \mathbb{R}^d$. \\ 
 
 \if{     In $d=3$, the Laplace-Beltrami
      operator has a natural interpretation in quantum mechanics, where it corresponds to (minus) the square of
      the angular momentum operator
      $-L^2$ (in position space)~\cite{sakurai_napolitano_2017}.
      For instance, the components of the $\harm{p}$s --- that are
      usually called spherical harmonics --- are the angular part of
      the atomic orbitals of hydrogenic atoms. In this setting, the
      integer $p$ indexing the
      decomposition~\eqref{harmonicTensorDecomp} is the quantum number
      of the orbital angular momentum of the electron.}\fi

          \item \textit{Eigenvectors of the tumbling operator on the unit sphere}:\\
              Thanks to the orthogonality relation~\eqref{harmonicOrthog}, the harmonic tensors are also the eigenvectors of the evolution  operator induced by the tumbles:
          \begin{equation}
              \mathcal{T} \harm{p} \equiv - \alpha \harm{p} +\frac{\alpha}{\Omega} \int d\bfu \harm{p} = -\alpha (1-\delta_{p,0}) \harm{p} 
          \end{equation}

      \if{
      Being eigenvectors of $L^2$, harmonic tensors of each order make
      up an irreducible representation of the rotation group
      $SO(d)$: any rotation $\mathcal{R} \in SO(d)$ maps a harmonic tensor $\bT$ back into another harmonic tensor $\bT'$ ~\cite{zee2016group,sakurai_napolitano_2017}.
      }\fi

    \item \textit{Parity}:\\
      A rank-$p$ harmonic tensor has a well-defined parity $(-1)^p$:
      \begin{equation}
        \bfu \> \to \> -\bfu \quad \Longrightarrow \quad \harm{p} \> \to \> (-1)^p \> \harm{p}
        \label{eq:parity}
      \end{equation}
      In particular, this implies that if $f(-\bu)=f(\bu)$ then all
odd moments in expansion~\eqref{harmonicTensorDecomp} vanish.

\if{
\item \textit{Transformation under rotation.} The vector space of $d$-dimensional harmonic tensors of order $p$, to which $\wh{\bu^{\otimes p}}$ belongs, supports a so-called ``irreducible representation'' of the rotation group $SO(d)$. As such, the tensorial harmonic expansion~\eqref{harmonicTensorDecomp} is perfectly fitted to integrate out an orientational degree of freedom, the slow components of the expansion being imposed by the large-scale orientational symmetry of the system (see appendix~\ref{app:harmTensor}).}\fi
\end{enumerate}

\subsubsection{Decomposition of $\cP(\bfr,\bfu)$ on the harmonic tensor basis}
As in the two-dimensional case, we start from the
dynamics of $\mathcal{P}(\bfr, \bfu)$---the probability of finding one
particle at position $\bfr$ with a given orientation $\bfu$---and to integrate out the orientational degree of
freedom. Our goal is to obtain the marginalized probability of finding one
particle at position $\bfr$, irrespective of its orientation. 
If we expand $\mathcal{P}(\bfr,\bfu)$ over the basis of harmonic tensors:
\begin{equation}
  \mathcal{P}(\bfr, \bfu) = \frac{1}{\Omega} \sum_{p=0}^\infty \frac{(d-2+2p)!!}{p!(d-2)!!} \bfa^p(\bfr) \cdot \harm{p}\;  .
\label{ABP-RTP_dDim:HarmTensDecomp}
\end{equation}
the harmonic components  $\bfa^p(\bfr)$ of $\cP$ turn out to be physically meaningful objects. Indeed: 
\begin{equation}\label{eq:apofP}
  \bfa^p(\br) = \int_{\mbS^{d-1}} \cP(\bfr,\bfu) \harm{p} \: \rmd \bfu = {\llangle \harm{p} \middle| \cP \rrangle}
\end{equation}
corresponds to the average of the spherical harmonic tensor of order $p$ over the orientation of the particles. In particular:
\begin{eqnarray}
  \bfa^{0}(\br) = \int_{\mbS^{d-1}} \cP(\bfr,\bfu) \: \rmd \bfu 
\end{eqnarray}
 is the marginalized probability of finding a particle at position
 $\bfr$ we are looking for---which is also the rotational--invariant part of $\mcP$. Furthermore,
 higher-order components give us:
\begin{eqnarray}
   \bfa^{1} (\br) &=&  {\llangle \bfu \middle| \cP \rrangle} \qquad \equiv \mathbf{m}(\bfr) \;, \\
   \bfa^{2} (\br) &=&  {\llangle \harm{2} \middle| \cP \rrangle} \>\> \equiv \mathbf{Q}(\bfr) \;,
\end{eqnarray}
where $\mathbf{m}(\bfr)$ and $\mathbf{{Q}}(\bfr)$ are the polar and nematic order parameters at $\br$, respectively. This relation between the $\{\ba^p\}$ and the order parameters of rotational symmetry breaking---which can be traced back to the fact that each space of $p^{th}$--order harmonic tensor make up an irreducible representation of $SO(d)$---can be further generalized, as detailed in Appendix~\ref{app:harmTensor}. In practice, Eq.~\eqref{ABP-RTP_dDim:HarmTensDecomp} is the starting point of our coarse-graining method.

\subsubsection{Diffusive Limit}\label{subsec:difflim}
The master equation that yields the time evolution of $\mathcal{P}(\mathbf{r},\bfu,t)$  is given by:
\begin{equation}
\partial_t\cP(\br,\bu,t) = -\nabla_\br\cdot \left[ v \bu \cP - D_t \nabla_\br \cP \right] - \alpha \cP + \frac{1}{\Omega} \int_{\mbS^{d-1}} \alpha \cP \rmd \bfu +\lapu \Gamma \mcP \; .
\label{ABP-RTP_dDim:FP}
\end{equation}
We now mutilply Eq.~\eqref{ABP-RTP_dDim:FP} by $\harm{p}$ and integrate over the sphere to determine the time evolution of the components $\bfa^p(\bfr)$ entering the expansion~\eqref{ABP-RTP_dDim:HarmTensDecomp}. 
For $p=0$, this yields:
\begin{equation}
    \partial_t {\llangle 1 | \cP \rrangle} = -\nabla_\bfr \cdot \left[ v_0 {\llangle \bu | \cP \rrangle} - v_1\gradr c \cdot {\llangle \bu^{\otimes 2} \middle| \cP \rrangle}  - D_t\nabla_\bfr {\llangle 1 \middle| \cP \rrangle} \right] \ .
    \label{ABP-RTP_dDim:a0_01}
\end{equation}
Using the definition~\eqref{eq:apofP} of $\bfa^p$ as well as the fact that $\bfu^{\otimes 2}=\wh{\bfu^{\otimes 2}}+\bI/d$, Eq.~\eqref{ABP-RTP_dDim:a0_01} reads
\begin{equation}
    \partial_t \bfa^0 = -\nabla_\bfr \cdot \left[ v_0 \bfa^1 - v_1 \gradr c \cdot \left( \bfa^2 + \frac{\bI}{d}\bfa^0\right)  - D_t\nabla_\bfr \bfa^0 \right] \ . \label{ABP-RTP_dDim:a0Dyn} 
\end{equation}
Eq.~\eqref{ABP-RTP_dDim:a0Dyn} is not closed since it involves the harmonics $\bfa^1$ and $\bfa^2$. 

To determine the dynamics of $\bfa^1$, we now multiply Eq.~\eqref{ABP-RTP_dDim:FP} by $\hu{1}=\bfu$ and integrate with respect to $\bfu$, which leads to:
\begin{eqnarray}
    \partial_t \bfa^1 &=& -\divr \left[ v_0{\llangle \bu^{\otimes 2} \middle| \cP \rrangle} - v_1 \gradr c \cdot {\llangle \bu^{\otimes 3} \middle| \cP\rrangle} -D_t \gradr \bfa^1  \right] \notag\\ 
    & & - \alpha_0 \bfa^1  - \alpha_1 \gradr c \cdot  {\llangle \bu^{\otimes 2}\middle|\cP \rrangle}
     +  {\llangle  \bu \middle| \lapu\left( \Gamma_{0} + \Gamma_{1}\bu\cdot \gradr c\right)\cP \rrangle} \ ,
     \label{ABP-RTP_dDim:a1_01} 
\end{eqnarray}
The fact that the tensors $\hu{p}$ are eigenvectors of the Laplacian $\lapu$, Eq.~\eqref{eq:eigvec_laplacian}, 
together with the expressions of $\hu{2}$ and $\hu{3}$ from Eq.~\eqref{eq:harmonic2}--\eqref{eq:harmonic3}, allows us to re--write
Eq.~\eqref{ABP-RTP_dDim:a1_01} as
\begin{eqnarray}
    \partial_t \bfa^1 &=& -\divr \left[ v_0\left( \bfa^2 + \frac{\bI}{d}\bfa^0 \right) - v_1 \gradr c \cdot \left( \bfa^3 + \frac{3}{d+2}\bfa^1\odot \bI  \right) -D_t \gradr \bfa^1  \right] \notag\\ 
    & & - \left[\alpha_0+(d-1)\Gamma_{0})\right] \bfa^1  - \left[\alpha_1+(d-1)\Gamma_{1}\right] \gradr c \cdot \left( \bfa^2 + \frac{\bI}{d}\bfa^0 \right) \ .
    \label{ABP-RTP_dDim:a1Dyn} 
\end{eqnarray}

Similarly, one can get the dynamics of the second-order harmonic moment $\bfa^2$ by multiplying Eq.~\eqref{ABP-RTP_dDim:FP} by $\wh{\bfu^{\otimes 2}}$ and integrating with respect to $\bfu$:
\begin{eqnarray}
    \partial_t \bfa^2 &=& -\divr \left[ v_0 {\llangle \bu\otimes\wh{\bu^{\otimes 2}} \middle| \cP \rrangle} - v_1 \gradr c \cdot  {\llangle  \bu^{\otimes 2}\otimes\wh{\bu^{\otimes 2}}\middle|\cP\rrangle} -D_t \gradr \bfa^2  \right] \notag\\ 
    & & - \alpha_0 \bfa^2  - \alpha_1 \gradr c \cdot  {\llangle  \bu\otimes\wh{\bu^{\otimes 2}}\middle|\cP \rrangle}
     +  {\llangle  \wh{\bu^{\otimes 2}} \middle| \lapu\left( \Gamma_{0} + \Gamma_{1}\bu\cdot \gradr c\right)\cP \rrangle} \ .
     \label{ABP-RTP_dDim:a2_01} 
\end{eqnarray}
Using the expression~\eqref{eq:harmonic2}--\eqref{eq:harmonic4}, one gets that
\begin{eqnarray}
\bfu\otimes\wh{\bfu^{\otimes 2}} = \bfu^{\otimes 3}-\frac{1}{d}\bfu\otimes\bI = \wh{\bfu^{\otimes 3}} + \frac{3}{d+2}\bfu\odot \bI -\frac{1}{d}\bfu\otimes\bI 
\end{eqnarray}
and
\begin{eqnarray}
    \bu^{\otimes 2}\otimes \hu{2} = \bu^{\otimes 4} - \frac{1}{d}\bu^{\otimes 2}\otimes \bI = \hu{4} + \frac{6}{d+4}\hu{2}\odot \bI + {\frac{3}{d(d+2)}\bI^{\odot 2}} - \frac{1}{d}\hu{2}\otimes\bI -\frac{1}{d^2}\bI^{\otimes 2} \ ,
    \label{eq:dev-u2u2}
\end{eqnarray}
so that Eq.~\eqref{ABP-RTP_dDim:a2_01} reads
\begin{eqnarray}
    \partial_t\bfa^2 &=& -\divr \Big[ v_0\left( \bfa^3 + \frac{3}{d+2} \bfa^1 \odot \bI -\frac{1}{d}\bfa^1\otimes \bI \right) 
    - v_1\gradr c \cdot \Big( \bfa^4 +\frac{6}{d+4}\bfa^2\odot\bI -\frac{1}{d}\bfa^2\otimes \bI \notag \\
    & & + \frac{3 \bI^{\odot 2}}{d(d+2)}\bfa^0 - \frac{\bI^{\otimes 2}}{d^2}\bfa^0 \Big) -D_t \divr \bfa^2\Big] -(\alpha_0 +2d \Gamma_{0})\bfa^2 \notag \\
    & &- (\alpha_1 + 2d\Gamma_{1})\gradr c \cdot \left[ \bfa^3 + \frac{3}{d+2}\bfa^1 \odot \bI - \frac{1}{d}\bfa^1\otimes \bI \right] \ .
    \label{ABP-RTP_dDim:a2Dyn} 
\end{eqnarray}
In general, projecting the master equation Eq.~\eqref{ABP-RTP_dDim:FP} onto higher-order harmonic modes $\harm{p}$ leads to the following dynamics for $\bfa^p$:
\begin{eqnarray}
\partial_t\bfa^p &=& -\divr \left[ v_0 {\llangle  \bfu\otimes \harm{p} \middle| \cP \rrangle} - v_1\gradr c \cdot {\llangle \bfu^{\otimes 2} \otimes \harm{p} \middle| \cP \rrangle} -D_t \divr \bfa^p \right] \notag \\
    & &  -\left[ \alpha_0 + p(p+d-2) \Gamma_{0} \right] \bfa^p - \left[\alpha_1 + p(p+d-2) \Gamma_{1} \right] \gradr c \cdot {\llangle  \bfu \otimes \harm{p}\middle| \cP \rrangle} \; .
    \label{ABP-RTP_dDim:apDyn} 
\end{eqnarray}
One can then obtain a closure of this hierarchy by observing from Eq.~\eqref{ABP-RTP_dDim:a0Dyn} that $\bfa^0$ is a conserved field (the marginal in space of $\cP$), so its relaxation time diverges with the system size. On the contrary, higher-order harmonics $\bfa^p$, $p\geq 1$, undergo both large-scale transport dynamics ($\sim \nabla_{\mathbf{r}}$) and fast exponential relaxations, with finite relaxation times $[\alpha_0 + p(p+d-2)\Gamma_{0}]^{-1}$. In the limit of large system sizes, we can therefore assume that, for all $p\geq 1$, $\bfa^p$ relaxes instantaneously to values enslaved to that of $\bfa^0(\mathbf{r},t)$. We thus set $\partial_t \bfa^{p}=0$ in Eq.~\eqref{ABP-RTP_dDim:apDyn}, which leads to:
\begin{eqnarray}
\bfa^p &=& \cO(\nabla_\bfr) \qquad \forall \> p > 2 \\
\label{ABP-RTP_dDim:a2DynClosed}
\bfa^2 &=& \cO(\nabla_\bfr^2) \\[0.2cm]
\bfa^1 &=& -\frac{\gradr (\> v_0\bfa^0 \>)}{d[\alpha_0 + (d-1)\Gamma_{0}]} - \bfa^0 \frac{\alpha_1 + (d-1)\Gamma_{1}}{d[\alpha_0+(d-1)\Gamma_{0}]} \gradr c + \mcO(\nabla_\bfr^2) \ .
\label{ABP-RTP_dDim:a1DynClosed}
\end{eqnarray}
Finally, we inject Eqs.~\eqref{ABP-RTP_dDim:a1DynClosed}--\eqref{ABP-RTP_dDim:a2DynClosed} into the dynamics of the zeroth-order harmonics, Eq.~\eqref{ABP-RTP_dDim:a0Dyn}. To close the hierarchy, we then truncate the expansion including terms up to $\mathcal{O}(\nabla_\bfr^2)$, as for the two-dimensional case.

All in all, this procedure leads to a Fokker-Planck equation for the marginalized probability density $\bfa^0(\mathbf{r}, t)$:
\begin{equation}
    \partial_t \bfa^0 = -\nabla_{\bfr} \cdot \left[ \mathbf{V} \bfa^0 - \cD \nabla_{\bfr} \bfa^0 \right]\;,
    \label{ABP-RTP_dDim::FokkerPlanck_meso}
\end{equation}
with the $d$-dimensional {drift velocity} $\mathbf{V}$ and diffusivity $\cD$:
\begin{equation}
    \mathbf{V} = -\frac{v_0 \nabla v_0}{d\left(\alpha_0  + (d-1)\Gamma_{0}\right)} - \frac{1}{d} \left[ v_1 + v_0 \frac{\alpha_1 +(d-1) \Gamma_{1}}{\alpha_0 + (d-1)\Gamma_{0}}\right] \nabla_{\bfr} c  \quad \text{and}\quad \cD = \frac{v_0^2}{d \left(\alpha_0  + (d-1)\Gamma_{0}\right)} + D_t \ .
    \label{ABP-RTP_dDim::drift_diff}
\end{equation}
As in two space dimensions, the large-scale dynamics of our ABP-RTP particle is approximated by the It\=o-Langevin equation associated with the Fokker-Planck equation~\eqref{ABP-RTP_dDim::FokkerPlanck_meso}:
\begin{equation}
    \dot{\bfr}_i = \mathbf{V}(\bfr_i) + \nabla_{\bfr_i} \cD(\bfr_i) + \sqrt{2  \cD(\bfr_i)} \bxi_i(t) \;.
    \label{eq:Langevin_ABP-RTP}
\end{equation}

\subsection{AOUPs in $d$ space dimensions}
\label{sec:AOUP_dDim}

We start from the dynamics of a single AOUP in $d$ dimensions, whose dynamics is given by:
\begin{eqnarray}\label{eq:AOUPssec}
    \dot{\br} &=& v\be+\sqrt{2D_t}\boldsymbol{\xi} \\
    \dot{\be} &=& - \frac{\be}{\tau} + \sqrt{\frac{2}{d \tau}}\bet \;,
\end{eqnarray}
{where $\be \in \mathbb{R}^d$ is an orientation vector introduced in Sec.~\ref{sec:AOUP_intro}. We stress that, contrary to the unit vector $\bfu$ in ABP-RTPs, the magnitude of $\be$ is allowed to fluctuate around its average value. }

{We consider a self--propulsion amplitude $v$ and a persistence time $\tau$ given by: }
\begin{eqnarray}
v(\br) = v_0(\br) - v_1\be\cdot\gradr c (\br)  \quad \text{and} \quad \tau^{-1}(\br) = \omega_0(\br) + \omega_1\be\cdot\gradr c(\br) \ .
\end{eqnarray}
To coarse-grain this dynamics on time scales much larger than $\tau$,
we consider the Fokker--Planck equation corresponding
to~\eqref{eq:AOUPssec}:
\begin{eqnarray}
\partial_t\cP  &=&-\divr \left[v_0\be\cP - v_1\gradr c \cdot \be^{\otimes 2} \cP -D_t\gradr \cP \right] \notag\\
& & - \dive\left[ -\omega_0\be\cP - \omega_1 \gradr c \cdot \be^{\otimes 2} \cP -  \frac{1}{d} \grade \left( \omega_0 \cP + \omega_1 \be\cdot \gradr c \cP \right)\right] \ ,
\label{AOUP_dDim:FP}
\end{eqnarray}
and we build the dynamics of the marginal of $\cP$ with respect to $\be$.

To do so, we introduce $\bm^p(\br)$, the $p^{\rm th}$  moment of $\cP(\br,\be)$
with respect to $\be$, which is defined as:
\begin{equation}
\bm^p(\br) \equiv \int_{\mathbb{R}^d} \be^{\otimes p} \cP(\br,\be) \mathrm{d}\be = {\llangle \be^{\otimes p} \middle| \cP \rrangle}\ ,
\end{equation}
{where $\rmd\be$ is the volume element in $\mathbb{R}^d$. Note that the product $\llangle \cdot \middle| \cdot \rrangle$ now involves an integral over the whole space $\mathbb{R}^d$, and not solely on the unit sphere $||\be||=1$.}
Integrating Eq.~\eqref{AOUP_dDim:FP} with respect to $\be$, we obtain
the following  dynamics for $\bm^0(\br)$:
\begin{eqnarray}
\partial_t\bm^0 =-\divr \left[ v_0\bm^1 - v_1 \gradr c \cdot \bm^2 -D_t\gradr \bm^0 \right] \ .
\label{AOUP_dDim:m0Dyn}
\end{eqnarray}
Equation~\eqref{AOUP_dDim:m0Dyn} is the first equation of a hierarchy
that determines the dynamics of the moments $\{\bm^p\}$.  In order to
obtain a closed equation for the spatial marginal $\bm^0$, we apply a strategy
akin to that developed in Sec.~\eqref{subsec:difflim} for the harmonic tensors.
Multiplying Eq.~\eqref{AOUP_dDim:FP} by $\be^{\otimes p}$ and
integrating with respect to $\be$ gives the dynamics of the $p$-th
moment as:
\begin{eqnarray}
\partial_t {\llangle \be^{\otimes p} \middle| \cP\rrangle} &=& -\divr \left[ v_0 {\llangle \be^{\otimes p+1}  \middle| \cP\rrangle} - v_1 \gradr c \cdot {\llangle \be^{\otimes p+2}  \middle| \cP \rrangle} -D_t \gradr {\llangle \be^{\otimes p}  \middle| \cP \rrangle} \right] \notag \\
& & +{\llangle \be^{\otimes p}  \middle| \dive \left[ (\omega_0 \be + \omega_1 \gradr c \cdot \be^{\otimes 2})\cP + \frac{1}{d}\grade \left[ (\omega_0  + \omega_1 \gradr c \cdot \be)\cP \right] \right] \rrangle} \ ,
\label{AOUP_dDim:pMoment01}
\end{eqnarray}
The second line of Eq.~\eqref{AOUP_dDim:pMoment01} is the
sum of the following four terms:
\begin{eqnarray}
(1) &\equiv& \omega_0 {\llangle \be^{\otimes p}  \middle| \dive \be \cP \rrangle} \\
(2) &\equiv& \omega_1 \gradr c \cdot {\llangle \be^{\otimes p}  \middle| \dive \be^{\otimes 2}\cP \rrangle} \\
(3) &\equiv& \frac{\omega_0}{d} {\llangle \be^{\otimes p}  \middle| \lape\cP \rrangle} \\
(4) &\equiv& \frac{\omega_1}{d} {\llangle \be^{\otimes p} \middle| \lape (\gradr c \cdot \be \cP) \rrangle} 
\end{eqnarray}
that we now compute.

Let us start with the coordinate $\alpha_1...\alpha_p$ of (1) in the
canonical basis of $\mbR^d$:
\begin{eqnarray}
(1)_{\alpha_1...\alpha_p}=\omega_0\int \left[\be^{\otimes p}\right]_{\alpha_1...\alpha_p} \dive \left[\be\cP\right] \mathrm{d}\be = \omega_0\int \left( \prod_{i=1}^p e_{\alpha_i}\right) \sum_{k=1}^d \partial_{e_k} \left[e_k\cP\right] \mathrm{d}\be \ .
\end{eqnarray}
Performing an integration by parts yields
\begin{eqnarray*}
(1)_{\alpha_1...\alpha_p}
&=& - \omega_0\sum_{k=1}^d \int \left[ \sum_{j=1}^p \left( \prod_{i=1,i\neq j}^p e_{\alpha_i} \right) \partial_{e_k} e_{\alpha_j} \right] e_k\cP \mathrm{d}\be \\
&=& - \omega_0\sum_{j=1}^p \int \left( \prod_{i=1,i\neq j}^p e_{\alpha_i} \right) e_{\alpha_j}\cP \mathrm{d}\be \\
&=& - p\, \omega_0  \int \left[\be^{\otimes p}\right]_{\alpha_1...\alpha_p}\cP \mathrm{d}\be \ ,
\end{eqnarray*}
\ie 
\begin{equation}
(1) = -p \, \omega_0\bm^{p} \ .
\end{equation}
A similar computation leads to:
\begin{equation}
    (2) = -p \, \omega_1 \gradr c \cdot \bm^{p+1} \ .
\end{equation}
We now turn to the computation of (3) ---setting aside the prefactor
$\omega_0/d$. Integrating by parts twice turns (3) into
\begin{eqnarray*}
\int \left[\be^{\otimes p}\right]_{\alpha_1...\alpha_p} \lape \cP \mathrm{d}\be &=& \int \sum_{k=1}^d \partial_{e_k}^2\left[ \prod_{i=1}^p e_{\alpha_i}\right]  \cP \mathrm{d}\be \\
&=& \sum_{k=1}^p \int \partial_{e_k} \left[ \sum_{j=1}^p \delta_{k\alpha_j} \prod_{i=1, i\neq j}^p e_{\alpha_i}  \right]  \cP \mathrm{d}\be \\
&=& \sum_{k=1}^p \int \left[ \sum_{j=1}^p \sum_{l=1,l\neq j}^p \delta_{k\alpha_j}\delta_{k\alpha_l} \prod_{i=1, i\neq j, l}^p e_{\alpha_i}  \right] \cP \mathrm{d}\be \\
&=& \sum_{j=1}^p \sum_{l=1,l\neq j}^p \int \left[  \prod_{i=1, i\neq j, l}^p e_{\alpha_i} \right] \delta_{\alpha_j\alpha_l} \cP \mathrm{d}\be  \ .
\end{eqnarray*}
To obtain a coordinate free expression of (3), we first note that, in
this last expression, the double sum contains as many terms with $j>l$
as $j<l$. Thanks to the symmetry of the term between brackets under
permutation $j\leftrightarrow l$, we get:
\begin{equation}
\int \left[\be^{\otimes p}\right]_{\alpha_1...\alpha_p} \lape \psi \mathrm{d}\be = \int 2\sum_{j=1}^p \sum_{l = j+1}^p  \left[ \be^{\otimes p-2} \right]_{\alpha_1...\alpha_{j-1}\alpha_{j+1}...\alpha_{l-1}\alpha_{l+1}...\alpha_p} \delta_{\alpha_j\alpha_l} \psi \mathrm{d}\be \ .
\label{AOUP_dDim:aSymTensProduct1}
\end{equation}
The tensor on the right-hand side of eq.~\eqref{AOUP_dDim:aSymTensProduct1} is proportional to the symmetric tensor product of $\be^{\otimes p-2}$ by the identity tensor $\bI$, $\be^{\otimes p-2}\odot \bI$. To see this let us denote by $\cG_p$ the group of permutation of $p$ elements. The tensor $\be^{\otimes p-2}\odot \bI$ reads
\begin{eqnarray}
\left[\be^{\otimes p-2}\odot\bI\right]_{\alpha_1...\alpha_p} \equiv   \frac{1}{p!}\sum_{\sigma\in\cG_p} e_{\alpha_{\sigma(1)}}...e_{\alpha_{\sigma(p-2)}}\delta_{\alpha_{\sigma(p-1)}\alpha_{\sigma(p)}} \ .
\end{eqnarray}
We now denote by $\cH$ the subgroup of $\cG_p$ that leaves  $\be^{\otimes p-2}\otimes \bI$ invariant and by $\cG_p/H$ the set of cosets of $\cH$ in $\cG_p$. Each element of $\cG_p/\cH$ is an equivalence class of elements of $\cG_p$ that are equal up to a permutation in $\cH$. We can then rewrite  $\be^{\otimes p-2}\odot \bI$ as follows:
\begin{eqnarray}
\left[\be^{\otimes p-2}\odot\bI\right]_{\alpha_1...\alpha_p}  = \frac{|\mcH|}{p!}\sum_{\sigma\in\cG_p/\mcH} e_{\alpha_{\sigma(1)}}...e_{\alpha_{\sigma(p-2)}}\delta_{\alpha_{\sigma(p-1)}\alpha_{\sigma(p)}} \ ,
\end{eqnarray}
where $|\cH|$ is the cardinal of $\cH$.
Since $\be^{\otimes p-2}$ and $\bI$ are both completely symmetric, the permutations that leaves (the coordinates of) $\be^{\otimes p-2}\otimes \bI$ invariant are the ones that permutes independently the coordinates of $\be^{\otimes p-2}$ on one side, and those of $\bI$ on the other. In other words $\mcH = \cG_{p-2}\times \mbZ_2$, so that \begin{equation}
\left[\be^{\otimes p-2}\odot\bI\right]_{\alpha_1...\alpha_p} = \frac{2}{p(p-1)}\sum_{\sigma\in\cG_p/\mcH} e_{\alpha_{\sigma(1)}}...e_{\alpha_{\sigma(p-2)}}\delta_{\alpha_{\sigma(p-1)}\alpha_{\sigma(p)}} \ .
\label{AOUP_dDim:aSymTensProduct2}
\end{equation}
{Finally, note that for each class in $\cG_p/\mcH$, there is a
  unique permutation $\sigma$ such that the sequences
  $\sigma(1),...,\sigma(p-2)$ and $\sigma(p-1),\sigma(p)$} are
respectively arranged in ascending order, which means that the tensors
appearing in Eqs.~\eqref{AOUP_dDim:aSymTensProduct1} and
\eqref{AOUP_dDim:aSymTensProduct2} are proportional. More precisely:
\begin{equation}
\int \be^{\otimes p} \lape \cP \mathrm{d}\be = p(p-1) \int \be^{\otimes p-2} \odot \bI \cP \mathrm{d}\be = p(p-1) \bm^{\otimes p-2} \odot \bI \ ,
\end{equation}
\ie
\begin{equation}
    (3) = p(p-1) \, \frac{\omega_0}{d} \bm^{\otimes p-2} \odot \bI \ .
\end{equation}
Lastly, note that everything we did to show that ${\llangle
\be^{\otimes p}  \middle| \lape \cP \rrangle = p(p-1)\llangle \be^{\otimes
  p-2}\odot \bI  \middle| \mcP\rrangle}$ holds if replace $\mcP$ by any other
function of $\be$. In particular:
\begin{eqnarray*}
 {\llangle \be^{\otimes p}  \middle| \lape (\gradr c \cdot \be \cP) \rrangle} &=&p(p-1){\llangle \be^{\otimes p-2}\odot \bI  \middle| (\gradr c \cdot \be )\cP \rrangle} \\
 &=&p(p-1){\llangle (\gradr c \cdot \be^{\otimes p-1})\odot \bI  \middle| \cP \rrangle} \ ,
\end{eqnarray*}
which gives
\begin{equation}
    (4) = \frac{\omega_1}{d} p(p-1) (\gradr c \cdot \bm^{p-1})\odot \bI \ .
\end{equation}
We are now able to rewrite Eq.~\eqref{AOUP_dDim:pMoment01} as follows 
\begin{eqnarray}
\partial_t \bm^p &=& -\divr \left[ v_0\bm^{p+1} - v_1 \gradr c \cdot \bm^{p+2} - D_t \gradr \bm^p \right] - p \, \omega_0\bm^p - p \, \omega_1 \gradr c\cdot \bm^{p+1} \notag\\
& & + p(p-1) \frac{\omega_0}{d}\bm^{p-2}\odot \bI +p(p-1) \frac{\omega_1}{d}(\gradr c \cdot \bm^{p-1})\odot \bI \ .
\label{AOUP_dDim:pMomentDyn}
\end{eqnarray}
For all $p\geq 1$, $\bm^p$ relaxes exponentially fast with a
characteristic time $1/(p\omega_0)$, whereas $\bm^0$ is a slow field
whose relaxation time diverges with the system size. Thus we can use a
fast--variable approximation for all $\bm^p$, $p\geq 1$, setting
$\partial_t\bm^p$ to zero in Eq.~\eqref{AOUP_dDim:pMomentDyn}, to get
\begin{eqnarray}
 p \, \omega_0\bm^p &=& -\divr \left[ v_0\bm^{p+1} - v_1 \gradr c \cdot \bm^{p+2} - D_t \gradr \bm^p \right] - p \, \omega_1 \gradr c\cdot \bm^{p+1} \notag\\
& & + p(p-1) \frac{\omega_0}{d}\bm^{p-2}\odot \bI +p(p-1) \frac{\omega_1}{d}(\gradr c \cdot \bm^{p-1})\odot \bI \ .
\label{AOUP_dDim:pMomentFastApprox}
\end{eqnarray}
In turn, Eq.~\eqref{AOUP_dDim:pMomentFastApprox} provides a bound on
the scaling of the moments $\bm^p$ in a gradient expansion:
\begin{equation}
    \forall p \in \mbN \ , \quad \bm^{2p}=\mcO(1) \quad \text{while} \quad \bm^{2p+1} = \mcO(\gradr) \ ,
\end{equation}
as well as the more precise scalings of $\bm^2$:
\begin{equation}
    \bm^2 = \frac{1}{d}\bm^0 \bI + \mcO(\gradr ) \ ,
    \label{AOUP_dDim:m2Scale}
\end{equation}
and $\bm^1$:
\begin{equation}
    \bm^1 = -\frac{1}{d\omega_0}\gradr (v_0 \bm^0) - \frac{\omega_1}{d\omega_0}\bm^0 \gradr c +\mcO(\gradr^2)\ .
    \label{AOUP_dDim:m1Scale}
\end{equation}
Inserting Eqs.~\eqref{AOUP_dDim:m2Scale}-- \eqref{AOUP_dDim:m1Scale}
into Eq.~\eqref{AOUP_dDim:m0Dyn} and truncating to the second order in
gradient gives the diffusive limit of the active Ornstein--Uhlenbeck
particle
\begin{equation}
    \partial_t \bm^0 = -\nabla_{\bfr} \cdot \left[ \mathbf{V} \bm^0 - \cD \nabla_{\bfr} \bm^0 \right]
    \label{AOUP_dDim::FokkerPlanck_meso}
\end{equation}
with the $d$-dimensionaldrift velocity $\mathbf{V}$ and diffusivity $\cD$:
\begin{equation}
    \mathbf{V} = -\frac{v_0 \nabla v_0}{d\omega_0} - \frac{1}{d} \left[ v_1 + v_0\frac{\omega_1}{\omega_0} \right] \nabla_{\bfr} c  \quad \text{and}\quad \cD = \frac{v_0^2}{d \omega_0} + D_t \ .
    \label{AOUP_dDim::drift_diff}
\end{equation}
As for ABP-RTPs, we can write the It\=o-Langevin equation for the dynamics of $\bfr_i$ associated with the Fokker-Planck equation~\eqref{AOUP_dDim::FokkerPlanck_meso}:
\begin{equation}
    \dot{\bfr}_i = \mathbf{V}(\bfr_i) + \nabla_{\bfr_i} \cD(\bfr_i) + \sqrt{2  \cD(\bfr_i)} \bxi_i(t) \;.
    \label{eq:Langevin_AOUP}
\end{equation}
We note that, while our derivations are based on moment expansions for
both ABP-RTPs and AOUPs, only the first moment contributes in the
drift-diffusion approximation of ABP-RTPs whereas we need the second
moment in the AOUP case.

\subsection{Fluctuating hydrodynamics}
\label{sec:Dean}
{In
Sec.~\ref{subsec:difflim} and~\ref{sec:AOUP_dDim}, we employed a diffusion-drift approximation to describe the dynamics of RTPs, ABPs and AOUPs on scales much larger than their persistence length, in terms of the
It\=o-Langevin Eqs.~\eqref{eq:Langevin_ABP-RTP},
\eqref{eq:Langevin_AOUP}.}  We now turn to build the
time-evolution of the fluctuating density field:
\begin{equation}
    \rho(\mathbf{r},t) = \sum_{i=1}^N \delta(\mathbf{r} - \mathbf{r}_{i}(t)) \;.
    \label{eq:def_density}
\end{equation}
To do so we follow the standard approach introduced by
Dean~\cite{dean_langevin_1996} and later generalized to the case of
multiplicative noise~\cite{solon_active_2015}. Applying the It\=o
formula to Eq.~\eqref{eq:def_density}, one gets
    \begin{equation}
        \frac{d }{dt}\rho(\bfr,t) = \sum_{i=1}^{N} \left[\nabla_{\bfr_i} \delta(\mathbf{r} - \mathbf{r}_i(t)) \cdot \dot{\bfr}_i + \cD(\bfr_i)\nabla_{\bfr_i}^2 \delta(\mathbf{r} - \mathbf{r}_i(t))\right]
        \label{eq:B9}
    \end{equation}
    The first term in Eq.~\eqref{eq:B9} can be re-expressed as:
    \begin{eqnarray}
    \sum_{i=1}^{N} \nabla_{\bfr_i} \delta(\mathbf{r} - \mathbf{r}_i(t)) \cdot \dot{\bfr}_i &=& \sum_{i=1}^{N}  \nabla_{\bfr_i} \delta(\mathbf{r}-\mathbf{r}_i) \cdot \biggl(\> \mathbf{V}(\bfr_i) + \nabla_{\bfr_i} \cD (\bfr_i) + \sqrt{2 \cD (\bfr_i)} \> \bxi_i \>\biggr) \notag \\ 
    &=& - \sum_{i=1}^{N} \nabla_\bfr \delta(\mathbf{r}-\mathbf{r}_i) \cdot \biggl(\> \mathbf{V}(\bfr_i) + \nabla_{\bfr_i} \cD (\bfr_i) + \sqrt{2 \cD (\bfr_i)} \> \bxi_i \>\biggr) \notag \\
    &=& - \sum_{i=1}^{N} \nabla_\bfr \cdot \biggl[ \delta(\mathbf{r}-\mathbf{r}_i) \biggl( \mathbf{V}(\bfr_i) + \nabla_{\bfr_i} \cD (\bfr_i) + \sqrt{2 \cD (\bfr_i)} \> \bxi_i \biggr) \>\biggr] \notag \\
    &=& - \sum_{i=1}^{N} \nabla_\bfr \cdot \biggl[ \delta(\mathbf{r}-\mathbf{r}_i) \biggl( \mathbf{V}(\bfr) + \nabla_{\bfr} \cD (\bfr) + \sqrt{2 \cD (\bfr)} \bxi_i \biggr) \>\biggr]\label{eq:N-1} \\
    &=& - \nabla_\bfr \cdot \biggl[ \rho(\bfr,t) \biggl( \mathbf{V}(\bfr) + \nabla_{\bfr} \cD (\bfr) \biggr) + \sqrt{2 \cD \rho (\bfr,t)} \mathbf{\Lambda}(\bfr,t) \>\biggr]\label{eq:N}\;.
    \label{eq:nabla1}
    \end{eqnarray}
    To go from Eq.~\eqref{eq:N-1} to~\eqref{eq:N}, we have introduced
    a centered Gaussian white noise field with unit variance,
    $\mathbf{\Lambda}(\bfr,t)$, such that:
    \begin{equation}
        \langle \bfLL(\bfr,t) \rangle = 0, \quad \langle \bfLL(\bfr,t) \otimes \bfLL(\bfr',t') \rangle = \bI \delta(\bfr-\bfr') \delta(t-t') \;,
    \end{equation}
    and used that the probability laws of the noise fields $- \nabla_\bfr \cdot \left[ \sqrt{2 \cD \rho (\bfr,t)}
    \mathbf{\Lambda}(\bfr,t) \right]$ and $- \sum_{i=1}^{N} \nabla_\bfr \cdot \left[ \sqrt{2 \cD (\bfr)} \delta(\bfr - \bfr_i) \> \bxi_i(t) \right]$ at equal time $t$, conditioned on the value of the fluctuating density $\rho(\bfr,t)$ at $t$, are the same, hence the two noises generate the same fluctuating hydrodynamics. For the sake of completeness, this equivalence is detailed in Appendix~\ref{app:Deannoise}. 
    
    Similarly, the second term in Eq.~\eqref{eq:B9} can be rewritten as:
    \begin{eqnarray}
    \sum_{i=1}^{N} \cD (\bfr_i) \nabla_{\bfr_i}^2 \delta(\mathbf{r} - \mathbf{r}_i(t)) &=& \sum_{i=1}^{N} \cD (\bfr_i) \nabla_{\bfr}^2 \delta(\mathbf{r} - \mathbf{r}_i(t)) 
    = \sum_{i=1}^{N} \nabla_{\bfr}^2 \left[ \delta(\mathbf{r} - \mathbf{r}_i(t)) \cD (\bfr_i) \right] \notag \\
    &=& \sum_{i=1}^{N} \nabla_{\bfr}^2 \left[ \delta(\mathbf{r} - \mathbf{r}_i(t)) \cD (\bfr) \right]
    = \nabla_{\bfr}^2 \bigl[\> \rho(\bfr,t) \cD (\bfr) \> \bigr] \;.
    \label{eq:nabla2}
    \end{eqnarray}
    Finally, we insert the expressions~\eqref{eq:nabla1}, \eqref{eq:nabla2} into Eq.~\eqref{eq:B9} to get the fluctuating hydrodynamics of the density field:
   \begin{equation}
  \partial_t \rho = - \nabla_{\mathbf{r}} \cdot \biggl\{  \mathbf{V}(\mathbf{r}) \rho - \cD (\mathbf{r})\nabla_{\mathbf{r}} \rho + \sqrt{2 \cD(\mathbf{r}) \rho} \>\> \mathbf{\Lambda}(\mathbf{r},t) \biggr\}\;,
  \label{eq:macro_field_theory}
\end{equation}
where we remind that the expressions of $\mathbf{V}$ and $\mathcal{D}$ are given by Eq.~\eqref{ABP-RTP_dDim::drift_diff} for RTP-ABPs and Eq.~\eqref{AOUP_dDim::drift_diff} for AOUPs. This step completes the coarse-graining process, establishing a connection between the microscopic dynamics of active particles and the macroscopic evolution of the density fields. {We note that Eq.~\eqref{eq:macro_field_theory} is expected to describe the relaxation and the fluctuations of the density field on time and space scales much larger than the persistence time and length, where the diffusion-drift approximation of the single-particle dynamics is expected to hold.}

\subsection{Numerical test for the coarse-grained theory of non-interacting particles}
\label{sec:micro-simul-noninteracting}

While Eq.~\eqref{eq:macro_field_theory} cannot yet be used to study
the collective behaviors of active particles, it can already predict
the steady-state position distribution $p_s(x)$ of
active particles in motility-regulating fields. We do so below,
exploring both the case in which $\{\gamma(x)\}$ vary over length
scales much larger than $\ell_p$, where our coarse-grained theory is
expected to hold, and its possible breakdown as the variations of
$\{\gamma(x)\}$ occur on scales comparable to $\ell_p$.

We simulated the dynamics of RTPs, ABPs, and AOUPs in $2d$ boxes of
sizes $L_x \times L_y$ in two different cases:
\begin{enumerate}
    \item In the presence of a space-dependent self-propulsion speed
      $v_0(\bfr)$ but without translational noise, \textit{i.e.} with
      $D_t = 0$. As shown in Appendix~\ref{app:cg_sols}, for all types
      of particles and in arbitrary dimension $d$, the coarse-grained
      solution exactly coincides with the solution of the microscopic
      master Eqs.~\eqref{eq:MasterEq_ABP},~
      \eqref{eq:MasterEq_RTP},~\eqref{eq:MasterEq_AOUPs}, which reads:
    \begin{equation}
        p_s(\bfr) \propto \frac{1}{v_0(\bfr)}
        \label{eq:steady-state_v0r} \;.
    \end{equation}
    In this case, the diffusion-drift approximation
    is always valid in the steady state, no matter
    how large the persistence length is compared to the scale at which
    $v_0(\bfr)$ varies. This is shown in
    Fig.~\ref{fig:space-dep-motility}.  \if{
      Solution~\eqref{eq:steady-state_v0r} reveals that active
      particles accumulate where they go slower. In our simulations,
      we have chosen $v_0(\bfr)$ to be of the form:
    \begin{equation}
        v_0(\bfr) = \exp\left[ A \sin\left( q_nx\right)\right] \; , \hspace{1cm} q_n = \frac{2\pi n}{L_x}\;, \quad n \in \mathbb{N} \;, 
        \label{eq:v0_r}
        \end{equation}
    so that the corresponding stationary distribution reads:
        \begin{equation}
            \cP_{ss}(\bfr) \propto \exp\left[-A \sin\left( q_n x \right)\right] \;.
            \label{eq:stedy-state_v0_r}
        \end{equation}
    Solution~\eqref{eq:stedy-state_v0_r} is equal to the Boltzmann weight of a passive particle in a periodic, sinusoidal potential along the $x$ direction. Indeed, our microscopic simulations match the analytical result, as shown in Fig.~\ref{fig:space-dep-motility}.
    }\fi
    
\begin{figure}
\centering
    \includegraphics[width=\textwidth]{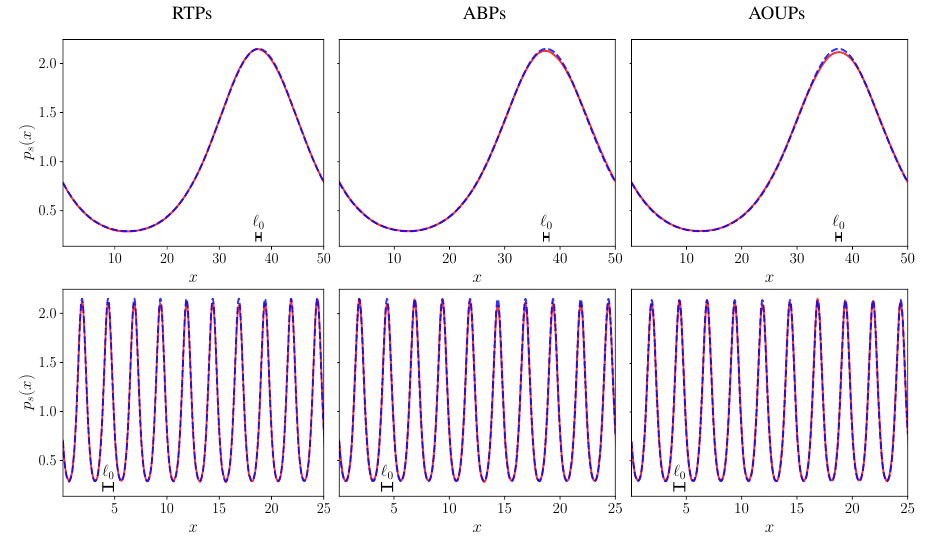}
    \caption{Stationary distribution $p_s(x)$ of RTPs (left), ABPs
      (center) and AOUPs (right) in $2d$ with a space-dependent
      propulsion speed $v_0(\bfr) = \bar{v} \exp[\sin(q_n x)]$, $q_n =
      2n \pi/L_x$, without translational noise. The blue dashed curve
      represents the theoretical
      prediction~\eqref{eq:steady-state_v0r} obtained from the
      coarse-grained theory; the solid red curve is obtained from
      sampling the particle's position in microscopic simulations. Top
      and bottom rows correspond to simulations with different values
      of $q_n$. The bare persistence length $l_0 \equiv \bar{v}
      \tau_0$ is shown in each panel for comparison with the scale
      $2\pi/q_x$ at which $v(\br)$ varies. Parameters: $L_x = L_y =
      50$, $ \bar{v} = 1$. RTPs: $\alpha_0 = 1$. ABPs: $\Gamma_0 =
      1$. AOUPs: $\omega_0 = 1$.}
    \label{fig:space-dep-motility}
\end{figure}

    \item In the presence of a space-dependent self-propulsion speed
      $v_0(\bfr)$ and with translational noise, \textit{i.e.} with $D_t
      > 0$. In this case, the presence of non-zero $D_t$ prevents us
      from finding an analytical solution to the microscopic master
      equations. As shown in Appendix~\ref{app:cg_sols}, the
      coarse-grained theory predicts:
    \begin{equation}
        \tilde{p}_s(\bfr) \propto \frac{1}{v_0(\bfr)} \dfrac{1}{\sqrt{1 + \dfrac{d D_t }{\tau_0 v_0^2(\bfr)}}}\;,\qquad \tau_0^{-1} = \begin{cases}
        & \alpha_0 \hspace{1.4cm} \text{ (RTP)} \\
        & (d-1) \Gamma_0  \quad \text{ (ABP)} \\
        & \omega_0 \hspace{1.4cm} \text{ (AOUP)} \\
        \end{cases} \label{eq:steady-state_finite_Dt}
    \end{equation}
    The validity of solution~\eqref{eq:steady-state_finite_Dt} now
    relies on the accuracy of the diffusion-drift approximation, and
    hence on the gradient expansion. To probe the validity of this
    approximation, we simulated the microscopic dynamics with a
    spatially periodic self-propulsion speed
    $v_0(x)=\bar{v}\exp[\sin(q x)]$. In all simulations, we keep the
    bare persistence length $\ell_0 \equiv \bar{v} \tau_0$ fixed and
    vary the wavevector $q$ in $v_0(x)$. Results of simulations for
    all types of particles are shown in Fig.~\ref{fig:finite_Dt}. At
    small persistence, the coarse-grained solution is in perfect
    agreement with the result of microscopic simulations. On the
    contrary, when $v_0(x)$ varies on scales comparable to the
    persistence length $\ell_0$, the coarse-grained description fails
    to capture the actual stationary distribution, as expected since
    the gradients of the fields are of order $\ell_0^{-1}$.
\begin{figure}
\centering
    \includegraphics[width=\textwidth]{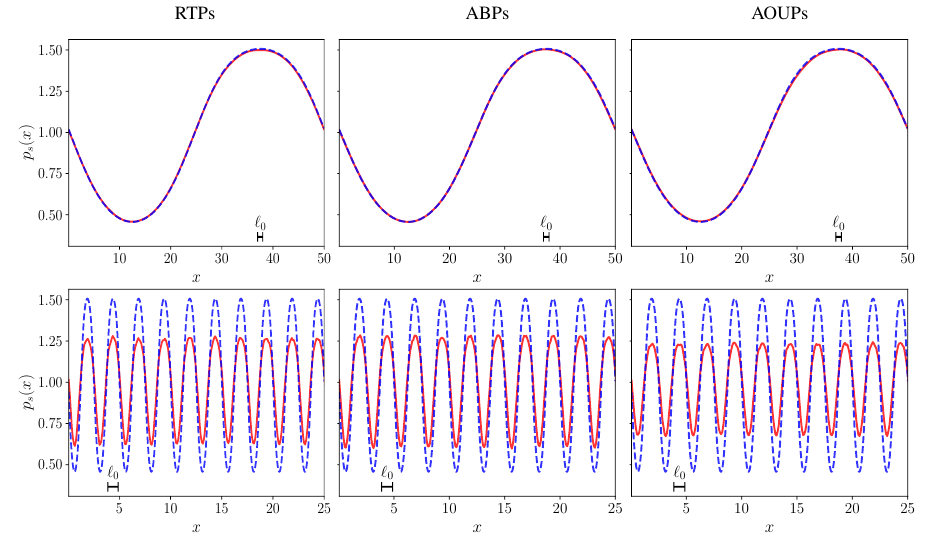}
    \caption{Stationary distribution $p_s(x)$ of RTPs (left), ABPs (center) and AOUPs (right) in $2d$ with space-dependent $v_0(\bfr) = \bar{v} \exp[\sin(q_n x)]$, $q_n = 2n \pi/L_x$, and finite translational diffusivity $D_t = 0.3$. The blue dashed curve represents the theoretical prediction~\eqref{eq:steady-state_finite_Dt} obtained from the coarse-grained theory; the solid red curve is obtained from sampling the particle's position in microscopic simulations. Top and bottom rows correspond to simulations with different wavevectors $q_n$ in $v_0(x)$. The bare persistence length $\ell_0 \equiv  \bar{v} \tau_0$ is shown for comparison at the bottom left of each panel. Parameters: $L_x = L_y = 50$, $\bar{v} = 1$. RTPs: $\alpha_0 = 1$. ABPs: $\Gamma_0 = 1$. AOUPs: $\omega_0 = 1$.}
    \label{fig:finite_Dt}
\end{figure}
\end{enumerate}

\section{Interacting active particles from micro to macro}
\label{sec:interacting}

So far, we have considered the dynamics of non-interacting particles
whose motility parameters depend on the position $\bfr$. We now turn
to the case in which this motility regulation is the result of
chemotactic or quorum-sensing interactions. As discussed in
Sec.~\ref{sec:interactions}, such interactions are in general nonlocal so that the
motility of a particle located at $\bfr$  admits a functional
dependence on the density field:
\begin{equation}
    \gamma(\bfr) \to \gamma(\bfr, [\rho]) \;.
 \end{equation}
The dynamics of each particle is then coupled to the others' via
complex $N$-body interactions. Nevertheless, since the density
$\rho(\br)$ is a conserved field, its evolution is expected to occur
on a large, diffusive timescale $T \sim L^2$. On time scales $\tau \ll
\Delta t \ll L^2$, we thus expect the diffusive approximation to hold
while $\rho(\br)$ has not yet relaxed. Under this assumption of
time-scale separation, which we refer to as a \textit{frozen-field
  approximation}, the $N$-body problem is mapped back onto a system of
$N$ independent particles, whose motility parameters depend only on
position $\bfr$ (through the frozen field $\rho(\bfr)$). We can then
use the result of the coarse-graining procedures detailed in
Sec.~\ref{sec:ABP-RTP_d} and~\ref{sec:AOUP_dDim} to predict the dynamics
of $\rho$, which will occur on longer time scales.

In Sec.~\ref{sec:MSD}, we first test this idea at the single particle
level by computing the particle mean-squared displacement (MSD) in a
homogeneous system at density $\rho_0$. We then discuss how to
generalize the It\=o-Langevin dynamics describing the evolution of the
density field to the case with interactions in
Sec.~\ref{sec:spuriousdrift}.  {Then, starting from the resulting}
stochastic field theory, we derive in Sec.~\ref{sec:correlations} the
structure factor, the pair correlation function, and the intermediate
scattering function for a system of active particles interacting via
QS. Our analytical predictions are then tested against microscopic
simulations of the interacting system.

\subsection{Mean squared displacement}
\label{sec:MSD}
According to the diffusion-drift approximation, the motion of an
active particle with motility regulation can be mapped to a passive
Langevin dynamics at mesoscopic scales
$\ell_p \ll \Delta r \ll L$, $\tau \ll \Delta t \ll L^2$: 
\begin{equation}
    \dot{\bfr}_i = \mathbf{V}(\bfr_i,[\rho]) + \nabla_{\bfr_i} \cD(\bfr_i,[\rho]) + \sqrt{2  \cD(\bfr_i,[\rho]) } \bxi_i(t) \;,
    \label{eq:mesoscopicLangevin-rho}
\end{equation}
where the $\{\bxi_i(t)\}$ are delta-correlated, centered Gaussian
white noises. Note that, at this stage, we have re-inserted the
density dependences, since we are now looking at the system at a scale
which is way larger than $\ell_p$. The question now is whether the transport coefficients in Eq.~\eqref{eq:mesoscopicLangevin-rho} can be related to the properties of particle trajectories at this scale, such as the mean-squared displacement.

\begin{figure}
\centering
\includegraphics[width=0.47\textwidth]{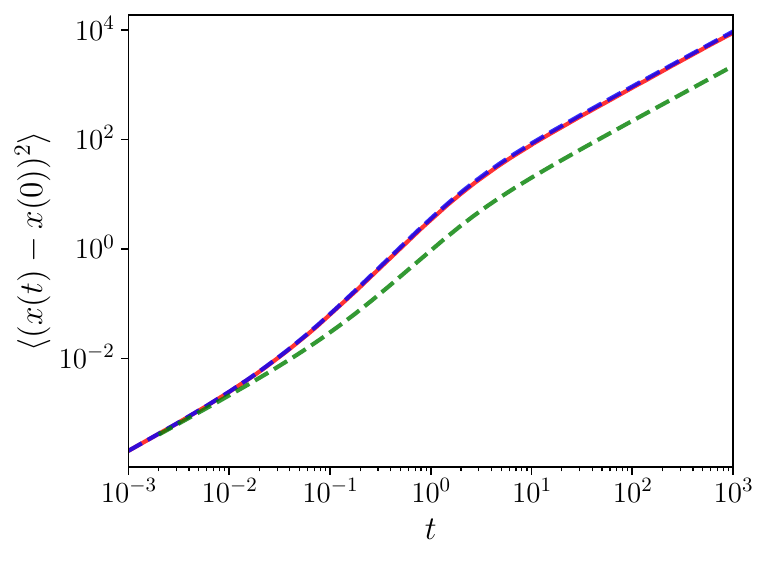}\hspace{0.5cm}
\includegraphics[width=0.47\textwidth]{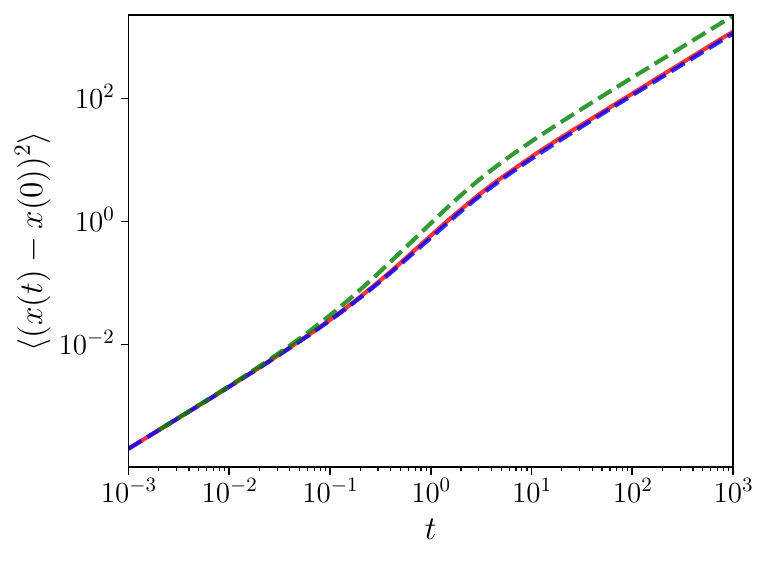}
\caption{Mean-square displacement for a homogeneous system of RTPs in
  $1d$ interacting via QS through $v(\rho) = v_0 \exp\left\{\kappa
  \tanh\left[(\rho-\rho_m)/{\varphi}\right]\right\}$. The panels show
  comparisons between numerical measurements (solid red lines) and
  diffusion-drift theory~\eqref{eq:MSD-micro_meso} (dashed blue
  lines). The non-interacting prediction for the MSD ($v \equiv v_0$)
  is also reported for comparison (dashed green lines). (\textit{Left})
  motility enhancement, $\kappa=1$. (\textit{Right}) Motility
  inhibition, $\kappa=-0.50$. Note that the impact of interactions is smaller in this case due to the smaller value of $|\kappa|$ that we used to prevent the occurrence of MIPS. Parameters of the simulation:
  $\rho_0 = 70, \rho_m = 50, \varphi = 20, \alpha_0=1, v_0=1, D_t=0.10$. Size of
  the simulation domain $L_x = 100$.}
\label{fig:MSD}
\end{figure}

To test this idea, we focus on the case of QS interactions that
regulate the self-propulsion speed. We take into account finite translational diffusivity $D_t>0$ and assume that the
interactions can be considered as local, \textit{i.e.} $v_i =
v(\rho(\bfr_i))$ for particle $i$, which we abbreviate as $v(\bfr_i)$. Integrating the microscopic
dynamics, we obtain the trajectory of particle $i$ as:
\begin{equation}
    \bfr_i(t) = \int_0^t \left[ v(\bfr_i(t')) \bfu_i(t') + \sqrt{2 D_t} \bxi_i(t') \right] \rmd t'+ \bfr_i(0) \;,
\end{equation}
where we remind that $\langle \bxi_i(t) \rangle = 0$ and $\langle \bxi_i(t) \otimes \bxi_j(t') \rangle = \delta_{ij} \delta(t-t') \bI$.
In $d$ space dimensions, the MSD is given by:
\begin{eqnarray}
    \langle \Delta r^2(t) \rangle &=& \langle |\bfr_i(t) - \bfr_i(0)|^2 \rangle = \llangle \left\{ \int_0^t \left[v(\bfr_i(t')) \bfu_i(t') + \sqrt{2 D_t} \bxi_i(t')\right] \rmd t' \right\}^2 \rrangle \notag \\[0.2cm]
    &=& 2 d D_t t + \underbrace{\int_0^t \rmd t' \int_0^t \rmd t''  \llangle v(\bfr_i(t')) v(\bfr_i(t'')) \bfu_i(t') \cdot \bfu_i(t'') \rrangle}_{(\star)} \notag \\[0.2cm] 
    && + 2 \sqrt{2 D_t} \underbrace{\int_0^t \rmd t' \int_0^t \rmd t'' \llangle v(\bfr_i(t')) \bfu_i(t') \cdot \bxi_i(t'') \rrangle}_{(\square)} \;.
\end{eqnarray}
Since the orientations decorrelate over a typical time $\tau$ as
$\langle \bfu_i(t) \cdot \bfu_i(0) \rangle = e^{-t/\tau}$, we need to
compute the correlations of $v(\br)$ over spatial scales of the order
of $\ell_p$. Assuming that the density field varies smoothly over
space, we expand $v(\bfr_i(t))-v(\bfr_i(0))$ on scales of order
$\ell_p$ as
\begin{equation}
    v(\bfr_i(t)) \simeq v(\bfr_i(0)) + \left. \frac{\partial v}{\partial \rho} \nabla_{\bfr} \rho \right|_{t=0, \bfr = \bfr_i(0)}\cdot \left[ \bfr_i(t) - \bfr_i(0) \right] = v(\bfr_i(0)) + \mathcal{O}\left(\frac{\ell_p}{L}\right) \;,
\end{equation}
where we used the fact that $\nabla_\bfr \rho \sim \mathcal{O}(1/L)$. Besides, when the system has relaxed to a homogeneous state with density
$\rho_0$, we write $v(\bfr_i(0)) \simeq
v(\rho_0)$ up to some corrections $\sim v'(\rho_0) \sqrt{\rho_0}$ which become negligible at large densities. We then compute $(\square)$ as:
\begin{eqnarray}
    (\square) = \int_0^t \rmd t' \int_0^t \rmd t'' v(\rho_0) \llangle \bfu_i(t') \cdot \bxi_i(t'') \rrangle + \cO\left( \frac{\ell_p}{L} \right) = 0 + \cO\left( \frac{\ell_p}{L} \right)\;,
\end{eqnarray}
since $\bfu_i, \bxi_i$ are independent and $\langle \bxi_i \rangle = 0$. We thus find:
\begin{eqnarray}
    \langle \Delta r^2(t) \rangle &=& 2 d D_t t + v(\rho_0)^2 \int_0^t \rmd t' \int_0^t \rmd t'' \llangle \bfu_i(t') \cdot \bfu_i(t'') \rrangle + \mathcal{O}\left(\frac{\ell_p}{L}\right) \notag \\[0.2cm]
    &=& 2 d D_t t + v(\rho_0)^2 \int_0^t \rmd t' \int_0^t \rmd t'' e^{-|t'-t''|/\tau} + \mathcal{O}\left(\frac{\ell_p}{L}\right) \notag\\[0.2cm]
    &=& 2 d t\left[\frac{v(\rho_0)^2 \tau}{d} + D_t \right] + 2 v(\rho_0)^2 \tau^2 \left[ e^{-t/\tau} - 1\right] + \mathcal{O}\left(\frac{\ell_p}{L}\right) \;.
    \label{eq:MSD-micro_meso}
\end{eqnarray}
Note that Eq.~\eqref{eq:MSD-micro_meso} is
expected to hold for all times $t \ll L^2$. At the mesoscopic scale we
also have $t \gg \tau$, so that the second contribution in
Eq.~\eqref{eq:MSD-micro_meso} is subleading. We then obtain the large-scale MSD
in terms of the mesoscopic diffusivity
\begin{equation}
    \mathcal{D}(\rho_0) = \frac{1}{2d t} \> \lim_{\substack{ t/\tau \to \infty \\ t/L^2 \to 0}} \langle \Delta r^2(t) \rangle= \frac{v(\rho_0)^2 \tau}{d} + D_t\;,
\end{equation}
corresponding to the result~\eqref{ABP-RTP_dDim::drift_diff} obtained from the diffusion-drift approximation. If the homogeneous profile is stable we expect this equality to hold also at steady-state, hence also for diffusive times $t \sim \mathcal{O}(L^2)$. This is indeed observed in simulations, as reported in Fig.~\ref{fig:MSD}.

\subsection{From mesoscopic to macroscopic description}
\label{sec:spuriousdrift}
Starting from the Langevin description~\eqref{eq:mesoscopicLangevin-rho} at mesoscopic scale we now want to derive the dynamics of the density field similarly to what we have done for the non-interacting case. However, one complication of the interacting case is that the operator $\nabla_{\bfr_i}$ applied to $\cD(\bfr_i, [\rho])$ now acts both on the first variable and on the field $[\rho]$,
the latter because $\rho$ is affected by a change in $\mathbf{r}_{i}$:
\begin{equation}
  \nabla_{\mathbf{r}_{i}} \cD (\mathbf{r}_{i}, [\rho]) 
  \if{
  = \nabla_1 \cD (\mathbf{r}_{i}, [\rho]) + \left[\nabla_{\mathbf{r'}} \frac{\delta \cD(\bfr_i)}{\delta \rho(\mathbf{r'})} \right]_{\bfr'=\bfr_{i}} 
  }\fi 
  =\nabla_1 \cD (\mathbf{r}_{i}, [\rho]) + \int \rmd^d \bfr' \frac{\delta \cD(\bfr_i)}{\delta \rho(\bfr')} \nabla_{\bfr_i} \rho(\bfr') \;.
  \label{eq:spurious_drift}
\end{equation}
where $\nabla_1$ denotes the derivative with respect to the first
variable. Nonetheless, it can be shown that the second term of
Eq.~\eqref{eq:spurious_drift} vanishes in many cases of interest
\cite{solon_active_2015}, which allows us to reproduce the derivation of Sec.~\ref{sec:Dean} for the fluctuating hydrodynamics. In particular, this occurs whenever $\cD$ is
a function of an effective density $\tilde{\rho}$, obtained by
convolution of $\rho(\bfr)$ with a \textit{symmetric} kernel
$K(\bfr)$:
\begin{equation}
    \cD(\bfr, [\rho]) = \cD(\tilde{\rho}(\bfr)) \;, \qquad \tilde{\rho}(\bfr) \equiv (K \circledast \rho) (\bfr) = \int \rmd^d\bfr' \> K(\bfr-\bfr') \rho(\bfr') \;.
\end{equation}
To see this, we use the definition of $\rho(\bfr')$:
\begin{equation}
    \nabla_{\bfr_i} \rho(\bfr') = \nabla_{\bfr_i} \sum_j \delta(\bfr'-\bfr_j) = \nabla_{\bfr_i} \delta(\bfr'-\bfr_i) = -\nabla_{\bfr'} \delta(\bfr'-\bfr_i) \; 
\end{equation}
and re-write Eq.~\eqref{eq:spurious_drift} as:
\begin{eqnarray}
    \nabla_{\bfr_i} \cD &=& \nabla_1 \cD - \int \rmd^d \bfr' \frac{\delta \cD(\bfr_i)}{\delta \rho(\bfr')} \nabla_{\bfr'} \delta(\bfr'-\bfr_i) = \nabla_1 \cD + \int \rmd^d \bfr'  \left[\nabla_{\bfr'} \frac{\delta \cD(\bfr_i)}{\delta \rho(\bfr')}\right]  \delta(\bfr'-\bfr_i) \notag\\[0.2cm]
    &=& \nabla_1 \cD + \left[\nabla_{\bfr'} \frac{\delta \cD(\bfr_i)}{\delta \rho(\bfr')}\right]_{\bfr'=\bfr_i} \;. \label{app:spurious2}
\end{eqnarray}
Applying the chain rule for functional derivatives:
\begin{eqnarray}
    \nabla_{\bfr'} \frac{\delta \cD(\bfr_i)}{\delta \rho(\bfr')} &=& \nabla_{\bfr'} \int \rmd \bfr'' \frac{\delta \cD(\bfr_i)}{\delta \tilde{\rho}(\bfr'')} \frac{\delta \tilde{\rho}(\bfr'')}{\delta \rho(\bfr')} = \int \rmd \bfr'' \frac{\partial \cD}{\partial \tilde{\rho}} \delta(\bfr_i - \bfr'') \> \nabla_{\bfr'} \frac{\delta \tilde{\rho}(\bfr'')}{\delta \rho(\bfr')} \;,
\end{eqnarray}
where in the last passage we have expanded the functional derivative $\delta \cD(\bfr_i) / \delta \tilde{\rho}(\bfr'')$. Integrating over $\bfr''$ then gives:
\begin{eqnarray}
    \nabla_{\bfr'} \frac{\delta \cD(\bfr_i)}{\delta \rho(\bfr')} &=& \frac{\partial \cD}{\partial \tilde{\rho}} \nabla_{\bfr'} \frac{\delta \tilde{\rho}(\bfr_i)}{\delta \rho(\bfr')} = \frac{\partial \cD}{\partial \tilde{\rho}} \nabla_{\bfr'} K(\bfr_i - \bfr')\; .
\end{eqnarray}
Finally, since $K(\bfr)$ is symmetric around the origin, we conclude:
\begin{equation}
\left[\nabla_{\bfr'} \frac{\delta \cD(\bfr_i)}{\delta \rho(\bfr')}\right]_{\bfr'=\bfr_i} = -\frac{\partial \cD}{\partial \tilde{\rho}} \nabla_\bfr K(0) = 0 \; .
\end{equation}
We note that such a symmetric $K$ is expected in the case where the
interactions are mediated by a diffusive field, as suggested by
Eq.~\eqref{eq:Greens}. In the following, we thus neglect this contribution to
the It\=o drift.

Subsequently, we can reproduce the computation of Sec.~\ref{sec:Dean} to go from our mesoscopic Langevin description to the
fluctuating hydrodynamics of $\rho$:
\begin{equation}
  \partial_t \rho = - \nabla_{\mathbf{r}} \cdot \biggl\{  \mathbf{V}(\mathbf{r},[\rho]) \rho - \cD (\mathbf{r},[\rho])\nabla_{\mathbf{r}} \rho + \sqrt{2 \cD(\mathbf{r},[\rho]) \rho} \>\> \mathbf{\Lambda}(\mathbf{r},t) \biggr\}\;,
  \label{eq:macro_field_theory2} 
\end{equation}
where we remind that the expressions of $\mathbf{V}$ and $\mathcal{D}$
are given by Eq.~\eqref{ABP-RTP_dDim::drift_diff} for RTP-ABPs and
Eq.~\eqref{AOUP_dDim::drift_diff} for AOUPs. Note that, at this stage,
all the functional dependencies in $\mathbf{V}$ and $\cD$ are to be
understood with respect to the \textit{fluctuating} density field
$\rho$.

The fluctuating-hydrodynamic Eq.~\eqref{eq:macro_field_theory2} can be
used to study the emergence of collective phenomena in scalar active
matter. For instance, it allows one to predict the onset of phase
separation in scalar active matter by studying the stability of
homogeneous phases in the presence of quorum-sensing and chemotactic
interactions. In particular, the expression of the transport
coefficients given in Eqs.~\eqref{ABP-RTP_dDim::drift_diff}
and~\eqref{AOUP_dDim::drift_diff} highlight the similarity between the
motility-induced phase separation observed in the presence of
quorum-sensing
interactions~\cite{tailleur2008statistical,cates2015motility} and the
phase separation induced by
chemoattractant~\cite{brenner1998physical,obyrne_lamellar_2020,zhang2021active}. In
the following Section, we focus on the static and dynamical properties
of homogeneous steady states to test the predictions of the
fluctuating hydrodynamics~\eqref{eq:macro_field_theory2}.

\subsection{Correlation functions in interacting particle systems}
\label{sec:correlations}
{The stochastic hydrodynamics~\eqref{eq:macro_field_theory2} has been obtained from the mesoscopic Langevin equation~\eqref{eq:mesoscopicLangevin-rho} without any approximation. Consequently, we expect Eq.~\eqref{eq:macro_field_theory2} to correctly account for fluctuations and correlations of the density profile on scales where the diffusion-drift approximation provides a good approximation of the microscopic active dynamics}. 
To test this idea we devote this section to the
derivation of the static structure factor $S(\bfq)$, spatial
correlation function $G(\bfr)$ and intermediate scattering function
$F(\bfq, t)$ for an active system with motility regulation. {The method presented here relies on the linearization of the fluctuating hydrodynamics around a homogeneous profile, a technique that has proved to be succesful at predicting correlation functions and transport coefficients in a variety of equilibrium~\cite{demery2014generalized} and nonequilibrium systems~\cite{fily2012athermal,jardat2022diffusion,benois2023enhanced,ghimenti2024irreversible}.}
Our analytical prediction are then tested against numerical simulations in Sec.~\ref{sec:simulations_correlations}.

As a microscopic model, we take the case of RTPs interacting via QS with:
\begin{equation}
  \begin{cases}
    & v(\bfr, [\rho]) = v(\tilde{\rho}(\bfr)) \\
    & \alpha(\bfr, [\rho]) = \alpha(\tilde{\rho}(\bfr)) \\    
  \end{cases}
  \quad \qquad \tilde{\rho}(\bfr) = (K \circledast \rho)(\bfr) = \int d^d \bfr' \; K(\bfr - \bfr') \rho(\bfr') \;,
\end{equation}
where $\tilde{\rho}$ represents an effective density at point $\bfr$, obtained by weighing the contribution of each particle by a kernel $K(\bfr)$. We take $K(\bfr)$ to be normalized and isotropic.

We then consider the corresponding stochastic field
theory~\eqref{eq:macro_field_theory2} and study the dynamics of
density fluctuations $\delta\rho = \rho-\rho_0$ around a stable
homogeneous profile $\rho_0$ at steady state. To do so, we first
derive the linearized hydrodynamics of our system. Expanding the
diffusive and drift terms in Eq.~\eqref{eq:macro_field_theory2} we
obtain:
\begin{eqnarray}
  \cD \nabla_\bfr \delta\rho &=& \frac{v^2(\boldsymbol{r}, [\rho])}{d \alpha(\boldsymbol{r}, [\rho])} \nabla_\bfr \delta\rho = \frac{v^2(\rho_0)}{d \alpha(\rho_0)} \nabla_\bfr \delta\rho + \mathcal{O}(\delta\rho^2) \equiv \cD_0 \nabla_\bfr \delta\rho + \mathcal{O}(\delta\rho^2) \;, \\[0.5cm]
   \boldsymbol{V} \rho &=& - \frac{v(\boldsymbol{r}, [\rho]) \nabla_\bfr v(\boldsymbol{r}, [\rho])}{d \alpha(\boldsymbol{r}, [\rho])} (\rho_0 + \delta\rho)  = - \frac{v v'(\tilde{\rho}(\bfr))}{d \alpha(\tilde{\rho}(\bfr))} (\rho_0 + \delta\rho) \nabla_\bfr \delta\tilde{\rho}(\bfr) \notag \\[0.2cm]
   &=& - \cD_0 \rho_0 \frac{d\log v_0}{d\tilde{\rho}} \nabla_\bfr (K  \circledast \delta \rho)(\bfr) + \mathcal{O}(\delta\rho^2) \;.
\end{eqnarray}
To treat the conserved noise term $\nabla_{\bfr} \cdot \left[ \sqrt{2 \cD(\bfr, [\rho])}\boldsymbol{\Lambda}(\bfr,t)  \right]$, we expand:
\begin{equation}
    2 \cD(\tilde{\rho}(\bfr)) \rho(\bfr) = 2 \cD_0 \rho_0 \left[1 + \frac{\rho_0}{\cD_0} \left. \frac{d \cD_0}{d \tilde{\rho}}  \right|_{\tilde{\rho}=\rho_0} \left(K \circledast \frac{\delta\rho}{\rho_0}\right)(\bfr) + \frac{\delta\rho}{\rho_0} \right] + \cO\left( \delta \rho^2\right) \;.
\end{equation}
When the homogeneous profile $\rho_0$ is stable, we expect density
fluctuations $\delta\rho$ to scale as $\sqrt{\rho_0}$. At large
densities, we can thus retain only the leading-order contribution to
the noise, which becomes additive and delta-correlated:
\begin{equation}
    \nabla_{\bfr} \cdot \left[ \sqrt{2 \cD(\bfr, [\rho])} \boldsymbol{\Lambda}(\bfr,t) \right] \approx \nabla_\bfr \cdot \left[ \sqrt{2\cD_0 \rho_0} \boldsymbol{\Lambda}(\bfr,t) \right] \;.
    \label{eq:add_noise-approx}
\end{equation}
All in all, we obtain the linear dynamics of $\delta\rho$ as:
\begin{equation}
    \partial_t \delta\rho =  \cD_0  \nabla_\bfr^2 \left[ \frac{d\log v_0}{d\tilde{\rho}} \rho_0 \;   (K \circledast \delta\rho) +  \delta\rho  \right] + \nabla_\bfr \cdot \left[ \sqrt{2\cD_0 \rho_0} \boldsymbol{\Lambda}(\bfr,t) \right] \; .
    \label{eq:transport-deltarho-struct}
\end{equation}
Next, we write Eq.~\eqref{eq:transport-deltarho-struct} in Fourier space. In a finite volume $L^d$, we adopt the following convention for the Fourier transform:
\begin{equation}
    f(\bfr) = \frac{1}{L^d} \sum_{n=-\infty}^{+\infty} e^{i \bfq_n \cdot \bfr} \hat{f}_n \>, \qquad \hat{f}_n = \int_{L^d} e^{-i \bfq_n \cdot \bfr} f(\bfr) \> d^d \bfr \>, \qquad \bfq_n = \frac{2\pi}{L} (n_1, n_2, \dots, n_d) \;.
\end{equation}
The dynamics in Fourier space then reads:
\begin{equation}
  \partial_t \delta\hat{\rho}_n = - q_n^2 \cD_0 \left(\frac{d \log v_0}{d\tilde{\rho}} \rho_0 \; \hat{K}_n + 1 \right)\delta \hat{\rho}_n - \sqrt{2 \cD_0 \rho_0} \; i \bfq_n \cdot \hat{\bfLL}_n \;,
  \label{FT-eq-nonlocal}
\end{equation}
where the Gaussian white noise in Fourier space satisfies: $\langle
\hat{\bfLL}_n(t) \hat{\bfLL}_m(t') \rangle = L^d \delta_{n,-m}
\delta(t-t')$. We note that Eq.~\eqref{FT-eq-nonlocal} leads to an
exponential growth of $\delta \hat{\rho}_n$ when:
\begin{equation}
-\cD_0 \left(\rho_0  \frac{d \log v_0}{d\tilde{\rho}_0} \hat{K}_n + 1\right) q_n^2 > 0 \;\;.
\end{equation}
Otherwise, the homogeneous configuration is (linearly) stable and
density fluctuations are damped. For hydrodynamic modes where $\bfq_n
\to 0$, $K_n \to 1$ and the homogeneous profile is linearly unstable
when:
\begin{equation}
\frac{d\log v(\tilde{\rho})}{d\tilde{\rho}} < -\frac{1}{\rho_0}\;,
\label{eq:spinodal-stability-MIPS}
\end{equation}
corresponding to the condition for a spinodal stability in
QS-MIPS~\cite{cates2015motility}. In the following we choose
parameters such that~\eqref{eq:spinodal-stability-MIPS} is far from
being satisfied, so that $\delta \hat\rho_n$ relaxes and its dynamics
is well described by Eq.~\eqref{FT-eq-nonlocal}. As a final remark,
note that $ \delta\hat{\rho}_{n=0}(t) = 0$ at any time $t$, due to
mass conservation:
\begin{equation}
  \delta\hat{\rho}_0(t) =  \int_{L^d} (\rho(\bfr,t)-\rho_0) = 0 \;.
\end{equation}

\subsubsection{Equal-time correlations: Structure factor and pair correlation function}
At steady state, the static structure factor is defined as:
\begin{equation}
    S(\bfq_n) \equiv \frac{1}{N} \langle \delta\hat{\rho}_{n} \delta\hat{\rho}_{-n} \rangle \;.
    \label{eq:S_qn}
\end{equation}
{where $\langle \cdot \rangle$ denotes the average over noise.}
Using the It\=o chain rule together with the dynamics~\eqref{FT-eq-nonlocal} we compute:
\begin{eqnarray}
    \partial_t \langle \delta\hat{\rho}_n \delta\hat{\rho}_m \rangle && = \langle ( \partial_t  \delta\hat{\rho}_n ) \delta\hat{\rho}_m \rangle + \langle \delta\hat{\rho}_n  ( \partial_t  \delta\hat{\rho}_m )  \rangle - 2 \bfq_n \cdot \bfq_m \cD_0\rho_0 L^d  \delta_{n,-m} \notag \\[0.2cm]
    && \begin{split}
        = -\cD_0 \left[ \left(\rho_0  \frac{d \log v_0}{d\tilde{\rho}_0} \hat{K}_n + 1\right) q_n^2 + \left(\rho_0 \frac{d \log v_0}{d\tilde{\rho}} \hat{K}_m + 1 \right) q_m^2 \right] \langle \delta\hat{\rho}_n \delta\hat{\rho}_m \rangle \\
        -2 \bfq_n \cdot \bfq_m \cD_0 \rho_0 L^d \delta_{n,-m} \;.
    \end{split}
    \label{eq:Ito-evolution-modes}
\end{eqnarray}
When the spinodal instability condition is
violated~\eqref{eq:spinodal-stability-MIPS}, the linear dynamics for
$\langle \delta\hat{\rho}_n \delta\hat{\rho}_m \rangle$ admits a
stationary solution. At steady state, Eq.~\eqref{eq:Ito-evolution-modes} leads to:
\begin{eqnarray}
  \langle \delta\hat{\rho}_n \delta\hat{\rho}_m \rangle =
  \begin{cases}
    & \dfrac{-2 \bfq_n \cdot \bfq_m \rho_0 L^d}{\left(\rho_0  \dfrac{d \log v_0}{d\tilde{\rho}}  \hat{K}_n + 1\right) q_n^2 + \left(\rho_0 \dfrac{d \log v_0}{d\tilde{\rho}}  \hat{K}_m + 1 \right) q_m^2} \delta_{n,-m} \qquad n, m \neq 0 \\[1cm]
    & 0 \hspace{9cm} n=m=0
    \end{cases}
    \label{eq:FT-correlation-struct}
\end{eqnarray}
The only non-zero contributions to the correlations thus come from $\bfq_n = -\bfq_m$:
\begin{equation}
    \langle \delta\hat{\rho}_n \delta\hat{\rho}_{-n} \rangle = \frac{\rho_0 L^d}{1 + \rho_0  \dfrac{d \log v_0}{d\tilde{\rho}} \hat{K}_n } (1 - \delta_{n,0})
    \label{eq:steady-state-rho_n,m}
\end{equation}
We can finally write the expression of the structure factor:
\begin{equation}
    S(\bfq_n) = \frac{\langle \delta\hat{\rho}_n \delta\hat{\rho}_{-n} \rangle}{N} = \frac{1 - \delta_{n,0}}{1 + \rho_0  \dfrac{d \log v_0}{d\tilde{\rho}} \hat{K}_n}\;.
    \label{eq:S_q}
\end{equation}
From the knowledge of $\langle \delta\hat{\rho}_n \delta\hat{\rho}_m \rangle$ at steady state we then compute the spatial correlation function as:
    \begin{equation}
      G(\bfr, \bfr') \equiv \langle \delta\rho(\bfr) \; \delta\rho(\bfr') \rangle\;.
    \end{equation}
By decomposing $\delta \rho(\bfr)$ into Fourier modes, we find:
\begin{eqnarray}
        G(\bfr, \bfr') &=& \frac{1}{L^{2d}} \sum_{n, m} e^{i \bfq_n \cdot \bfr} e^{i \bfq_m \cdot \bfr'}  \langle \delta\hat{\rho}_n \delta\hat{\rho}_{m} \rangle = \frac{1}{L^{2d}} \sum_{n} e^{i \bfq_n \cdot (\bfr - \bfr')} \rho_0 L^d S(\bfq_n) \notag \\[0.3cm]
  &=&  \rho_0 \left[\frac{1}{L^{d}}\sum_{n}  e^{i \bfq_n \cdot (\bfr - \bfr')}  S(\bfq_n)\right]\;.
  \label{eq:S_G0}
\end{eqnarray}
Since $S(\bfq_n) \to 1$ for $\bfq_n \to \infty$, it is convenient to shift it by a constant $-1$ to perform the Fourier transform. Eventually, this gives:
\begin{eqnarray}
        \frac{G(\bfr, \bfr')}{\rho_0} &=&  \frac{1}{L^{d}}\sum_{n}  e^{i \bfq_n \cdot (\bfr - \bfr')}  \left[S(\bfq_n) - 1\right] + \delta(\bfr - \bfr')\;.
  \label{eq:S_G}
\end{eqnarray}
For $\bfr \neq \bfr'$, the inverse Fourier transform of $[S(\bfq)-1]$ thus corresponds to the spatial correlation function $G(\bfr,\bfr')/\rho_0$.

\subsubsection{Dynamics: Intermediate scattering function}
Finally, we derive the expression of the intermediate scattering function:
\begin{equation}
  F(\bfq, t) = \lim_{\tau \to \infty} \frac{1}{N} \langle \delta\hat{\rho}_{\bfq}(\tau+t) \; \delta\hat{\rho}_{-\bfq}(\tau) \rangle \;.
\end{equation}
The function $F(\bfq, t)$ provides information on the relaxation mechanisms of density modes at steady-state. The computation and
measurement of $F(\bfq,t)$ has recently attracted attention in the community on both experimental~\cite{wilson2011differential,martinez2012differential,kurzthaler2022characterization,zhao2024quantitative} and theoretical~\cite{martens2012probability,chakraborty2022anomalous,chakraborty2023time} levels.

By definition, the intermediate scattering function at $t=0$ coincides with the static structure factor $S(\bfq)$. 
Starting from the linearized dynamics of $\delta\rho$ in Fourier space~\eqref{FT-eq-nonlocal}, we compute:
\begin{eqnarray}
    \partial_t F(\bfq_n,t) &=& \partial_t\frac{\langle \delta\hat{\rho}_n (\tau+t) \; \delta\hat{\rho}_{-n}(\tau) \rangle}{N} = \frac{\langle \partial_t  \delta\hat{\rho}_{n} (\tau+t) \;  \delta\hat{\rho}_{-n}(\tau)   \rangle}{N} \notag \\[0.2cm]
    &=& - q_n^2 \cD_0 \left(1 + \frac{d \log v_0}{d\tilde{\rho}} \rho_0 \; \hat{K}_n\right) F(\bfq_n,t) \;.
\end{eqnarray}
Solving for $F(\bfq_n)$ with the initial condition $F(\bfq_n,t=0) = S(\bfq_n)$, we obtain:
\begin{equation}
  F(\bfq_n,t) = S(\bfq_n) \; \exp\left[- q_n^2 \cD_0  \left(1 + \frac{d \log v_0}{d\tilde{\rho}} \rho_0 \; \hat{K}_n\right) t\right] \;.
\end{equation}
Finally, in the limit of local interactions $K(\bfr) \equiv \delta(\bfr)$, the intermediate scattering function decays as:
\begin{equation}
  F(\bfq,t) = S(\bfq) \; e^{- q^2 \cD_{\text{eff}} \; t} \; , \qquad \cD_{\text{eff}} \equiv \cD_0 \left(1 + \frac{d \log v_0}{d\tilde{\rho}} \rho_0 \;\right)\;.
  \label{eq:F_q-decay}
\end{equation}
It can be instructive to compare this result with $F(\bfq,t)$ in an ideal gas. In the latter, it is known \cite{chaikin_principles_1995} that the intermediate scattering function decays exponentially in time with a rate $\cD_0 q^2$, where $\cD_0$ is the diffusivity of the gas. This scaling is found also in the active case, as shown in Eq.~\eqref{eq:F_q-decay}; however, the value of the effective diffusivity $\cD_{\text{eff}}$ in the active gas is renormalized by QS interactions.

\subsubsection{Simulations}
\label{sec:simulations_correlations}
To test the predictions of our field
theory~\eqref{eq:macro_field_theory}, we perform particle-based
simulations with QS-RTPs in $1d$ moving according to the
dynamics~\eqref{eq:RTPs_dyn}. We consider a self-propulsion speed
regulated as:
\begin{eqnarray}
    \label{eq:QS_v_rho}
    v(\tilde{\rho}) &=& v_0 \exp \left[ \kappa \tanh\left( \frac{\tilde{\rho}-\rho_m}{\varphi} \right) \right] \; , \\[0.2cm]
    \tilde{\rho}(x) &=& (K \circledast \rho)(x) \;, \qquad
    K(x) = \frac{1}{Z} \exp\left(-\frac{r_0^2}{r_0^2-x^2}\right) \theta(r_0-|x|) \;,
    \label{eq:rho_tilde_def}
\end{eqnarray}
where $\circledast$ denotes the convolution product, and $Z$ is a
normalization factor for the bell-shaped convolution kernel $K$.  The
tumbling rate $\alpha_0$ is kept constant, and the QS-interaction radius $r_0 = 1$.

\begin{figure}
\centering
\includegraphics[width=0.47\textwidth]{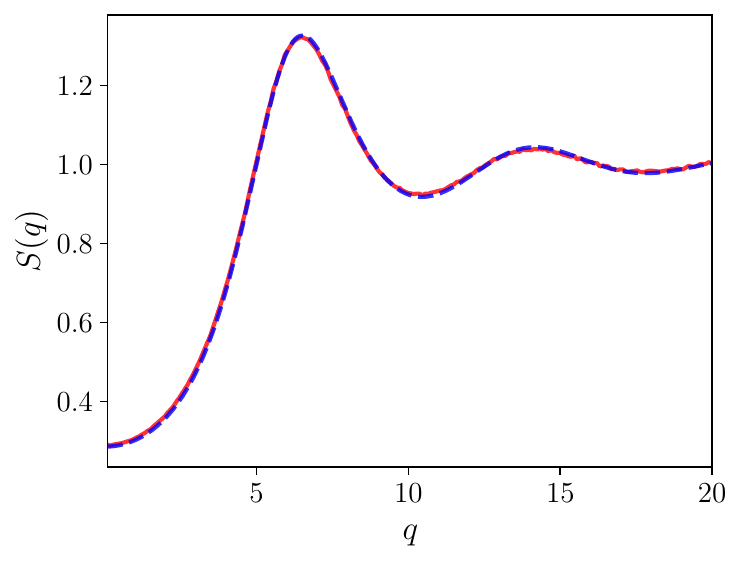}\hspace{0.5cm}
\includegraphics[width=0.47\textwidth]{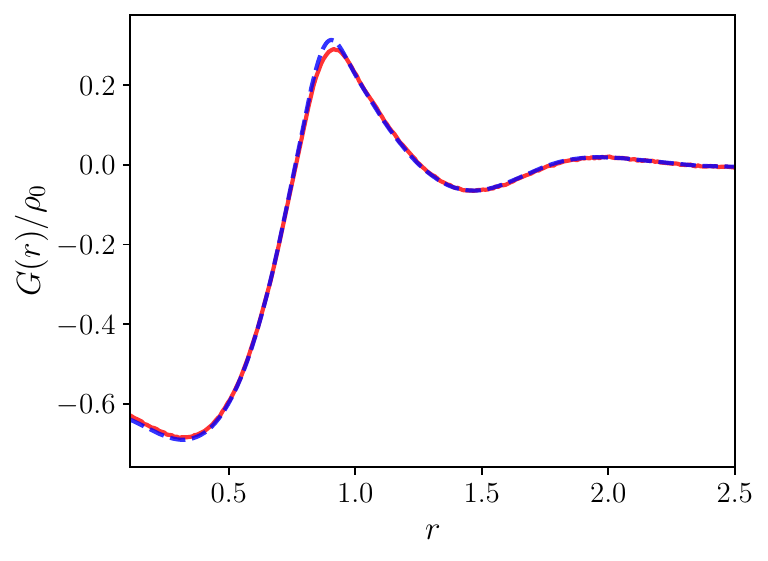}
\caption{Structure factor $S(q)$ (\textit{left}) and spatial correlation function $G(r)$ (\textit{right}) for a homogeneous gas of RTPs interacting via QS according to Eq~\eqref{eq:QS_v_rho}: comparison between numerical measurements (solid line) and theoretical predictions from Eqs.~\eqref{eq:S_q}, \eqref{eq:S_G} (dashed blue line). Parameters of the simulation: $\rho_0 = 50, \rho_m = 50, \varphi = 20, \alpha_0=2, v_0=1, \kappa=1$. Size of the simulation domain $L_x = 1000$.}
\label{fig:S_q-G_r}
\end{figure}

\begin{figure}
\centering
\includegraphics[width=0.47\textwidth]{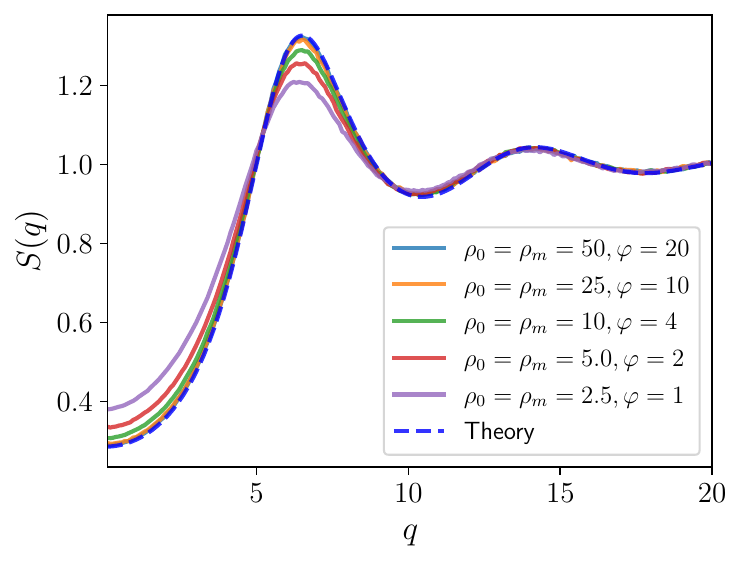}\hspace{0.5cm}
\includegraphics[width=0.47\textwidth]{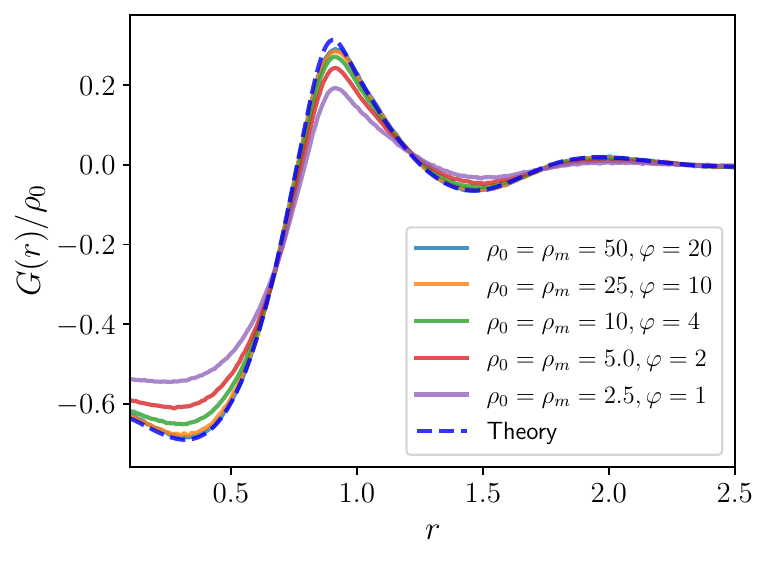}
\caption{Structure factor $S(q)$ (\textit{left}) and spatial correlation function $G(r)$ (\textit{right}) for a homogeneous gas of RTPs interacting via QS according to Eq~\eqref{eq:QS_v_rho} for different values of the density $\rho_0$: comparison between numerical measurements (solid lines) and theoretical predictions from Eqs.~\eqref{eq:S_q}, \eqref{eq:S_G} (dashed blue line). Parameters of the simulation: $\rho_0 = \rho_m, \varphi = 2/5~\rho_m, \alpha_0=2, v_0=1, \kappa=1$. Size of the simulation domain $L_x = 1000$.}
\label{fig:S_q-G_r_different-rhos}
\end{figure}

We simulate our system for values of the parameters where the
steady-state configuration is homogeneous and measure both the
structure factor $S(q)$ and the correlation function $G(r)$. As shown
in Fig.~\ref{fig:S_q-G_r}, the agreement between theory and
simulations is remarkable at sufficiently large densities, without any
fit parameter. We note that, at smaller densities,
discrepancies between our final predictions and numerical simulations
are both expected and observed, due to the failure of the
additive-noise approximation, Eq.~\eqref{eq:add_noise-approx}. To show the convergence at high density, we perform simulations at increasing values of $\rho_0=h \bar \rho$, with $\bar\rho=2.5$ and $h\in\{1,2,4,10,20\}$. To collapse our predictions, $\rho_m$ and $\varphi$ in Eq.~\eqref{eq:QS_v_rho} are also rescaled by a factor of $h$. That way, the theoretical structure factor $S(q)$  predicted by Eq.~\eqref{eq:S_qn} does not depend on $h$, while the corrections due to the additive-noise approximation are expected to decay with $h$. This is indeed the behavior reported in 
Fig.~\ref{fig:S_q-G_r_different-rhos}.

\begin{figure}
\centering
\includegraphics[width=0.47\textwidth]{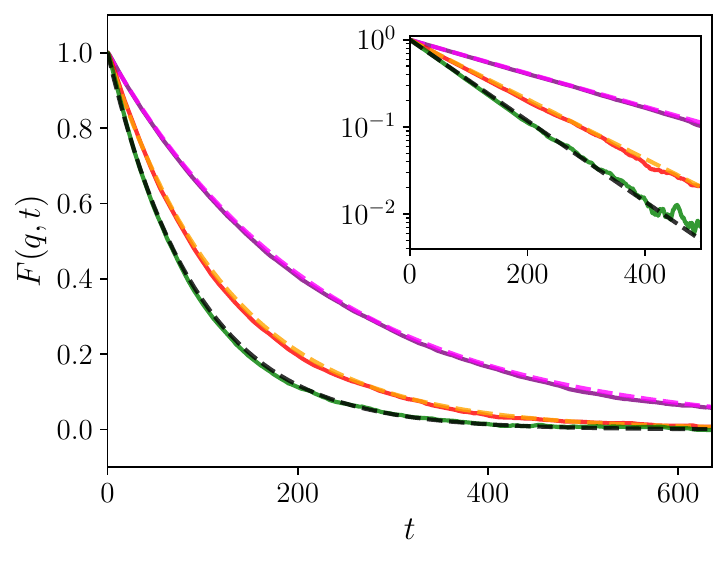}\hspace{0.5cm}
\includegraphics[width=0.47\textwidth]{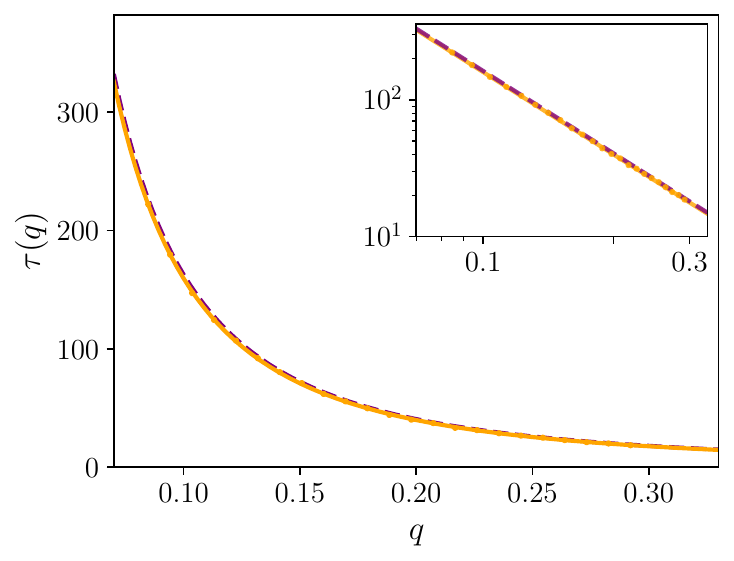}
\caption{Measurements of the intermediate scattering function for a homogeneous gas of RTPs interacting via QS according to Eq~\eqref{eq:QS_v_rho} in $1d$. \textit{(Left)}: intermediate scattering function $F(q,t)$ for three different Fourier modes ($q=0.085$ in magenta, $q=0.113$ in orange, $q=0.132$ in green). The solid curves represent the theoretical predictions from Eq.~\eqref{eq:F_q-decay}, while the dashed lines correspond to $F(q,t)$ measured from simulations. The same curves are represented in semi-log scale in the inset, to highlight the exponential behaviour. \textit{(Right)}: Decay time $\tau(q)$ of $F(q,t)$ as a function of the wave-vector $q$. Data-points are obtained by measuring $\tau$ from the decay of $F(q,t)$ at a given $q$; the errorbars are so small that they are not visible ($\sim 10^{-3}$). The fitted curve (orange) is then compared with our theoretical prediction (purple) from Eq.~\eqref{eq:F_q-decay}. The inset shows the same curves in log-log scale, to highlight the power-law behaviour.
Parameters of the simulation: $\rho_0 = 30, \rho_m = 50, \varphi = 20, \alpha_0=1, v_0=1, \kappa=0.5$. Size of the simulation domain $L_x = 2000$.}
\label{fig:F_q}
\end{figure}

We then turn to the measurement of the intermediate scattering function $F(q,t)$. We consider a system with motility enhancement ($\kappa > 0$) and fix the density $\rho_0$. For a range of Fourier modes, we measure the decay time $\tau(q)$ of the corresponding $F(q,t)$ from simulations. We then fit the curve $\tau(q)$ and determine the value of the effective diffusivity $\cD_{\text{eff}}$, defined as in Eq.~\eqref{eq:F_q-decay}. By comparing the measured effective diffusivity with its theoretical value, we are able to test our analytical predictions for $F(q,t)$. 
For an average density of $\rho_0 = 30$, the theoretical value of $\cD_{\text{eff}} = 0.614$ is in good agreement with the result obtained from the fit: $\cD_{\text{eff}}^{\> \text{fit}} = 0.626 \pm 0.001$. In the left panel of Figure~\ref{fig:F_q}, we report three examples for different modes of the exponential decay of $F(q,t)$. In the right panel, we plot $\tau(q)$ as a function of the inverse wavelength $q$, comparing the results from our simulations with the analytical predictions.

\section{Active mixtures}\label{sec:mixtures}
In this Section, we show how the methods described in the previous
sections can be generalized to active mixtures. Multi-component active
systems have drawn increasing attention in recent
years~\cite{saha2019pairing,agudo2019active,saha2020scalar,you2020nonreciprocity,frohoff2021localized,frohoff2021suppression,duan2023dynamical,brauns2024nonreciprocal},
due to the rich phenomenology they exhibit both at the static and
dynamical level: from the demixing of two \textit{E. Coli}
strains~\cite{curatolo2020cooperative} to run-and-chase dynamics in
bacterial mixtures~\cite{dinelli2023non,duan2023dynamical} and
emergent chiral phases in two species of aligning
particles~\cite{fruchart2021non,kreienkamp2022clustering}. To
understand--and possibly control--the wealth of phenomena that emerge
in these systems we thus need to bridge between microscopic and
macroscopic dynamics of active mixtures. Here we consider a system of
$N$ particles belonging to $S$ different species; we label each
species with an index $\mu \in \{1, \dots, S\}$, and assume that there
are a total of $N_\mu$ particle of type $\mu$. Each particle will thus
be identified by a pair of indices $(i,\mu)$, with $i \in
\{1,\dots,N_\mu\}$. Finally, we denote by $\rho_\mu$ the density field
associated with species $\mu$, defined as:
\begin{equation}
  \rho_\mu (\bfr,t) = \sum_{i = 1}^{N_\mu} \delta(\bfr - \bfr_{i,\mu}(t)) \;.
\end{equation}
Each $(i,\mu)$-particle undergoes motility regulation through QS via: 
\begin{equation}
 \gamma_\mu = \gamma_{0\mu}(\bfr_{i,\mu}, [\{\rho_\nu\}]) \;,
 \end{equation}
where $\gamma_\mu$ stands for any motility parameter (persistence
time, self-propulsion speed...). As regards to chemotaxis, we consider
the general case where the bias on particle $(i,\mu)$ is generated by
the gradients of $n$ different chemical fields $\{c_h(\bfr)\}$:
\begin{equation}
 \gamma_\mu = \gamma_\mu(\bfr_{i,\mu}, \{\nabla_\bfr c_h(\bfr)\}) \;.
 \end{equation}
For tactic interactions, the chemical fields are taken to be
functionals of the density fields: $c_h(\bfr,\{[\rho_\nu]\})$. All in
all, we express the effect of QS and chemotaxis on motility as:
\begin{equation}
    \gamma_\mu(\bfr_{i,\mu},\{[\rho_\nu]\}) = \gamma_{0\mu}(\bfr_{i,\mu},\{[\rho_\nu]\}) + \bfu_{i,\mu} \cdot \sum_{h=1}^n \gamma_{1\mu}^h \nabla_{\bfr_{i,\mu}} c_h(\bfr_{i,\mu},\{[\rho_\nu]\}) \;.
\end{equation}
As in Sec.~\ref{sec:interacting}, we assume a separation of time
scales between the fast microscopic degrees of freedom and the slowly diffusing density fields. Hence, we map the $N$-body microscopic
dynamics into a non-interacting problem through the frozen-field
approximation, thanks to which the motility parameters become
position-dependent functions. We are thus able to write down the
master equation for the probability $\cP_\mu(\bfr,\bfu)$ of finding a
particle of type $\mu$ in position $\bfr$ with orientation
$\bfu$.

For a mixture of RTPs and ABPs, the $S$-species master equation
generalizes the single-species one, Eq.~\eqref{ABP-RTP_dDim:FP}, as:
\begin{equation}
\partial_t\cP_\mu(\br,\bu) = -\nabla_\br \cdot \left[ v_\mu \bu \cP_\mu - D_{t\mu} \nabla_\br \cP_\mu \right] - \alpha_\mu \cP_\mu + \frac{1}{\Omega} \int \alpha_\mu \cP_\mu \rmd \bfu +\lapu \Gamma_\mu \mcP_\mu \;,
\label{ABP-RTP_dDim_mixture:FP}
\end{equation}
where the motility parameters are given by:
\begin{eqnarray}
\label{eq:v_mixture}
  v_\mu      &=& v_{0\mu}(\bfr) -  \bfu \cdot \sum_{h=1}^n v_{1\mu}^h \> \nabla_\bfr c_h(\bfr) \\[0.2cm]
\label{eq:alpha_mixture}
  \alpha_\mu &=& \alpha_{0\mu}(\bfr) + \bfu \cdot \sum_{h=1}^n \alpha_{1\mu}^h \> \nabla_\bfr c_h(\bfr) \\[0.2cm]
\label{eq:gamma_mixture}
  \Gamma_\mu &=& \Gamma_{0\mu}(\bfr) + \bfu \cdot \sum_{h=1}^n \Gamma_{1\mu}^h \> \nabla_\bfr c_h(\bfr)\;.
\end{eqnarray}
Similarly, for a mixture of AOUPs, the single-species master equation~\eqref{AOUP_dDim:FP} is generalized to:
\begin{eqnarray}
\partial_t\cP_\mu(\br,\be) &=&-\divr \left[v_{\mu} \be\cP_\mu -D_{t\mu}\gradr \cP_\mu \right] - \dive\left[ -\tau_\mu^{-1}\be\cP_\mu -\frac{1}{d}\grade (\tau_\mu^{-1} \cP_\mu)\right] \;,
\label{AOUP_dDim_mixture:FP}
\end{eqnarray}
with:
\begin{eqnarray}
\label{eq:v_mixtureAOUP}
  v_\mu      &=& v_{0\mu}(\bfr) -  \bfe \cdot \sum_{h=1}^n v_{1\mu}^h \> \nabla_\bfr c_h(\bfr) \\[0.2cm]
  \label{eq:tau_mixture}
  \tau_\mu^{-1}  &=& \omega_{0\mu}(\bfr) +  \bfe \cdot \sum_{h=1}^n \omega_{1\mu}^h\> \nabla_\bfr c_h(\bfr) \;.
\end{eqnarray}
Starting from the microscopic
dynamics~\eqref{ABP-RTP_dDim_mixture:FP},~\eqref{AOUP_dDim_mixture:FP},
one then follows the same steps as in the single-species case
presented in Sec.~\ref{sec:ABP-RTP_d} to obtain an effective Langevin
description at the mesoscopic scale. For the sake of completeness, we
report the full computation in Appendix~\ref{app:cg_mixtures}. Under
the diffusion-drift approximation, the large-scale dynamics of
particle $i$ of type $\mu$ eventually follows an It\=o-Langevin
equation:
\begin{equation}
    \dot{\bfr}_{i,\mu} = \mathbf{V}_\mu(\bfr_{i,\mu},[\{\rho_\nu\}]) + \nabla_{\bfr_{i,\mu}} \cD_\mu(\bfr_{i,\mu},[\{\rho_\nu\}]) + \sqrt{2  \cD_\mu(\bfr_{i,\mu},[\{\rho_\nu\}]) } \bxi_{i,\mu}(t)
    \label{eq:Langevin_meso_mixture}
\end{equation}
where the $\{\bxi_{i,\mu}(t)\}$ are delta-correlated, centred Gaussian white noise terms, and: 
\begin{eqnarray}
\label{ABP-RTP_dDim_mixture:drift_diff}
\text{ABP-RTP}: &\quad&
\begin{cases}
    & \mathbf{V}_\mu = -\dfrac{v_{0\mu} \nabla v_{0\mu}}{d\left[\alpha_{0\mu}  + (d-1)\Gamma_{0\mu}\right]} - \dfrac{1}{d} \sum_{h=1}^n \left[ v_{1\mu}^h + v_{0\mu} \dfrac{\alpha_{1\mu}^h +(d-1) \Gamma_{1\mu}^h}{\alpha_{0\mu} + (d-1)\Gamma_{0\mu}}\right] \nabla_{\bfr} c_h \\[0.6cm]
    & \cD_\mu = \dfrac{v_{0\mu}^2}{d \left[\alpha_{0\mu}  + (d-1)\Gamma_{0\mu}\right]} + D_{t\mu} \ .
    \end{cases} \\[0.5cm]
\text{AOUP}: &\quad&
\begin{cases}
    & \mathbf{V}_\mu = -\dfrac{v_{0\mu} \nabla v_{0\mu}}{d\omega_{0\mu}} - \dfrac{1}{d} \sum_{h=1}^n \left[ v_{1\mu}^h + v_{0\mu} \dfrac{\omega_{1\mu}^h}{\omega_{0\mu}}\right] \nabla_{\bfr} c_h \\[0.6cm]
    & \cD_\mu = \dfrac{v_{0\mu}^2}{d \omega_{0\mu}} + D_{t\mu} \ .
\end{cases}
\label{AOUP_dDim_mixture:drift_diff}
\end{eqnarray}

\subsection{Coupled fluctuating hydrodynamics for active mixtures}
Starting from the stochastic dynamics~\eqref{eq:Langevin_meso_mixture} we now derive the time-evolution of the density field of species $\mu$:
\begin{equation}
    \rho_\mu(\mathbf{r},t) = \sum_{i=1}^{N_\mu} \delta(\mathbf{r} - \mathbf{r}_{i,\mu}(t))
    \label{eq:SI_def_density_mixture}
\end{equation}
where the sum is taken over all particles of species $\mu$. This is a generalization of the single-species case of Sec.~\ref{sec:Dean}, which we detail here for the sake of completeness. Applying the It\=o formula to Eq.~\eqref{eq:SI_def_density_mixture}, one gets
    \begin{equation}
        \frac{d }{dt}\rho_\mu(\bfr,t) =  \sum_{i=1}^{N_\mu} \left[\nabla_{\bfr_{i,\mu}} \delta(\mathbf{r} - \mathbf{r}_{i,\mu}(t)) \cdot \dot{\bfr}_{i,\mu} + \cD_\mu(\bfr_{i,\mu}, [\{\rho_\nu\}])\nabla_{\bfr_{i,\mu}}^2 \delta(\mathbf{r} - \mathbf{r}_{i,\mu}(t)) \right]
        \label{eq:SI_B9_mixture}
    \end{equation}
   In the next passages we omit the $[\{\rho_\nu\}]$ dependence in $\cD_\mu, \mathbf{V_\mu}$, which is implicitly assumed throughout the derivation. The first term in Eq.~\eqref{eq:SI_B9_mixture} can be re-expressed as:
    \begin{eqnarray}
    \tiny
    \sum_{i=1}^{N_\mu} \nabla_{\bfr_{i,\mu}} \delta(\mathbf{r} - \mathbf{r}_{i,\mu}(t)) \cdot \dot{\bfr}_{i,\mu} &=& \sum_{i=1}^{N_\mu}  \nabla_{\bfr_{i,\mu}} \delta(\mathbf{r}-\mathbf{r}_{i,\mu}) \cdot \biggl(\> \mathbf{V}_\mu(\bfr_{i,\mu}) + \nabla_{\bfr_{i,\mu}} \cD_\mu (\bfr_{i,\mu}) + \sqrt{2 \cD_\mu (\bfr_{i,\mu})} \> \bxi_{i,\mu} \>\biggr) \notag \\ 
    &=& - \sum_{i=1}^{N_\mu} \nabla_\bfr \delta(\mathbf{r}-\mathbf{r}_{i,\mu}) \cdot \biggl(\> \mathbf{V}_\mu(\bfr_{i,\mu}) + \nabla_{\bfr_{i,\mu}} \cD_\mu (\bfr_{i,\mu}) + \sqrt{2 \cD_\mu (\bfr_{i,\mu})} \> \bxi_{i,\mu} \>\biggr) \notag \\
    &=& - \sum_{i=1}^{N_\mu} \nabla_\bfr \cdot \biggl[ \delta(\mathbf{r}-\mathbf{r}_{i,\mu}) \biggl( \mathbf{V}_\mu(\bfr_{i,\mu}) + \nabla_{\bfr_{i,\mu}} \cD_\mu (\bfr_{i,\mu}) + \sqrt{2 \cD_\mu (\bfr_{i,\mu})} \> \bxi_{i,\mu} \biggr) \>\biggr] \notag \\
    &=& - \sum_{i=1}^{N_\mu} \nabla_\bfr \cdot \biggl[ \delta(\mathbf{r}-\mathbf{r}_{i,\mu}) \biggl( \mathbf{V}_\mu(\bfr) + \nabla_{\bfr} \cD_\mu (\bfr) + \sqrt{2 \cD_\mu (\bfr)} \bxi_{i,\mu} \biggr) \>\biggr]\label{eq:N-1_mixture} \\
    &=& - \nabla_\bfr \cdot \biggl[ \rho_\mu(\bfr,t) \biggl( \mathbf{V}_\mu(\bfr) + \nabla_{\bfr} \cD_\mu (\bfr) \biggr) + \sqrt{2 \cD_\mu \rho_\mu (\bfr,t)} \mathbf{\Lambda}_\mu(\bfr,t) \>\biggr]\label{eq:N_mixture}\;.
    \label{eq:SI_nabla1_mixture}
    \end{eqnarray}
    Similarly to what we did in Sec.~\ref{sec:Dean}, to go from Eq.~\eqref{eq:N-1_mixture} to~\eqref{eq:N_mixture} we have introduced $S$ centered Gaussian white noise fields $\mathbf{\Lambda}_\mu(\mathbf{r},t)$ with:
    \begin{equation}
    \langle \mathbf{\Lambda}_\mu(\mathbf{r},t) \rangle = 0 \; , \quad  \langle \Lambda_{\mu,i} (\mathbf{r},t) \Lambda_{\nu,j}(\mathbf{r}',t') \rangle = \delta_{ij} \delta_{\mu\nu} \delta(t-t') \delta(\bfr - \bfr')
    \end{equation}
    where $\mu,\nu$ are the usual species indices and $i,j$ indicate
    spatial components. Similarly, the second term in
    Eq.~\eqref{eq:SI_B9_mixture} can be rewritten as:
    \begin{eqnarray}
    \sum_{i=1}^{N_\mu} \cD_\mu (\bfr_{i,\mu}) \nabla_{\bfr_{i,\mu}}^2 \delta(\mathbf{r} - \mathbf{r}_{i,\mu}(t)) &=& \sum_{i=1}^{N_\mu} \cD_\mu (\bfr_{i,\mu}) \nabla_{\bfr}^2 \delta(\mathbf{r} - \mathbf{r}_{i,\mu}(t)) 
    = \sum_{i=1}^{N_\mu} \nabla_{\bfr}^2 \left[ \delta(\mathbf{r} - \mathbf{r}_{i,\mu}(t)) \cD_\mu (\bfr_{i,\mu}) \right] \notag\\
    &=& \sum_{i=1}^{N_\mu} \nabla_{\bfr}^2 \left[ \delta(\mathbf{r} - \mathbf{r}_{i,\mu}(t)) \cD_\mu (\bfr) \right]
    = \nabla_{\bfr}^2 \bigl[\> \rho_\mu(\bfr,t) \cD_\mu (\bfr) \> \bigr] \;.
    \label{eq:SI_nabla2_mixture}
    \end{eqnarray}
    Finally, we insert the expressions~\eqref{eq:SI_nabla1_mixture}, \eqref{eq:SI_nabla2_mixture} into Eq.~\eqref{eq:SI_B9_mixture} to get the fluctuating hydrodynamics of the density fields:
   \begin{equation}
  \partial_t \rho_\mu = - \nabla_{\mathbf{r}} \cdot \biggl\{  \mathbf{V}_\mu(\mathbf{r}, [\{\rho_\nu\}]) \rho_\mu - \cD_\mu (\mathbf{r}, [\{\rho_\nu\}])\nabla_{\mathbf{r}} \rho_\mu + \sqrt{2 \cD_\mu(\mathbf{r}, [\{\rho_\nu\}]) \rho_\mu} \>\> \mathbf{\Lambda}_\mu(\mathbf{r},t) \biggr\}\;.
\label{eq:macro_field_theory_SSpecies}
\end{equation}
Eq.~\eqref{eq:macro_field_theory_SSpecies} can now be used to
described the large-scale collective behaviors of $S$ species of ABPs,
RTPs, or AOUPs interacting via QS or taxis.

\section{Conclusion and discussion}
In this work, we have bridged the microscopic dynamics and the
large-scale behaviors of dry scalar active systems. We have studied
three distinct types of microscopic dynamics, namely run-and-tumble
(RTP), active Brownian (ABP) and active Ornstein-Uhlenbeck (AOUP) and
considered motility regulation both by external spatial modulation and
by density-dependent interactions like quorum sensing and
chemotaxis.

In all cases, we have mapped the microscopic dynamics of these systems
into an effective Langevin description via a diffusive approximation,
valid at large spatial and temporal scales. Finally, we have derived
the associated fluctuating dynamics for the density modes. We have
tested the results of the coarse-grained theory against particle-based
simulations for both the non-interacting and interacting cases; in the
latter, we have managed to compute correlation functions starting from
the stochastic hydrodynamics, obtaining a significant agreement with
measurements from microscopic simulations. Finally, we have extended
the coarse-graining machinery to active mixtures, \textit{i.e} active
systems made up of many components.

Establishing a connection between the microscopic and macroscopic
behavior of active systems is a problem of paramount importance to
achieve a fine control over the rich emergent phenomenology of these
systems, with crucial implications both for biology across scales and
for the engineering of smart materials. While symmetry-based
phenomenological theories can capture the qualitative features of the
macroscopic dynamics, this approach is limited by the lack of
connection with an explicit microscopic model, hence the need for a
solid coarse-graining framework for active systems.

In this article, we have thus proposed a general approach to
coarse-graining in scalar active matter by considering different types
of microscopic dynamics and interactions. The methods described here
are not exclusive to dry active matter, but bear strong analogies with
the ones in the literature of kinetic theories for wet active
systems~\cite{saintillan2008instabilities,weady2022thermodynamically}. Our
hope is that this work can offer a basic toolbox that can be employed
beyond the cases considered here.

Obviously, even within the context of dry scalar active matter, this
work is far from offering a complete overview. For instance, it would
be worth studying the interplay between motility-regulation and steric
repulsion, which is especially relevant for dense active systems. {While there already exist well-established coarse-graining methods in the literature for active particles with pairwise forces~\cite{fily2012athermal,solon_generalized_2018,te2023derive}, how these interactions compete with motility regulation remains poorly characterized.} In
addition to that, the recent years have seen an upsurge of interest
for proliferating active matter~\cite{hallatschek2023proliferating},
characterized by a non-conserved number of particles. These may
include, for instance, active systems with birth-and-death dynamics,
prey-predator interactions, chemical reactions and so forth. The role
of population dynamics has been investigated through phenomenological
field theories, for instance by showing that it can lead to arrested
phase separation and wavelength
selection~\cite{cates2010arrested}. Providing a solid theoretical
framework to bridge from microscopic to macroscopic descriptions in
proliferating active matter would thus be an exciting research
direction to pursue in the future.

Finally, we note that our coarse-graining strongly relies on the
diffusion approximation. How to go beyond that approximation to
capture the leading order correction in $\ell_p/L$ is a fascinating
open challenge on which progress has been done
recently~\cite{duan2023dynamical}. {Research in this direction could also be relevant for the study of critical density fluctuations, as recent works~\cite{paoluzzi2016critical,gnan2022critical,maggi2022critical,paoluzzi2023noise} have shown how a colored noise in the dynamics of the density field can be used to compute critical exponents in active field theories. Whether such colored noise could naturally emerge from a refined coarse-grained theory would be an interesting question to address.}

\textbf{Acknowledgements.} The authors thank F. Ghimenti, G. Spera and P. Muzzeddu for
useful discussions and A. Curatolo for early involvement in this
work. JT acknowledges the support of ANR grant THEMA. AD acknowledges
an international fellowship from Idex Universite de Paris.

\textbf{Data availability statement.} The data that support the
findings of this study are available upon request from the authors.

\textbf{Copyright.} This Accepted Manuscript is available for reuse
under a CC BY-NC-ND licence after the 12 month embargo period provided
that all the terms and conditions of the licence are adhered to.

\printbibliography

\newpage

\appendix{}

\section{Spherical harmonics, harmonic tensors and order parameters}
\label{app:harmTensor}

\paragraph{Generalized Fourier series.} Square integrable, real--valued functions on the unit sphere $\mbS^{d-1}$ of $\mbR^d$ are notoriously  decomposable onto the eigenfunctions of the Laplacian of $\mbS^{d-1}$, here denoted by $\Delta_\bu$. Formally, this is written as the Hilbert direct--sum decomposition 
\begin{equation}
\mbL^2(\mbS^{d-1})=\bigoplus_{n\in\mbN} \mcH_n(\mbS^{d-1}) \ ,
\label{eq:Hilbert_sum}
\end{equation}
where $\mcH_n$ is the eigenspace of $\Delta_\bu$ with eigenvalue
$-n(n+d-1)$, the dimension of which is $\dim(\mcH_n) =
\binom{n+d-1}{n} - \binom{n+d-3}{n-2}$. Since the operator
$\Delta_\bu$ is self--adjoint for the canonical scalar product of
$\mbL^2(\mbS^{d-1})$, the spaces $\mcH_n$ are two--by--two orthogonal, and
their elements are called the $n^{th}$ order spherical
harmonics. In practice, this decomposition, which generalizes the
Fourier decomposition to higher dimensions, is done by choosing an
orthonormal basis $(Y_{n\ell})_\ell$ for each $\mcH_n$. One then
decomposes any function $f$ as
\begin{equation}
f(\bu) = \sum_{n\in\mbN}  f_n(\bu) = \sum_{n\in\mbN} \sum_{\ell =1}^{\dim \mcH_n} c_{n\ell}Y_{n\ell}(\bu) \ ,
\label{eq:harmFunct}
\end{equation} 
where $f_n$ is the component of $f$ in $\mcH_n$ and the coefficients $c_{n\ell}$ are obtained by taking the scalar products between $f$ and the corresponding basis elements $Y_{n\ell}$. Note that, in general, the choice of the $Y_{n\ell}$ is done relatively to a previous, \textit{arbitrary} choice of an orthonormal basis of $\mbR^n$. 

\paragraph{Order parameter.}
Rotational invariance is a fundamental symmetry of the laws of
nature. Consequently, the disordered phase of numerous many--body systems respect this symmetry. Nevertheless, as a control
parameter is changed, this symmetry can be spontaneously broken, leading to the system being invariant only under the action of a subgroup $G$
of $SO(d)$. To account for the symmetry-change of the system, one then needs to introduce an order parameter. The latter is a function(al) of the probability distribution $f$ which has to:
\begin{enumerate}[(i)]
\item vanish when $f$ is invariant under the full $SO(d)$ group ;
\item be invariant under $G$ in the `low-temperature' phase, when $f$ is not invariant under $SO(d)$ but is $G$ invariant;
\item be ``as simple as possible'', in a loose sense. In practice, most order parameters belong to a tensor space on which $SO(d)$ acts linearly and whose dimension is chosen as small as possible. 
\end{enumerate}

To construct the order parameter, one can use the
decomposition~\eqref{eq:Hilbert_sum}.
Indeed, the subspaces $\mcH_n$ are so--called
irreducible representations of $SO(d)$\footnote{We recall that a representation of a group consists in a vector space on
which this group acts through linear transformations.}: each harmonic component $f_n$ in Eq.~\eqref{eq:harmFunct} remains in $\mcH_n$
after an arbitrary rotation is applied to $f$ and, furthermore,  the
spaces $\mcH_n$ are minimally stable, i.e. it
is not possible to further decompose them into
smaller subspaces that are rotationally stable. The
components $f_n$ are thus good candidates to act as order parameters. Note that, in dimension $d=2$ and $d=3$, it can even be shown that they are the only ones, in the sense that any irreducible representation of $SO(d)$ is isomorphic to one of the $\mcH_n$.

In practice, one constructs the order parameter as follows: Consider all the $\mcH_n$ that contains a non-trivial $G$--invariant subspace and choose the smallest corresponding value $n_0$. Then $f_{n_0}$ satisfies all the requirements to play the role of an order parameter:
\begin{itemize}
\item[$\rightarrow$] condition (i) is satisfied since any rotationally invariant function $f$ on $\mbS^{d-1}$ is such $f_n=0$, for all $n>0$. Thus, $f_{n_0} = 0$ in the rotationally invariant phase.
\item[$\rightarrow$] condition (ii) since, if $f$ is $G$-invariant, so are all the $f_n$, and $f_{n_0}$ in particular. 
\item[$\rightarrow$] condition (iii) since $f_{n_0}$ transforms linearly under $SO(d)$ (because it belongs to one of its representations) and the dimension of $\mcH_{n_0}$ is minimal by definition.
\end{itemize}

\if{
In addition, among all the potential candidates in our procedure, it is intuitively obvious that we chose the simplest one we chose the simplest one in the by choosing the minimum degree $n_0$---all the other being higher order polynomials---as further explained in the next section.

In practice, an order parameter obtained by the procedure described above is not very convenient yet.
Indeed, to describe the dynamics of an inhomogeneous---\textit{e.g} nematic or polar---system, it is not very convenient to model its evolution by means of a field that assigns to each point in space an element of $\mcH_{n_0}$, \textit{i.e.} a $\mbR$--valued function on the unit sphere. We would usually rather like to be able to locally describe the system by arrays of numbers. In addition, these arrays of numbers should transform nicely under change of coordinates. To do so, one could try to pick an arbitrary basis of $\mcH_{n_0}$, and stack the coordinates of the local $\mcH_{n_0}$--value of the field into a column vector. Unfortunately, the resulting array would generically transform in an awful way under change of basis. Luckily, there exists a slightly more subtle, basis--invariant, way of associating to any element of $\mcH_{n_0}$ a ``nice array of numbers'', \ie a tensor, as described in the following paragraph.
}\fi

\paragraph{From harmonic scalar functions to harmonic tensors.}
At this stage, our order parameter is a real-valued function $f_{n}$ of $\bu\in\mbS^{d-1}$. (From now on, we drop the subscript $0$ for clarity.) As any function of $\mcH_{n}$, it is the restriction of a homogeneous, harmonic
polynomial\footnote{We recall that a polynomial on $\mbR^d$ is said to be harmonic whenever it lies in the kernel of the Laplacian $\Delta$ of $\mbR^d$~\cite{atkinson2012spherical}.} of order $n$ on $\mbR^d$. In particular, for each
element $Y_{n\ell}(\bu)$ of the arbitrarily--chosen basis appearing in
Eq.~\eqref{eq:harmFunct}, we denote this polynomial on $\mbR^d$
by $Y_{n\ell}(\br)$.  Then, we recall that for, any $n^{\rm
  th}$--order homogeneous polynomial on $\mbR^d$, there exists a
unique $n^{\rm th}$--order symmetric tensor $\bY_{n\ell}$ such that
\begin{equation}
Y_{n\ell}(\bu)= \bY_{n\ell}\cdot \bu^{\otimes n} = \sum_{i_1,\dots, i_n =1} ^d Y^{i_1\dots i_n}u_{i_1}\dots u_{i_n} \ .
\end{equation}
Note that the harmonicity of $Y_{n\ell}(\br)$ is equivalent to the vanishing of the trace of the associated tensor: $\mathrm{tr}(\bY_{n\ell})\equiv \sum_j Y^{j j i_3\dots i_n} =0$.

Since $f_n$ is a homogeneous polynomial of order $n$, there exists a unique corresponding harmonic tensor $\bA_n$ such that
\begin{equation}
f(\bu) = \sum_{n\in\mbN}\bA_n \cdot \bu^{\otimes n} = \sum_{n\in\mbN}\left(\sum_{\ell =1}^{\dim \mcH_n} c_{n\ell} \bY_{n\ell} \right) \cdot \bu^{\otimes n} \ .
\label{eq:decompositionAn}
\end{equation}
It can then be shown that 
\begin{equation}
\bA_n = \frac{1}{\Omega}\frac{(2n+d-2)!!}{n!!(d-2)!!}\int_{\mbS_{d-1}} f(\bu)\widehat{\bu^{\otimes n}} \; \rmd\bu \ ,
\end{equation}
which is, together with~\eqref{eq:decompositionAn} and up to a rescaling of the $\bA_n$, the decomposition described in the section~\ref{sec:harmonic-tensors} of the main text. 

In addition to being very convenient for computing the diffusive limits of ABPs and RTPs in a coordinate--free manner as done in this article, the tensors $\bA_n$ are (isomorphic to) the usual order parameters chosen to describe phase transitions accompanied by a spontaneous symmetry breaking as $SO(d)\rightarrow G\subset SO(d)$ (\textit{e.g.} transitions from isotropic to polar, nematic, or hexatic phases).

\newpage

\section{From microscopic to macroscopic noise}
\label{app:Deannoise}
In this section we show in which sense the noise fields
\begin{equation} 
    - \sum_{i=1}^{N} \nabla_\bfr \cdot \left[ \sqrt{2 \cD (\bfr)} \delta(\bfr - \bfr_i) \> \bxi_i \right] \quad\text{and}\quad  -\nabla_\bfr \cdot [ \sqrt{2 \mathcal{D}(\bfr) \rho(\bfr,t)}
    \mathbf{\Lambda}(\bfr,t) ] \label{eq:noises}
\end{equation}
are equivalent. We first note that Eq.~\eqref{eq:B9} is a Markovian dynamics so that the dynamics in $[t,+\infty[$ is entirely determined by the values of $\rho(\bfr,t)$ and by the noise realizations for $s\geq t$.  To show that the noise fields appearing in Eq.~\eqref{eq:noises} leads to the same fluctuating hydrodynamics for the density field, it is thus sufficient to show that they generate the same noise statistics at time $t$, conditioned on the value of $\rho(\bfr,t)$.

To be more precise, we first rewrite Eq.~\eqref{eq:macro_field_theory} in discrete time, using It\=o discretization:
\begin{equation}
    \rho(\bfr, t_{n+1}) = \rho(\bfr, t_n) - \nabla_{\bfr} \cdot \left[A(\bfr, t_n) \Delta t + \sqrt{2 \cD(\bfr)} \sum_{i=1}^N \delta(\bfr - \bfr_i(t_n)) \Delta\bxi_i(t_n) \right] \;, \quad t_n = n \Delta t\;,
    \label{eq:almost}
\end{equation}
where $A(\bfr,t_n) \equiv \mathbf{V}(\bfr) \rho(\bfr,t_n) - \cD(\bfr) \nabla_{\bfr} \rho(\bfr,t_n)$ is the deterministic part of the stochastic equation, while the $\{\Delta\bxi_i\}$ are microscopic Gaussian white noises that satisfy:
\begin{equation}
    \langle \Delta\bxi_i(t_n) \rangle = \mathbf{0}\;, \qquad \langle \Delta\bxi_i(t_n) \otimes \Delta\bxi_j(t_m) \rangle = \bI \delta_{nm} \delta_{ij} \Delta t  \;.
    \label{eq:microstat}
\end{equation}
We remind that $i,j$ stand for particle indices whereas $n$ and $m$ refer to the discrete time values.
The fact that the noises $\Delta\bxi_i(t_n) $ are delta-correlated in time ensures that the dynamics is Markovian. To show that the two noises in Eq.~\eqref{eq:noises} lead to the same fluctuating hydrodynamics, it is thus sufficient to show that the random noises
\begin{equation}
   {\Delta \zeta(\bfr,t_n)} \equiv   -    \sum_{i=1}^N \nabla_{\bfr} \cdot \delta(\bfr - \bfr_i(t_n)) {\Delta \bxi_i(t_n)}\;\quad \text{and} \quad - \nabla_{\bfr} \cdot \left[\sqrt{ \rho(\bfr,t_n) }  {\Delta\bLambda(\bfr,t_n)} \right]
    \label{appeq:processes}
\end{equation}
have the same statistics at fixed $t_n$ and $\rho(\bfr,t_n)$, where the $\Delta\bLambda(\bfr,t_n)$ are Gaussian noise fields such that:
\begin{equation}
    \langle \Delta\bLambda(\bfr,t_n) \rangle = \mathbf{0} \;, \qquad \langle  \Delta\bLambda(\bfr,t_n)  \otimes \Delta\bLambda(\bfr',t_m) \rangle = \bI \Delta t \delta(\bfr-\bfr') \delta_{nm}  \;.
      \label{eq:lambda-stat}
\end{equation}
First of all, we note that the two processes in Eq.~\eqref{appeq:processes} are Gaussian, as they result from the application of linear operators to Gaussian noises. Therefore, to prove that they are equivalent, we only need to show the equality of the first two moments at fixed time.

We start by computing the average of the two noises. Thanks to It\=o discretization, $\Delta \bxi_i(t_n)$ and $\Delta\bLambda(\bfr,t_n)$ are independent of the values of $\bfr_i(t_n)$ and $\rho(\bfr,t_n)$ so that: 
\begin{eqnarray}
  \langle  \Delta \zeta(\bfr,t_n)\rangle&=&-    \sum_{i=1}^N \nabla_{\bfr} \cdot \delta(\bfr - \bfr_i(t_n)) \langle{\Delta \bxi_i(t_n)}\rangle=0\\
  \langle- \nabla_{\bfr} \cdot [\sqrt{ \rho(\bfr,t_n) }  \Delta\bLambda(\bfr,t_n) ]\rangle&=&- \nabla_{\bfr} \cdot [   \sqrt{ \rho(\bfr,t_n) } \langle  \Delta\bLambda(\bfr,t_n) \rangle]=0\;,
\end{eqnarray}
where we remind that the brackets correspond to averages conditioned on the value of $\rho(\bfr,t_n)$. Let us now compute the equal-time second cumulant of $\Delta \zeta(\bfr,t_n)$:
\begin{eqnarray}
  \langle  \Delta \zeta(\bfr,t_n) \Delta \zeta(\bfr',t_n)\rangle&=&
  \llangle \nabla_{\bfr} \cdot \left[ \sum_{i=1}^N \delta(\bfr - \bfr_i(t_n)) \Delta \bxi_i(t_n) \right] \nabla_{\bfr'} \cdot \left[ \sum_{j=1}^N \delta(\bfr' - \bfr_j(t_n)) \Delta \bxi_j(t_n) \right] \rrangle \notag \\[0.2cm]
  &=& \sum_{i,j} \llangle \nabla_{\bfr} \cdot \left[ \delta(\bfr - \bfr_i(t_n)) \Delta \bxi_i(t_n) \right] \nabla_{\bfr'} \cdot \left[ \delta(\bfr' - \bfr_j(t_n)) \Delta \bxi_j(t_n) \right] 
  \rrangle \notag \\[0.2cm]
  &=& \sum_{i,j}  \nabla_{\bfr}\delta(\bfr - \bfr_i(t_n)) \otimes\nabla_{\bfr'} \delta(\bfr' - \bfr_j(t_n)) \cdot  \llangle \Delta \bxi_i(t_n) \otimes \Delta \bxi_j(t_n) \rrangle\;,
  \label{eq:step1corr}
\end{eqnarray}
where we used the It\=o convention to separate the average on the noises from the rest. We remind that, if $\bT$ and $\bS$ are rank-2 tensors, the notation $\bT \cdot \bS$ stands for tensor contraction: $\bT \cdot \bS = \sum_{\alpha=1}^d \sum_{\beta=1}^d T_{\alpha\beta} S_{\alpha\beta}$. From the statistics of the microscopic noise~\eqref{eq:microstat} we get:
\begin{eqnarray}
     \langle  \Delta \zeta(\bfr,t_n) \Delta \zeta(\bfr',t_n)\rangle &=& \sum_{i,j} \delta_{ij} \bI \cdot  \nabla_{\bfr}\delta(\bfr - \bfr_i(t_n)) \otimes\nabla_{\bfr'} \delta(\bfr' - \bfr_j(t_n))
      \notag \\[0.2cm]
     &=& \Delta t \sum_{i=1}^N \bI \cdot  \nabla_{\bfr}\delta(\bfr - \bfr_i(t_n)) \otimes\nabla_{\bfr'} \delta(\bfr' - \bfr_i(t_n))  \notag\\[0.2cm]
     &=& \Delta t  \sum_{i=1}^N \sum_{\alpha=1}^d \partial_{r_\alpha}  \partial_{r'_\alpha} \delta(\bfr - \bfr_i(t_n)) \delta(\bfr' - \bfr_i(t_n)) 
      \notag \\[0.2cm]
     &=& \Delta t \sum_{\alpha=1}^d \partial_{r_\alpha}  \partial_{r'_\alpha}  \sum_{i=1}^N \delta(\bfr - \bfr_i(t_n)) \delta(\bfr' - \bfr_i(t_n)) 
      \notag \\[0.2cm]
     &=& \Delta t \sum_{\alpha=1}^d \partial_{r_\alpha}  \partial_{r'_\alpha} \left[ \delta(\bfr - \bfr')  \sum_{i=1}^N  \delta(\bfr' - \bfr_i(t_n))  \right] \notag \\[0.2cm]
     &=& \Delta t \sum_{\alpha=1}^d \partial_{r_\alpha}  \partial_{r'_\alpha} \left[ \delta(\bfr - \bfr') \rho(\bfr,t_n) \right] 
     \label{eq:step2corr}\;.
\end{eqnarray}
The equal-time correlation~\eqref{eq:step2corr} of $\Delta \zeta(t_n)$ is then equal to the one obtained from the noise $-\nabla_{\bfr} \cdot \left[\sqrt{\rho(\bfr,t_n)} \Delta\bfLL(\bfr,t_n) \right]$ for a given value of $\rho(\bfr, t_n)$, since:
\begin{eqnarray}
    \llangle \nabla_{\bfr} \cdot \left[\sqrt{\rho(\bfr,t_n)} \Delta\bfLL(\bfr,t_n) \right] \nabla_{\bfr'} \cdot \left[\sqrt{\rho(\bfr',t_n)} \Delta\bfLL(\bfr',t_n) \right] \rrangle &=& \nabla_{\bfr} \otimes \nabla_{\bfr'} \cdot \llangle \sqrt{\rho(\bfr,t_n) \rho(\bfr',t_n)} \Delta\bfLL(\bfr,t_n) \otimes \Delta\bfLL(\bfr',t_n) \rrangle \notag \\[0.2cm]
    &=& \nabla_{\bfr} \otimes \nabla_{\bfr'} \cdot  \sqrt{\rho(\bfr,t_n) \rho(\bfr',t_n)} \llangle \Delta\bfLL(\bfr,t_n) \otimes \Delta\bfLL(\bfr',t_n) \rrangle \notag \\[0.2cm]
    &=& \Delta t \nabla_{\bfr} \otimes \nabla_{\bfr'} \cdot \left[ \sqrt{\rho(\bfr,t_n) \rho(\bfr',t_n)}  \bI \delta(\bfr-\bfr') \right] \notag \\[0.2cm]
    &=& \Delta t \sum_{\alpha=1}^d \partial_{r_\alpha} \partial_{r'_\alpha} \left[\sqrt{\rho(\bfr,t_n) \rho(\bfr',t_n)} \delta(\bfr-\bfr') \right] \notag \\[0.2cm]
    &=& \Delta t \sum_{\alpha=1}^d \partial_{r_\alpha} \partial_{r'_\alpha} \left[\delta(\bfr-\bfr') \rho(\bfr,t_n) \right] \;.
\end{eqnarray}
The two stochastic increments appearing in~\eqref{appeq:processes} thus have equal first and second moment; being Gaussian, these processes are equal in law. Therefore, in the limit $\Delta t \to 0$, the two noises~\eqref{eq:noises} generate the same fluctuating hydrodynamics. 

\if{
More generally, one can also look at higher-order moments at equal time of the noise $\Delta \zeta$. For the $p$-th order moment, we get:
\begin{eqnarray}
    \llangle \prod_{k=1}^p \Delta \zeta(\bfr^{(k)}, t_n) \rrangle &=& \llangle \prod_{k=1}^p \nabla_{\bfr^{(k)}} \cdot \left[ \sum_{i=1}^N \delta(\bfr^{(k)}-\bfr_i(t_n)) \Delta \bxi_i(t_n) \right] \rrangle \notag \\[0.2cm]
    &=& \sum_{i_1=1}^N \cdots \sum_{i_p=1}^N \llangle \prod_{k=1}^p  \Delta\bxi_{i_k}(t_n) \cdot \nabla_{\bfr^{(k)}}  \delta(\bfr^{(k)}-\bfr_{i_k}(t_n)) \rrangle \notag \\[0.2cm]
    &=& \sum_{i_1=1}^N \cdots \sum_{i_p=1}^N \sum_{\alpha_1=1}^d \cdots \sum_{\alpha_p=1}^d \llangle \prod_{k=1}^p  \Delta\xi_{i_k,\alpha_k}(t_n) \partial_{r^{(k)}_{\alpha_k}}  \delta(\bfr^{(k)}-\bfr_{i_k}(t_n)) \rrangle \notag \\[0.2cm]
    &=& \sum_{i_1=1}^N \cdots \sum_{i_p=1}^N \sum_{\alpha_1=1}^d \cdots \sum_{\alpha_p=1}^d \llangle \prod_{k=1}^p \Delta\xi_{i_k,\alpha_k}(t_n)  \rrangle \llangle \prod_{k=1}^p  \partial_{r^{(k)}_{\alpha_k}}  \delta(\bfr^{(k)}-\bfr_{i_k}(t_n)) \rrangle \;. 
    \label{eq:step1pthmoment}
\end{eqnarray}
In Eq.~\eqref{eq:step1pthmoment} we have used It\=o convention to isolate the average of the noise. Applying Wick's theorem, we conclude that this average is zero for all odd $p$, otherwise: 
\begin{equation}
    p \text{ even:} \quad \llangle \prod_{k=1}^p \Delta\xi_{i_k,\alpha_k}(t_n)  \rrangle = (\Delta t)^{p/2} \langle \xi^p \rangle \prod_{k=2}^p \delta_{i_1 i_k} \delta_{\alpha_1 \alpha_k} \;,
\end{equation}
where $\langle \xi^p \rangle$ is the $p^{\rm th}$-moment of the normal distribution $\mathcal{N}(0,1)$. Eq.~\eqref{eq:step1pthmoment} then simplifies dramatically and we get:
\begin{eqnarray}
    \llangle \prod_{k=1}^p \Delta \zeta(\bfr^{(k)}, t_n) \rrangle
    &=& (\Delta t)^{p/2} \langle \xi^p \rangle \sum_{i_1=1}^N \cdots \sum_{i_p=1}^N \sum_{\alpha_1=1}^d \cdots \sum_{\alpha_p=1}^d \llangle \prod_{k=1}^p \delta_{i_1 i_k} \delta_{\alpha_1 \alpha_k} \partial_{r^{(k)}_{\alpha_k}}  \delta(\bfr^{(k)}-\bfr_{i_k}(t_n)) \rrangle \notag \\[0.2cm] 
    &=& (\Delta t)^{p/2} \langle \xi^p \rangle  \sum_{i=1}^N \sum_{\alpha=1}^d \llangle \prod_{k=1}^p \partial_{r^{(k)}_{\alpha}}  \delta(\bfr^{(k)}-\bfr_{i}(t_n)) \rrangle \label{eq:step2pthmoment} \\[0.2cm] 
    &=& (\Delta t)^{p/2} \langle \xi^p \rangle  \sum_{\alpha=1}^d \left( \prod_{k=1}^p \partial_{r^{(k)}_{\alpha}} \right)\llangle  \sum_{i=1}^N \prod_{k=1}^p \delta(\bfr^{(k)}-\bfr_{i}(t_n)) \rrangle \label{eq:step3pthmoment}
\end{eqnarray} 
To go from Eq.~\eqref{eq:step2pthmoment} to Eq.~\eqref{eq:step3pthmoment} we have used the fact that the derivatives are applied to different variables $\bfr^{(k)}_{\alpha_k}$, so the product of the derivatives of can be replaced by the application of all the derivatives on the product of the functions. The sum over $i$ can then be brought inside, and we can use the properties of Delta functions to get to the final result: 
\begin{eqnarray}
    \llangle \prod_{k=1}^p \Delta \zeta(\bfr^{(k)}, t_n) \rrangle
    &=& (\Delta t)^{p/2} \langle \xi^p \rangle  \sum_{\alpha=1}^d \left( \prod_{k=1}^p \partial_{r^{(k)}_{\alpha}} \right)\left[ \rho(\bfr^{(1)}, t_n) \prod_{k=2}^p \delta(\bfr^{(1)}-\bfr^{(k)}) \right] \;.
    \label{eq:step4pthmoment}
\end{eqnarray}
Finally, we can compare Eq.~\eqref{eq:step4pthmoment} with the $p$-th order moment of the noise $-\nabla_{\bfr} \cdot \left[\sqrt{\rho(\bfr,t_n)} \Delta\bfLL(\bfr,t_n) \right]$ conditioned on the value of $\rho(\bfr, t_n)$. We have:
\begin{eqnarray}
    && \llangle \prod_{k=1}^p \nabla_{\bfr^{(k)}} \cdot \left[\sqrt{\rho(\bfr^{(k)},t_n)} \Delta\bfLL(\bfr^{(k)},t_n) \right] \rrangle = \prod_{k=1}^p \llangle  \sum_{\alpha=1}^d \partial_{r^{(k)}_{\alpha}}  \left[ \sqrt{\rho(\bfr^{(k)},t_n)} \Delta\Lambda_{\alpha}(\bfr^{(k)},t_n) \right] \rrangle = \notag \\[0.2cm]
    && = \sum_{\alpha_1=1}^d \cdots \sum_{\alpha_p=1}^d \left(\prod_{k=1}^p \partial_{r^{(k)}_{\alpha_k}} \right) \llangle \prod_{k=1}^p \left[ \sqrt{\rho(\bfr^{(k)},t_n)} \Delta\Lambda_{\alpha_k}(\bfr^{(k)},t_n) \right] \rrangle \notag \\[0.2cm]
    && = \sum_{\alpha_1=1}^d \cdots \sum_{\alpha_p=1}^d \left(\prod_{k=1}^p \partial_{r^{(k)}_{\alpha_k}} \right) \left[ \llangle \prod_{k=1}^p  \sqrt{\rho(\bfr^{(k)},t_n)} \rrangle \llangle \prod_{k=1}^p  \Delta\Lambda_{\alpha_k}(\bfr^{(k)},t_n)  \rrangle \right] \notag \\[0.2cm]
    && =  \sum_{\alpha_1=1}^d \cdots \sum_{\alpha_p=1}^d \left(\prod_{k=1}^p \partial_{r^{(k)}_{\alpha_k}} \right) \left[ \prod_{k=1}^p  \sqrt{\rho(\bfr^{(k)},t_n)} \llangle \prod_{k=1}^p  \Delta\Lambda_{\alpha_k}(\bfr^{(k)},t_n)  \rrangle \right] 
    \label{eq:lambdapth1}\;.
\end{eqnarray}
In the last passage we have used the fact that $\rho(\bfr,t_n)$ is fixed. Once again we can use Wick's theorem on the Gaussian noises $\Delta \bfLL$ to write:
\begin{equation}
    \llangle \prod_{k=1}^p \Delta\Lambda_{\alpha_k}(\bfr^{(k)},t_n) \rrangle = (\Delta t)^{p/2}  \langle \xi^p\rangle \prod_{k=2}^p \delta_{\alpha_1 \alpha_k} \delta(\bfr^{(1)}-\bfr^{(k)}) \;,
    \label{eq:lambdapth2}
\end{equation}
where we remind that $\langle \xi^p \rangle$ is the $p$-th moment of the normal distribution. As a consequence, when $p$ is odd the $p$-th moment of is zero. Inserting Eq.~\eqref{eq:lambdapth2} inside~\eqref{eq:lambdapth2} we obtain:
\begin{eqnarray}
      && \llangle \prod_{k=1}^p \nabla_{\bfr^{(k)}} \cdot \left[\sqrt{\rho(\bfr^{(k)},t_n)} \Delta\bfLL(\bfr^{(k)},t_n) \right]  \rrangle = \notag \\[0.2cm]
      && = (\Delta t)^{p/2}  \langle \xi^p\rangle \sum_{\alpha=1}^d \left(\prod_{k=1}^p \partial_{r^{(k)}_{\alpha}} \right) \left[ \llangle \prod_{k=1}^p  \sqrt{\rho(\bfr^{(k)},t_n)} \rrangle \prod_{k=2}^p \delta(\bfr^{(1)}-\bfr^{(k)})  \right] \notag \\[0.2cm]
      && = (\Delta t)^{p/2}  \langle \xi^p\rangle \sum_{\alpha=1}^d \left(\prod_{k=1}^p \partial_{r^{(k)}_{\alpha}} \right) \left[ \rho(\bfr^{(1)},t_n)^{p/2}  \prod_{k=2}^p \delta(\bfr^{(1)}-\bfr^{(k)})  \right] \notag \\[0.2cm]
\end{eqnarray}
}\fi

\newpage

\section{Detail on numerical simulations}
\subsection{Microscopic simulations}
Microscopic simulations of RTPs, ABPs and AOUPs are carried out in $d=1$ and $d=2$ dimensions, in continuous space and with periodic boundary conditions. For non-interacting simulations, at each time-step $dt$ we first compute the space-dependent motility parameters $\{\gamma(\bfr_i)\}$ for each particle $i$. We then update the position $\bfr_i$ by an amount $v_i \bfu_{i} dt + \sqrt{2 D_t dt} \boldsymbol{\Delta\eta}_i$, where $\boldsymbol{\Delta\eta}_i$ is a vector of $d$ independent Gaussian random variables of unit variance and zero mean. Once the particle position has been updated, we update its orientation vector $\bfu_i$ depending on the specific dynamics:
\begin{itemize}
    \item For RTPs, we draw a random number $\delta$ uniformly distributed on $[0,1]$ and compare it with $\eps = \alpha(\bfr_i) dt$. If $\delta <\eps$, we draw a new orientation $\bfu_i'$ from the unit sphere $\mathbb{S}^{d-1}$, otherwise we keep the same orientation and $\bfu_i(t+dt) = \bfu_i(t)$. In $1d$, the orientation $\sigma \in \{\pm1\}$ flips, $\sigma \to -\sigma$, with probability $\eps = \frac{1}{2} \alpha(x_i) dt$.
    \item For ABPs in $2d$, we draw a random number from a Gaussian distribution $\mathcal{N}(0,1)$ and update the orientation angle as  $\theta_i(t+ dt) = \theta_i(t) + \sqrt{2 \Gamma(\bfr_i) dt} \mathcal{N}(0,1)$, with $\bfu_i = (\cos\theta_i, \sin\theta_i)$.
    \item For AOUPs in $2d$, we update the orientation by $\bfe_i(t+dt) = [1-\tau^{-1}(\bfr_i) dt] \bfe_i + \sqrt{\tau^{-1}(\bfr_i) dt} \boldsymbol{\Delta\xi}_i$, where $\boldsymbol{\Delta\xi}_i$ is a $d$-dimensional vector of independent, centered Gaussian random variables with unit variance.
\end{itemize}

\textbf{Quorum-sensing interactions}. For simulations of QS-RTPs in $1d$ we resort to spatial hashing: we divide the simulation domain into $L$ boxes of width $r_{QS} = 1$. To compute the local density $\tilde \rho(x)$ around a particle at position $x$ in box $i$, it then suffices to look at the particles that are in boxes $i-1$, $i$, and $i+1$ to evaluate the convolution entering Eq.~\eqref{eq:rho_tilde_def}. Spatial hashing thus reduces the computational complexity of the algorithm to $\cO(N)$. Once the local density $\tilde \rho(x)$ is known, we can evaluate the speed of the particle using Eq.~\eqref{eq:QS_v_rho}. 

Since the tumbling rate is a constant, we can use a better algorithm than the one described above to implement the tumbling events. At simulation time $t=0$, we sample the next time $t_i$ at which particle  $i$ will flip from an exponential distribution $p(t_i) \propto \exp(-\frac{\alpha_0}{2} t_i)$. The simulation time is then increased from $t$ to $t+dt$ as follows. For each particle $i$, if $t_i>t+dt$, the position of the particle is updated during the next time step, but not its orientation. For any particle $i$ such that $t<t_i<t+dt$, we update the particle position and orientation as follows:
\begin{enumerate}
    \item Until the time $t_i$, only the position of the particle evolves, according to $x_i \to x_i+ \Delta x_i$, with $\Delta x_i = v_i \sigma_i (t_i-t) + \sqrt{2 D_t (t_i-t)} \Delta \eta_i$, with $\Delta \eta_i$ a zero-mean unit-variance Gaussian random variable.
    \item We flip the orientation of the particle and we increment $t_i$ as $t_i \to t_i+ \delta t$, where $\delta t$ is sampled from $p(\delta t) \propto \exp(-\frac{\alpha_0}{2} \delta t)$. 
\end{enumerate}
If need be, we iterate steps 1 \& 2 until $t_i>t+dt$.

\subsection{Mean-squared displacement (MSD)}
To measure the MSD in $1d$ particle simulations, we first let the system relax during a time $\tau_{\rm relax}$, which also corresponds to the initial time of our measurements $t_0 = \tau_{\rm relax}$. Then, we iterate the following steps:
\begin{itemize}
    \item Store the initial position of all particles, $x_{0,i} \equiv x_i(t_0)$.
    \item To compute the particle displacements in a system with PBC, we keep track of the boundary crossings. Whenever a particle crosses the boundary at $x=0$ or $x=L$, we update a variable $n_{\rm cross}(i)$ by $-1$ or $+1$, respectively.
    \item Every $dt_{MSD}$ we measure the squared distance of the particle from its reference position as:
    \begin{equation*}
        \text{MSD}(t_k) = [x_i(t_k)+L n_{\rm cross}(i) - x_{0,i}]^2 \;, \qquad t_k = t_0 + k dt_{MSD} \;.
    \end{equation*}
    We then average this quantity over all particles and obtain the corresponding MSD($t_k$).
    \item After a time $t_{\rm max}$ has elapsed since $t_0$, we save the array MSD$(t_k)$ in a file.
    \item We wait an additional time $d\tau_{MSD}$, then set the current time $t$ as the new initial time $t_0$ for the next MSD measurement. We set both MSD and $n_{\rm cross}$ to 0 and restart the measurements.
\end{itemize}
We eventually average the MSD array over the successive time windows to get the final curve. 

Finally, to span the six decades shown in Fig.~\ref{fig:MSD}, we use two different timesteps: $dt=0.01$ and $dt=0.0001$. The former allows us to sample the long-time regime, and we use it to measure the MSD($t$) for $t>10$; the smaller timestep is used to measure the MSD($t)$ for $t<10$ with a finer resolution. The final curves presented in Sec.~\ref{sec:MSD} are then obtained by superposing the measurements corresponding to these two regimes. 

\subsection{Structure factor}
To measure the static structure factor $S(q)$ in $1d$ simulations, we define the Fourier modes $q_n = \frac{2 \pi}{L_x} n$, with $n \in \left\{1, \dots, N_q - 1\right\}$, where $L_x$ is the domain size. In simulations, we let the system relax for a time $\tau_{\rm relax}$ and then measure the structure factor at intervals of $d\tau_{S_Q}$. For each wavevector $q_n$ we compute the associated Fourier component of the density $\hat{\rho}_n(t)$ as:
\begin{equation}
    \hat{\rho}_n(t) = \sum_{i=1}^N e^{-i q_n x_i(t)} \;,
    \label{app:rho_n}
\end{equation}
where $i$ is the particle index and $x_i$ the particle's position. The structure factor sampled at time $t$ is then given by:
\begin{equation}
    S(q_n, t) = \frac{1}{N} |\hat{\rho}_n(t)|^2 (1-\delta_{n,0})\;. 
    \label{eq:S_q_sampled}
\end{equation}
Note that ($1-\delta_{n,0})$ ensures mass conservation, hence Eq.~\eqref{eq:S_q_sampled} coincides with the definition~\eqref{eq:S_qn} of $S(q)$ given in the main text. The final curve for $S(q_n)$ is obtained by averaging over all samples at different times $t_k=k d\tau_{S_Q}$: $S(q_n) = \langle S(q_n, t_k) \rangle_{t_k} \;$.

\subsection{Spatial correlation}
To compute the theoretical prediction for the spatial correlation function $G(r)$ we need to (inverse) Fourier transform $S(q_n)-1$ according to Eq.~\eqref{eq:S_G}. We remind that $S(q) \to 1$ for $q \to \infty$, hence the need of shifting it by a constant $-1$ before applying the Fourier transform. 

To measure the density-density correlation $G(r)$ in $1d$ simulations, we first measure the pair distribution function $g(r)$,  which gives the probability of finding a particle at distance $[r, r+dr]$ from a particle located at the origin. In a homogeneous and isotropic system at density $\rho_0$, we follow~\cite{hansen2013theory} to relate $G(r)$ and $g(r)$:
\begin{eqnarray}
    G(r) &=& \langle \delta\rho(r) \delta\rho(0) \rangle = \langle \rho(r) \rho(0) \rangle - \rho_0^2 \notag \\[0.2cm]
    &=& \llangle \sum_{i=1}^N \sum_{j\neq i}^N \delta(r-x_i) \delta(x_j)\rrangle + \llangle \sum_{i=1}^N \delta(r-x_i) \delta(x_i)\rrangle -\rho_0^2 \notag\\[0.2cm]
    &=& \rho_0^2 g(r) + \delta(r) \llangle \sum_{i=1}^N \delta(r-x_i) \rrangle -\rho_0^2 = \rho_0^2 \left[ g(r) - 1 \right] + \rho_0 \delta(r) \;.
    \label{eq:rel_g_G}
\end{eqnarray}

In simulations, we proceed as follows:
\begin{itemize}
    \item Choose an interval $[r_{\rm min}, r_{\rm max}]$ over which to sample $g(r)$. This interval is divided into bins of width $dr$.
    \item Choose a subset of $K = N/10$ reference particles. 
    \item Every $dt_{\rm meas}$:
    \begin{itemize}
        \item For each reference particle $k$, we look at its neighbors and measure $n_i^{(k)}$, the number of particles at distance $\Delta$, with $r_{\rm min} + i \> dr \leq \Delta \leq r_{\rm min} + (i+1)  dr$. In $d=1$, the $i^{th}$ bin includes particles located both to the left and to the right of the reference particle.   
        \item We average $n_i^{(k)}$ over the $K$ reference particles: $n_i = \frac{1}{K} \sum_{k=1}^K n_i^{(k)}$.
        \item In $1d$, the pair distribution function $g_i$ for the $i^{th}$ spatial bin can then be computed as~\cite{hansen2013theory}:
    \begin{equation}
        g_i = \frac{1}{\rho_0} \frac{n_i}{2 dr} \;,
    \end{equation}
    which corresponds to the ratio between the density at distance $r_i$ from a reference particle and the homogeneous density $\rho_0$.
    \end{itemize}
    \item We average $g_i$ over the different time measurements to obtain the final curve for $g$. Using Eq.~\eqref{eq:rel_g_G} we compute the associated correlation function $G(r)$.
\end{itemize}

\subsection{Intermediate scattering function}
To measure the intermediate scattering function $F(q,t)$ in our simulations, we first choose a subset of $n$ Fourier modes $q_n \in \{q_{\rm min}, q_{\rm min} + dq, \dots, q_{\rm min}+(n-1) dq \}$. We let the simulation relax for during a time $\tau_{\rm relax}$ and then measure $F(q,t)$ for each mode $q_n$ over a time window $[\tau_i, \tau_i + t_{\rm max}]$, with a time separation $dt_{\rm meas}$ between each measurement inside this interval. The quantity we measure is:
\begin{equation}
    F_i(q_n, t) = \hat{\rho}_n(\tau_i) \hat{\rho}_n(\tau_i+t)\;\;, \quad t \in \{0, dt_{\rm meas}, 2 dt_{\rm meas},\dots, t_{\rm max} \}\;,
    \label{eq:F_q_sampled}
\end{equation}
where $\hat{\rho}_n$ is defined in Eq.~\eqref{app:rho_n}.
After storing the matrix $F_i(q_n, t)$ for the $i^{th}$ time window, we start a new measurement at $\tau_{i+1}=\tau_i + \rmd\tau$. We repeat this procedure to collect samples of $F_i(q,t)$ over $K$ successive time windows. The intermediate scattering function is then given by averaging $F_i(q_n, t)$ over the different samples $i$:
\begin{equation}
    F(q_n, t) = \frac{1}{K} \sum_{i=1}^{K} F_i(q_n,t) \;.
\end{equation}

\newpage

\section{Stationary distributions obtained from coarse-grained equations}

In this Appendix we show how the coarse-grained equations~\eqref{ABP-RTP_dDim::FokkerPlanck_meso},~\eqref{AOUP_dDim::FokkerPlanck_meso} can be used to compute the steady-state distribution of active particles with space-dependent motility parameters. As explained in Sec.~\ref{sec:micro-simul-noninteracting}, the validity of these solutions is restricted to spatial modulations that occur over large spatial scales compared to the persistence length $\ell_p$.

\subsection{ABP-RTPs with position-dependent speed and translational noise}
\label{app:sec:ABP-RTP-Dt}
We consider the case of ABP-RTP with:
\begin{eqnarray*}
    v = v_0(\bfr)\;, \quad \Gamma = \Gamma_0\;, \quad \alpha = \alpha_0
\end{eqnarray*}
and translational noise $D_t$. 
When $D_t = 0$, the associated master equation~\eqref{ABP-RTP_dDim:FP} is exactly solvable at steady state and the corresponding solution reads~\cite{schnitzer_theory_1993}:
\begin{equation}
    \cP_s(\bfr, \bfu) \propto \frac{1}{v_0(\bfr)} \;.
    \label{eq:ABP-RTP_microsol}
\end{equation}
Whenever $D_t > 0$, instead, no exact solution for the microscopic dynamics~\eqref{ABP-RTP_dDim_mixture:FP} is known for generic $v_0(\bfr)$. However, we can look for an approximate solution $\tilde{p}_s(\bfr)$ using the coarse-grained dynamics:
\begin{equation}
    \partial_t \tilde{p}_s(\bfr) = -\nabla_{\bfr} \cdot \left[ \mathbf{V}  \tilde{p}_s(\bfr) - \cD \nabla_{\bfr}  \tilde{p}_s(\bfr) \right] = 0 \;,
    \label{eq:app-cg-ABP-RTP-2}
\end{equation}
with:
\begin{equation}
    \mathbf{V} = - \frac{v_0}{d \left(\alpha_0  + (d-1)\Gamma_{0}\right)} \nabla_\bfr v_0  \quad \text{and}\quad \cD = \frac{v_0^2}{d \left(\alpha_0  + (d-1)\Gamma_{0}\right)} + D_t\ .
\end{equation}
If we look for a flux-less stationary solution of Eq.~\eqref{eq:app-cg-ABP-RTP-2}:
\begin{eqnarray}
 0 &=& \mathbf{V} \tilde{p}_s - \cD \nabla_\bfr \tilde{p}_s \\
 0 &=& \frac{v_0}{d \left(\alpha_0  + (d-1)\Gamma_{0}\right)} \nabla_\bfr v_0 + \left[\frac{v_0^2}{d \left(\alpha_0  + (d-1)\Gamma_{0}\right)} + D_t\right] \nabla_\bfr \log \tilde{p}_s \\[0.2cm]
0 &=& \nabla_\bfr \frac{v_0^2}{2} + \left[v_0^2 +d \left(\alpha_0  + (d-1)\Gamma_{0}\right) D_t \right] \nabla_\bfr \log \tilde{p}_s \;.
 \label{eq:ABP-RTP-cg-general-2}
\end{eqnarray}
Setting $\tilde{p}_s(\bfr) = \exp[-\Phi(\bfr)]$ in Eq.~\eqref{eq:ABP-RTP-cg-general-2} we obtain:
\begin{equation}
   \nabla_\bfr \Phi(\bfr) = \frac{1}{2} \> \dfrac{\nabla_\bfr v_0^2(\bfr)}{v_0^2(\bfr) + d \left(\alpha_0  + (d-1)\Gamma_{0}\right) D_t }\;,
\end{equation}
which is solved by:
\begin{equation}
    \Phi(\bfr) = \frac{1}{2} \log\left[ v_0^2(\bfr) + d \left(\alpha_0  + (d-1)\Gamma_{0}\right) D_t  \right] \;.
\end{equation}
The coarse-grained stationary distribution thus reads:
\begin{equation}
    \tilde{p}_s(\bfr) \propto \frac{1}{v_0(\bfr)} \dfrac{1}{\sqrt{1 + \dfrac{d \left(\alpha_0  + (d-1)\Gamma_{0}\right) D_t }{v_0^2(\bfr)}}} \;.
\end{equation}
Note that for $D_t=0$ the coarse-grained solution recovers exactly the microscopic one~\eqref{eq:ABP-RTP_microsol}.

\bigskip

\subsection{AOUPs with position-dependent speed and translational noise}
\label{app:cg_sols}
We consider the case of an AOUP with $v = v_0(\bfr)$, constant $\omega = \omega_0$ and a constant translational noise $D_t$. 
For finite $D_t > 0$ the associated master equation~\eqref{AOUP_dDim:FP} is not exactly solvable, so we look for an approximate solution $\tilde{p}_s(\bfr)$ of the coarse-grained dynamics:
\begin{equation}
    \partial_t \tilde{p}_s(\bfr) = -\nabla_{\bfr} \cdot \left[ \mathbf{V}  \tilde{p}_s(\bfr) - \cD \nabla_{\bfr} \tilde{p}_s(\bfr) \right] = 0
    \label{eq:app-cg-AOUP-2}
\end{equation}
with:
\begin{equation}
    \mathbf{V} = - \frac{v_0}{d \omega_{0}} \nabla_\bfr v_0  \quad \text{and}\quad \cD = \frac{v_0^2}{d \omega_0 } + D_t\ .
\end{equation}
The stationary solution of Eq.~\eqref{eq:app-cg-AOUP-2} can be found following the same steps as in Sec.~\ref{app:sec:ABP-RTP-Dt}. Eventually one finds:
\begin{equation}
    \tilde{p}_s(\bfr) \propto \frac{1}{v_0(\bfr)} \dfrac{1}{\sqrt{1 + \dfrac{d \omega_0 D_t }{v_0^2(\bfr)}}} \;.
\end{equation}

\newpage

\section{Diffusion-drift approximation for active mixtures}
\label{app:cg_mixtures}

\subsection{ABP-RTP mixture: Single-particle coarse-graining}
In this Appendix we derive the mesoscopic Langevin equation for mixtures of ABP-RTPs starting from the associated master equation~\eqref{ABP-RTP_dDim_mixture:FP}. The derivation is carried out under the assumption of space-dependent motility parameters~\eqref{eq:v_mixture}-\eqref{eq:gamma_mixture}, but once again we remark its relevance for the interacting case, thanks to the frozen-field approximation. \\
To begin with, we expand $\cP_\mu(\bfr,\bfu)$ on the spherical harmonics basis:
\begin{equation}
\cP_\mu(\bfr,\bfu) = \sum_{p=0}^\infty \frac{1}{\Omega} \frac{(d-2+2p)!!}{p!(d-2)!!} \: \bfa_\mu^p(\br)\cdot \harm{p} \;, \qquad \bfa_\mu^p = \int_{\mathbb{S}^{d-1}} \rmd\bfu \cP_\mu \harm{p} = {\llangle \harm{p}  \middle| \cP_\mu \rrangle}\;.
\label{ABP-RTP_dDim:HarmTensDecomp-mixture}
\end{equation}
Integrating Eq.~\eqref{ABP-RTP_dDim_mixture:FP} over all possible orientations $\bfu$ on the unit sphere $\mathbb{S}^{d-1}$, we obtain:
\begin{equation}
    \partial_t {\llangle 1  \middle| \cP_\mu \rrangle} = - \nabla_\bfr \cdot \left[ v_{0\mu} {\llangle \bfu  \middle| \cP_\mu \rrangle}  - \left(\sum_{h=1}^n v_{1\mu}^h \nabla_\bfr c_h\right) \cdot {\llangle \bfu^{\otimes 2}  \middle| \cP_\mu \rrangle} - D_{t\mu} \nabla_\bfr {\llangle 1  \middle| \cP_\mu\rrangle} \right] \;.
\end{equation}
Using the definition~\eqref{ABP-RTP_dDim:HarmTensDecomp-mixture} of harmonic components $\bfa_\mu^p$ and the decomposition $\bfu^{\otimes 2} = \frac{\bI}{d} + \harm{2}$, we obtain:
\begin{equation}
    \partial_t \bfa^0_\mu = - \nabla_\bfr \cdot \left[ v_{0\mu} \bfa^1_\mu - \left(\sum_{h=1}^n v_{1\mu}^h \nabla_\bfr c_h\right) \cdot \left(\bfa^2_\mu + \frac{\bI}{d} \bfa^0_\mu \right) - D_{t\mu} \nabla_\bfr \bfa_\mu^0\right] \;,
    \label{eq:0order-dyn-mixt}
\end{equation}
Next, we multiply our master Eq.~\eqref{AOUP_dDim_mixture:FP} by $\harm{1} = \bfu$ and integrate over $\bfu$ to obtain the dynamics of the first harmonic component: 
\begin{eqnarray}
    \partial_t {\llangle \harm{1}  \middle| \cP_\mu \rrangle} =&&- \nabla_\bfr \cdot \left[ v_{0\mu} {\llangle \bfu^{\otimes 2}  \middle| \cP_\mu \rrangle}  - \left(\sum_{h=1}^n v_{1\mu}^h \nabla_\bfr c_h\right) \cdot {\llangle \bfu^{\otimes 3}  \middle| \cP_\mu \rrangle} - D_{t\mu} \nabla_\bfr {\llangle \harm{1} \middle| \cP_\mu \rrangle} \right] \notag \\[0.2cm]
    &&- \alpha_{0\mu} {\llangle \harm{1}  \middle| \cP_\mu\rrangle} - \left(\sum_{h=1}^n \alpha_{1\mu}^h \nabla_\bfr c_h\right) \cdot {\llangle \bfu^{\otimes 2}  \middle| \cP_\mu \rrangle} + \Gamma_0 {\llangle \harm{1}  \middle| \Delta_\bfu \cP_\mu \rrangle} \notag \\[0.2cm]
    &&+ \left(\sum_{h=1}^n \Gamma_{1\mu}^h \nabla_\bfr c_h\right) \cdot {\llangle \harm{1}  \middle| \Delta_\bfu (\bfu\cP_\mu) \rrangle}
\end{eqnarray}
To simplify this equation, we first write $\bfu^{\otimes 3}$ and $\bfu^{\otimes 2}$ as a combination of harmonic tensors, using Eqs.~\eqref{eq:harmonic0}-\eqref{eq:harmonic4}. Then, using the fact that $\harm{p}$ is an eigenvector of the Laplacian $\Delta_\bfu$ with eigenvalue $-p(p+d-2)$, we obtain:
\begin{eqnarray}
    \partial_t \bfa^1_\mu = &&-\nabla_\bfr \cdot \left[ v_{0\mu} \left(\bfa^2_\mu + \frac{\bI}{d} \bfa^0_\mu \right) - \left(\sum_{h=1}^n v_{1\mu}^h \nabla_\bfr c_h\right) \cdot \left(\bfa^3_\mu + \frac{3}{d+2} \bfa^1_\mu \odot \bI \right) - D_{t\mu} \nabla_\bfr \bfa_\mu^1\right] \notag \\[0.2cm]
    &&- \left[\alpha_{0\mu} + (d-1) \Gamma_{0\mu}\right] \bfa_\mu^1 - \left[\sum_{h=1}^n \left(\alpha_{1\mu}^h + (d-1) \Gamma_{1\mu}^h\right) \nabla_\bfr c_h \right] \cdot \left(\bfa^2_\mu + \frac{\bI}{d} \bfa^0_\mu \right) \;.
\end{eqnarray}
The dynamics of the second harmonic component $\bfa_\mu^2$ can be obtained in a similar way, by multiplying the master Eq.~\eqref{AOUP_dDim_mixture:FP} by $\harm{2}$ and integrating over $\bfu$: 
\begin{eqnarray}
    \partial_t {\llangle \harm{2}  \middle| \cP_\mu \rrangle} =&&- \nabla_\bfr \cdot \left[ v_{0\mu} {\llangle \bfu \otimes \harm{2}  \middle| \cP_\mu \rrangle}  - \left(\sum_{h=1}^n v_{1\mu}^h \nabla_\bfr c_h\right) \cdot {\llangle \bfu^{\otimes 2} \otimes \harm{2}  \middle| \cP_\mu \rrangle} - D_{t\mu} \nabla_\bfr {\llangle \harm{2}  \middle| \cP_\mu \rrangle} \right] \notag \\[0.2cm]
    &&-\alpha_{0\mu} {\llangle \harm{2}  \middle| \cP_\mu \rrangle} - \left(\sum_{h=1}^n \alpha_{1\mu}^h \nabla_\bfr c_h\right) \cdot {\llangle \bfu \otimes \harm{2}  \middle| \cP_\mu \rrangle} + \Gamma_0 {\llangle 
 \harm{2}  \middle| \Delta_\bfu \cP_\mu \rrangle} \notag \\[0.2cm] &&+\left(\sum_{h=1}^n \Gamma_{1\mu}^h \nabla_\bfr c_h\right) \cdot {\llangle \harm{2}  \middle| \Delta_\bfu (\bfu \cP_\mu ) \rrangle}
    \label{eq:harm2-dyn0-mixture}
\end{eqnarray}
Using the expressions~\eqref{eq:harmonic0}-\eqref{eq:harmonic4}, one gets:
\begin{eqnarray}
\bfu\otimes\wh{\bfu^{\otimes 2}} &=& \bfu^{\otimes 3}-\frac{1}{d}\bfu\otimes\bI \quad \>=\> \wh{\bfu^{\otimes 3}} + \frac{3}{d+2}\bfu\odot \bI -\frac{1}{d}\bfu\otimes\bI \;,\\[0.1cm]
\bu^{\otimes 2}\otimes \hu{2} &=& \bu^{\otimes 4} - \frac{1}{d}\bu^{\otimes 2}\otimes \bI \>=\> \hu{4} + \frac{6}{d+4}\hu{2}\odot \bI + \frac{3}{d(d+2)}\bI^{\odot 2} - \frac{1}{d}\hu{2}\otimes\bI -\frac{1}{d^2}\bI^{\otimes 2} \;,
\end{eqnarray}
which can be inserted in Eq.~\eqref{eq:harm2-dyn0-mixture} to obtain:
\begin{eqnarray}
    \partial_t\bfa_\mu^2 &=& -\divr \left[ v_{0\mu}\left( \bfa_\mu^3 + \frac{3}{d+2} \bfa_\mu^1 \odot \bI -\frac{1}{d}\bfa_\mu^1\otimes \bI \right) 
    - \left(\sum_{h=1}^n v_{1\mu}^h \nabla_\bfr c_h\right) \cdot \Big( \bfa_\mu^4 +\frac{6}{d+4}\bfa_\mu^2\odot\bI \right.\notag\\ 
    && \left. \hspace{1.2cm} -\frac{1}{d}\bfa_\mu^2\otimes \bI + \frac{3 \bI^{\odot 2}}{d(d+2)} \bfa^0_\mu- \frac{\bI^{\otimes 2}}{d^2}\bfa^0_\mu \Big) -D_{t\mu} \divr \bfa_\mu^2 \right]\notag \\
    & & -(\alpha_{0\mu} + 2d \Gamma_{0\mu})\bfa_\mu^2 -\left[\sum_{h=1}^n \left(\alpha_{1\mu}^h + 2d \Gamma_{1\mu}^h\right) \nabla_\bfr c_h \right] \cdot \left( \bfa_\mu^3 + \frac{3}{d+2}\bfa_\mu^1 \odot \bI - \frac{1}{d}\bfa_\mu^1\otimes \bI \right) \ .
    \label{ABP-RTP_dDim:a2Dyn-mixture} 
\end{eqnarray}
In general, projecting the master Eq.~\eqref{ABP-RTP_dDim_mixture:FP} onto $\harm{p}$ leads to the following dynamics for the harmonic component $\bfa^p$:
\begin{eqnarray}
\partial_t\bfa^p &=& -\divr \left[ v_{0\mu} {\llangle \bfu \otimes \harm{p} \middle| \cP_\mu \rrangle} - \left(\sum_{h=1}^n v_{1\mu}^h \nabla_\bfr c_h\right) \cdot  {\llangle \bfu^{\otimes 2} \otimes \harm{p}  \middle| \cP_\mu \rrangle} -D_{t\mu} \divr \bfa_\mu^p \right] \notag \\
    & &  -\left[ \alpha_{0\mu} + p(p+d-2) \Gamma_{0\mu} \right] \bfa_\mu^p - \left[\sum_{h=1}^n (\alpha_{1\mu}^h + p(p+d-2) \Gamma_{1\mu}^h )\nabla_\bfr c_h\right] \cdot {\llangle \bfu \otimes \harm{p} \middle| \cP_\mu \rrangle} \; .
    \label{ABP-RTP_dDim_mixture:apDyn} 
\end{eqnarray}
As in the single-species case, the $0^{\rm th}$-order harmonics is a conserved mode that evolves over a slow, diffusive timescale. On the contrary, all modes $p \geq 1$ undergo a fast exponential relaxation with finite relaxation times $[\alpha_{0\mu} + p(p+d-2)\Gamma_{0\mu}]^{-1}$. When studying the diffusive dynamics of $\bfa^0_\bfu$ we therefore assume $\partial_t \bfa_\mu^{p} = 0$ for $p \geq 1$, which leads to:
\begin{eqnarray}
\label{eq:pmode-mixture}
\bfa_\mu^p &=& \cO(\nabla_\bfr)\;, \hspace{1cm} p > 2\\[0.2cm]
\bfa_\mu^2 &=& \cO(\nabla_\bfr^2)\;,\\[0.2cm]
\bfa_\mu^1 &=& -\frac{\gradr (\> v_{0\mu}\bfa_\mu^0 \>) }{d[\alpha_{0\mu} + (d-1)\Gamma_{0\mu}]}- 
\bfa_\mu^0 \sum_{h=1}^n \frac{\alpha_{1\mu}^h + (d-1)\Gamma_{1\mu}^h}{d[\alpha_{0\mu} + (d-1)\Gamma_{0\mu}]} \gradr c_h + \mcO(\nabla_\bfr^2)\;.
\label{eq:1mode-mixture}
\end{eqnarray}
To conclude, we insert Eq.~\eqref{eq:pmode-mixture}, \eqref{eq:1mode-mixture} inside the dynamics~\eqref{eq:0order-dyn-mixt} of the zeroth-order harmonics $\bfa_\mu^0$. In the diffusion-drift approximation, we truncate the resulting equation including all terms up to $\cO(\nabla_\bfr^2)$. Finally, we obtain the dynamics of $\bfa^0_\mu(\bfr,t)$, i.e. the probability of finding a $\mu$-particle at position $\bfr$ as:
\begin{equation}
    \partial_t \bfa^0_\mu = -\nabla_{\bfr} \cdot \left[ \mathbf{V}_\mu \bfa_\mu^0 - \cD_\mu \nabla_{\bfr} \bfa^0_\mu \right] \;,
    \label{ABP-RTP_dDim_mixture:FokkerPlanck_meso}
\end{equation}
where the drift and diffusion coefficients read, respectively:
\begin{equation}
\begin{split}
    & \mathbf{V}_\mu = -\frac{v_{0\mu} \nabla v_{0\mu}}{d\left[\alpha_{0\mu}  + (d-1)\Gamma_{0\mu}\right]} - \frac{1}{d} \sum_{h=1}^n \left[ v_{1\mu}^h + v_{0\mu} \frac{\alpha_{1\mu}^h +(d-1) \Gamma_{1\mu}^h}{\alpha_{0\mu} + (d-1)\Gamma_{0\mu}}\right] \nabla_{\bfr} c_h \\
    & \cD_\mu = \frac{v_{0\mu}^2}{d \left[\alpha_{0\mu}  + (d-1)\Gamma_{0\mu}\right]} + D_{t\mu} \ .
    \label{ABP-RTP_dDim_mixture:drift_diff2}
    \end{split}
\end{equation}
One can then associate to Eq.~\eqref{ABP-RTP_dDim_mixture:FokkerPlanck_meso} the following It\=o-Langevin dynamics for particle $i$ of species $\mu$:
\begin{equation}
    \dot{\bfr}_{i,\mu} = \mathbf{V}_\mu(\bfr_{i,\mu},[\{\rho_\nu\}]) + \nabla_{\bfr_{i,\mu}} \cD_\mu(\bfr_{i,\mu},[\{\rho_\nu\}]) + \sqrt{2  \cD_\mu(\bfr_{i,\mu},[\{\rho_\nu\}]) } \bxi_{i,\mu}(t) \;,
    \label{app:Langevin_meso_mixture}
\end{equation}
where the $\{\bxi_{i,\mu}(t)\}$ are delta-correlated, centred Gaussian white noises.

\bigskip

\subsection{AOUP mixture: Single-particle coarse-graining}
In this Appendix we carry out the coarse-graining for AOUPs with spatial dependent motility parameters according to~\eqref{eq:v_mixtureAOUP},~\eqref{eq:tau_mixture}.
We start from the master equation for $\cP_\mu(\bfr, \be)$:
\begin{eqnarray}
\partial_t\cP_\mu(\br,\be) &=&-\divr \left[ v_{0\mu}(\bfr) \be\cP_\mu -  \be^{\otimes 2} \cdot \left(\sum_{h=1}^n v_{1\mu}^h \nabla_\bfr c_h(\bfr)\right) \cP_\mu  -D_{t\mu}\gradr \cP_\mu \right] \notag \\
&& + \dive \left[ \omega_{0\mu} (\bfr)\be\cP_\mu  + \be^{\otimes 2} \cdot \left( \sum_{h=1}^n \omega_{1\mu}^h \> \nabla_\bfr  c_h(\bfr)\right) \cP_\mu \right] \notag \\[0.2cm]
&& +\frac{1}{d}\Delta_\bfe \left[\omega_{0\mu} (\bfr) \cP_\mu + \be \cdot  \left( \sum_{h=1}^n \omega_{1\mu}^h \> \nabla_\bfr  c_h(\bfr)\right) \cP_\mu \right] \;.
\label{AOUP_dDim_mixture:masterEq}
\end{eqnarray}
As we did in Sec.~\ref{sec:AOUP_dDim}, we define the $p$-th order moment of $\cP_\mu$ as:
\begin{equation}
\bfm^p_\mu(\bfr) \equiv \int_{\mathbb{R}^d} \rmd \be \> \be^{\otimes p} \cP_\mu(\bfr, \be) = {\llangle \BFE{p}  \middle| \cP_\mu \rrangle} \; .
\end{equation}
In particular, we note that $\bfm_\mu^0(\bfr, t)$ corresponds to the probability of finding a particle of type $\mu$ at position $\bfr$ at time $t$, marginalized over all possible orientations $\bfe$. \\
Multiplying the master Eq.~\eqref{AOUP_dDim_mixture:masterEq} by $\be^{\otimes p}$ and integrating over $\bfe$, we obtain:
\begin{eqnarray}
\partial_t {\llangle \BFE{p}  \middle| \cP_\mu 
\rrangle} &=&-\divr \left[ v_{0\mu} {\llangle \BFE{p+1}  \middle| \cP_\mu \rrangle} -  \left(\sum_{h=1}^n v_{1\mu}^h \nabla_\bfr c_h\right) \cdot {\llangle \BFE{p+2}  \middle| \cP_\mu  \rrangle}  -D_{t\mu}\gradr {\llangle \BFE{p} \middle| \cP_\mu  \rrangle} \right] \notag \\
&& + {\llangle \BFE{p}  \middle| \dive \left[ \omega_{0\mu} \be\cP_\mu  + \be^{\otimes 2} \cdot \left( \sum_{h=1}^n \omega_{1\mu}^h \> \nabla_\bfr  c_h\right) \cP_\mu \right] 
\rrangle} \notag \\[0.2cm]
&&  +\frac{1}{d} {\llangle
 \> \BFE{p}  \middle|\Delta_\bfe \left[ \omega_{0\mu} \cP_\mu + \be \cdot \left( \sum_{h=1}^n \omega_{1\mu}^h \> \nabla_\bfr  c_h\right) \cP_\mu \right] \rrangle} \;.
\label{AOUP_dDim_mixture:mp}
\end{eqnarray}

Using the results obtained in Sec.~\ref{sec:AOUP_dDim}:
\begin{eqnarray}
  \omega_{0\mu} {\llangle  \BFE{p}  \middle| \dive (  \be\cP_\mu ) \rrangle} &=& -p  \, \omega_{0\mu} \bfm^p_\mu \\[0.2cm]
 \left( \sum_{h=1}^n \omega_{1\mu}^h \> \nabla_\bfr  c_h\right) \cdot {\llangle \BFE{p}  \middle| \dive ( \be^{\otimes 2}  \cP_\mu ) \rrangle} &=& -p  \left( \sum_{h=1}^n \omega_{1\mu}^h \> \nabla_\bfr  c_h\right) \bfm^{p+1}_\mu \\[0.2cm]
 \omega_{0\mu} {\llangle \BFE{p}  \middle| \Delta_\bfe \cP_\mu \rrangle}  &=& p(p-1) \omega_{0\mu} \bfm_\mu^{p -2} \odot \bI \\[0.2cm]
  \left( \sum_{h=1}^n \omega_{1\mu}^h \> \nabla_\bfr  c_h\right) \cdot {\llangle  \BFE{p}  \middle| \Delta_\bfe ( \be  \cP_\mu ) \rrangle} &=& p(p-1)  \left( \sum_{h=1}^n \omega_{1\mu}^h \> \nabla_\bfr  c_h\right) \bfm^{p-1}_\mu \odot \bI
\end{eqnarray}
we can eventually rewrite Eq.~\eqref{AOUP_dDim_mixture:mp} as:
\begin{eqnarray}
\partial_t \bfm^p_\mu &=&-\divr \left[ v_{0\mu}  \bfm^{p+1}_\mu  -  \left(\sum_{h=1}^n v_{1\mu}^h \nabla_\bfr c_h\right) \cdot \bfm^{p+2}_\mu  -D_{t\mu}\gradr \bfm^{p}_\mu \right]  -p \, \omega_{0\mu} \bfm^p_\mu-p  \left( \sum_{h=1}^n \omega_{1\mu}^h \> \nabla_\bfr  c_h\right) \bfm^{p+1}_\mu  \notag \\[0.2cm]
&&  + p(p-1) \frac{\omega_{0\mu}}{d} \bfm_\mu^{p -2} \odot \bI + p(p-1)  \frac{1}{d} \left( \sum_{h=1}^n \omega_{1\mu}^h \> \nabla_\bfr  c_h\right) \bfm^{p-1}_\mu \odot \bI
\label{AOUP_dDim_mixture:pMomentDyn}
\end{eqnarray}
From Eq.~\eqref{AOUP_dDim_mixture:pMomentDyn} we find that the zeroth-order mode is a slow conserved field, whose dynamics is given by:
\begin{eqnarray}
\partial_t \bfm^0_\mu &=&-\divr \left[ v_{0\mu}  \bfm^{1}_\mu  -  \left(\sum_{h=1}^n v_{1\mu}^h \nabla_\bfr c_h\right) \cdot \bfm^{2}_\mu  -D_{t\mu}\gradr \bfm^{0}_\mu \right]\;.
\label{AOUP_dDim_mixture:m0Dyn}
\end{eqnarray}
Conversely, all modes $\bm_\mu^p$ with $p \geq 1$ relax exponentially fast with a characteristic time $(p\,\omega_{0\mu})^{-1}$. 
We therefore use a fast--variable approximation for all $\bm_\mu^p$, $p\geq 1$, setting and set $\partial_t\bm_\mu^p$ for all $p \geq 1$ to zero in Eq.~\eqref{AOUP_dDim_mixture:pMomentDyn}:
\begin{eqnarray}
 p \, \omega_{0\mu} \bm^p &=& -\divr \left[ v_{0\mu}  \bfm^{p+1}_\mu  -  \left(\sum_{h=1}^n v_{1\mu}^h \nabla_\bfr c_h\right) \cdot \bfm^{p+2}_\mu  -D_{t\mu}\gradr \bfm^{p}_\mu \right]  -p  \left( \sum_{h=1}^n \omega_{1\mu}^h \> \nabla_\bfr  c_h\right) \bfm^{p+1}_\mu  \notag \\[0.2cm]
 &&  + p(p-1) \frac{\omega_{0\mu}}{d} \bfm_\mu^{p -2} \odot \bI + p(p-1)  \frac{1}{d} \left( \sum_{h=1}^n \omega_{1\mu}^h \> \nabla_\bfr  c_h \right) \bfm^{p-1}_\mu \odot \bI
\label{AOUP_dDim_mixture:pMomentFastApprox}
\end{eqnarray}
Eq.~\eqref{AOUP_dDim_mixture:pMomentFastApprox} provides a bound on the scaling of the moments $\bm_\mu^p$ in gradients:
\begin{equation}
    \forall l \in \mbN \ , \quad \bm_\mu^{2l}=\mcO(1) \quad \text{while} \quad \bm_\mu^{2l+1} = \mcO(\gradr) \ ,
\end{equation}
as well as the more precise scalings of $\bm_\mu^2$:
\begin{equation}
    \bm_\mu^2 = \frac{1}{d}\bm_\mu^0 \bI + \mcO(\gradr ) \ ,
    \label{AOUP_dDim_mixture:m2Scale}
\end{equation}
and $\bm_\mu^1$:
\begin{equation}
    \bm_\mu^1 = -\frac{1}{d\omega_{0\mu}}\gradr (v_{0\mu} \bm_\mu^0) - \left( \sum_{h=1}^n \omega_{1\mu}^h \> \nabla_\bfr  c_h \right) \bm_\mu^0 +\mcO(\gradr^2)\ .
    \label{AOUP_dDim_mixture:m1Scale}
\end{equation}
Finally, we  insert Eqs.~\eqref{AOUP_dDim_mixture:m2Scale}-- \eqref{AOUP_dDim_mixture:m1Scale} into Eq.~\eqref{AOUP_dDim_mixture:m0Dyn} to close the dynamics of $\bm_\mu^0$. We truncate the resulting equation at the second order in gradient, thus obtaining:
\begin{equation}
    \partial_t \bm_\mu^0 = -\nabla_{\bfr} \cdot \left[ \mathbf{V}_\mu \bm_\mu^0 - \cD_\mu \nabla_{\bfr} \bm_\mu^0 \right] \; ,
\label{AOUP_dDim_mixture:FokkerPlanck_meso}
\end{equation}
where the mesoscopic drifts and diffusivities are respectively given by:
\begin{equation}
    \mathbf{V}_\mu = -\frac{v_{0\mu} \nabla v_{0\mu}}{d\omega_{0\mu}} - \frac{1}{d} \sum_{h=1}^n \left[ v_{1\mu}^h + v_{0\mu}\frac{\omega_{1\mu}^h}{\omega_{0\mu}} \right] \nabla_{\bfr} c_h  \qquad \text{and}\qquad \cD_\mu = \frac{v_{0\mu}^2}{d \omega_{0\mu}} + D_{t\mu} \ .
    \label{AOUP_dDim_mixture:drift_diff2}
\end{equation}
One then associates to Eq.~\eqref{AOUP_dDim_mixture:FokkerPlanck_meso} the following It\=o-Langevin dynamics for particle $i$  species $\mu$:
\begin{equation}
    \dot{\bfr}_{i,\mu} = \mathbf{V}_\mu(\bfr_{i,\mu},[\{\rho_\nu\}]) + \nabla_{\bfr_{i,\mu}} \cD_\mu(\bfr_{i,\mu},[\{\rho_\nu\}]) + \sqrt{2  \cD_\mu(\bfr_{i,\mu},[\{\rho_\nu\}]) } \bxi_{i,\mu}(t) \;,
    \label{app:Langevin_meso_mixture-AOUP}
\end{equation}
where the $\{\bxi_{i,\mu}(t)\}$ are delta-correlated, centred Gaussian white noise terms.

\end{document}